\documentclass{aa}
\usepackage{graphicx}

\begin{document}
\title{Constraining the nature of High Frequency Peakers}
\subtitle{I. The spectral variability}
\author{
%To be written
M. Orienti\inst{1,2} \and
D. Dallacasa\inst{1,2} \and 
C. Stanghellini\inst{2}
}
\offprints{M. Orienti}
\institute{
Dipartimento di Astronomia, Universit\`a di Bologna, via Ranzani 1,
I-40127, Bologna, Italy \and 
Istituto di Radioastronomia -- INAF, via Gobetti 101, I-40129, Bologna,
Italy
}
\date{Received \today; accepted ?}

\abstract
%Context heading
{}
%Aims heading
{We investigate the spectral characteristics of 51 candidate High
  Frequency Peakers (HFPs), from the ``bright'' HFP sample, in order
  to determine the nature of each object, and to obtain a smaller sample of
  genuine young radio sources.}     
%Methods heading
{Simultaneous multi-frequency VLA observations carried out at
  various epochs have been used to detect flux density and spectral
  shape variability in order to pinpoint contaminant objects, since
  young radio sources are not expected to be significantly 
variable on such a short
  time-scale.}  
%Results heading
{From the analysis of the spectral variability we find 13 contaminant objects,
11 quasars, 1 BL Lac, and 1 unidentified object, which we have
rejected from the sample of candidate young radio
sources. The $\sim$6 years elapsed between the first and latest observing run
are not enough to detect any substantial evolution of the overall spectrum of
genuine, non variable, young radio sources. If we also consider 
the pc-scale information, we find that the total
radio spectrum we observe is the result of the superposition of the
spectra of different regions (lobes, hot-spots, core, jets),
instead of a single homogeneous radio component. This
indicates that the radio source structure plays a relevant role in
determining the spectral shape also
in the rather common case in which the morphology appears unresolved
even on high-resolution scales.
}
%Conclusion heading
{}
%}
\titlerunning{Constraining the nature of High Frequency Peakers}
\maketitle
\section{Introduction}
The origin and evolution of radio emission is one of the
greatest challenges in the study of Active Galactic Nuclei.
The onset of radio activity is often thought to be linked to
merger or accretion events in the host galaxy, which provide enough
fuel to feed the central AGN. As a consequence, at
least in its early stages, the radio emission 
evolves in a quite dense and possibly
inhomogeneous ambient
medium which can influence its growth.\\ 
The ideal targets to understand such phenomena are the young radio
sources, whose radio lobes still reside within the innermost region of
the host galaxy.\\
Among them, the smallest (i.e. youngest) objects are the most suitable
to investigate the role played by the host galaxy Interstellar Medium (ISM)
on the evolution/growth of the radio source.\\
The evolutionary stage of the powerful radio sources is related to
their linear sizes. Following self-similar evolution models, the most
compact sources will evolve into the extended radio
source population (Fanti et al. \cite{cf95}; Readhead et
al. \cite{rh96}; Snellen et al. \cite{sn00}). \\
However, it has also been claimed (Alexander \cite{alexander00}; 
Marecki et al. \cite{ma03}) that 
a fraction of young and compact radio sources may die in an early
stage before becoming large scale objects.\\
The early stages in the evolution scheme 
(typical ages $<$ 10$^{4}$ years, Polatidis \& Conway
\cite{pc03}; Murgia \cite{mm03})
are represented by the population of Compact
Symmetric Objects (CSOs). They are a scaled-down version of the large,
powerful radio sources (core-jets-lobes; Wilkinson et al. \cite{wil94}), 
with a well defined peak in their radio spectrum 
at frequencies between
$\sim$ 100 MHz to a few GHz.\\ 
In this framework, the anti-correlation found
between the intrinsic peak frequency and the source size (O'Dea \&
Baum \cite{odea97}) implies that
the youngest sources must be sought among those with spectral peaks
occurring at frequencies higher than 5 GHz, and  
termed ``High Frequency Peakers'' (HFPs,
Dallacasa \cite{dd03}).\\   
The selection of a statistically complete 
sample made of genuine young radio sources only is
essential to study in detail the physical nature of this class of
objects.\\
A sample of candidate HFPs (the ``bright'' sample; Dallacasa et
al. \cite{dd00}) was constructed on the basis of both the shape of the
simultaneous 
radio spectrum and the turnover frequency. However, such selection tools can
introduce a contamination by beamed radio sources. 
For example, blazar objects, although usually characterized by
a flat spectrum, can occasionally 
show a convex spectrum when their radio emission is
dominated by a flaring, self-absorbed component (e.g. Torniainen
  et al. \cite{torni05}).\\
During the majority of their lifetime blazars and young radio sources
display very different characteristics: the former posses significant
flux-density and spectral variability, and the emission,
often polarized, has a Core-Jet morphology on various scales (from
tens of kpc down to pc-scale). The latter have no 
variability and the radio emission, almost completely
unpolarized at least at low frequencies, has a ``Double/Triple''
structure (Orienti et al. \cite{mo06}).\\
In this paper we present the results of simultaneous multi-frequency
VLA observations, made at several epochs, 
in addition to
those already published by Dallacasa et al. (\cite{dd00}) and Tinti et
al. (\cite{st05}). These new observations were carried out 
to further identify blazars with
variability on time-scales longer than the time elapsed between the
previous two epochs (Tinti et al. \cite{st05}), and to study the
evolution of the radio spectrum of those HFPs already confirmed as
genuine young radio sources (Orienti et al. \cite{mo06}).\\ 
In general, they better characterize the properties of the sources of
the ``bright'' HFP sample.\\

\begin{table}
\begin{center}
\begin{tabular}{cccc}
\hline
Date&Conf&Obs Time&code\\
\hline
&&&\\
Sep 12 2003& AnB&240&a\\
Sep 13 2003& AnB&240&b\\
Sep 14 2003& AnB&240&c\\
Sep 15 2003& AnB&620&d\\
%Dec 12 2003& B  &120&e\\
Jan 22 2004& BnC&120&e\\
Jan 26 2004& BnC&180&f\\
Jan 28 2004& BnC&240&g\\
Jan 30 2004& BnC&120&h\\
Mar 21 2004& CnD&120&i\\
&&&\\
\hline
\end{tabular} 
\vspace{0.5cm}
\end{center}   
\caption{VLA observations and configurations. The total observing time
(Column 3) is inclusive of the scans on the HFPs from the ``faint'' sample.} 
\label{obslog}
\end{table}

\begin{figure*}
\begin{center}
\includegraphics{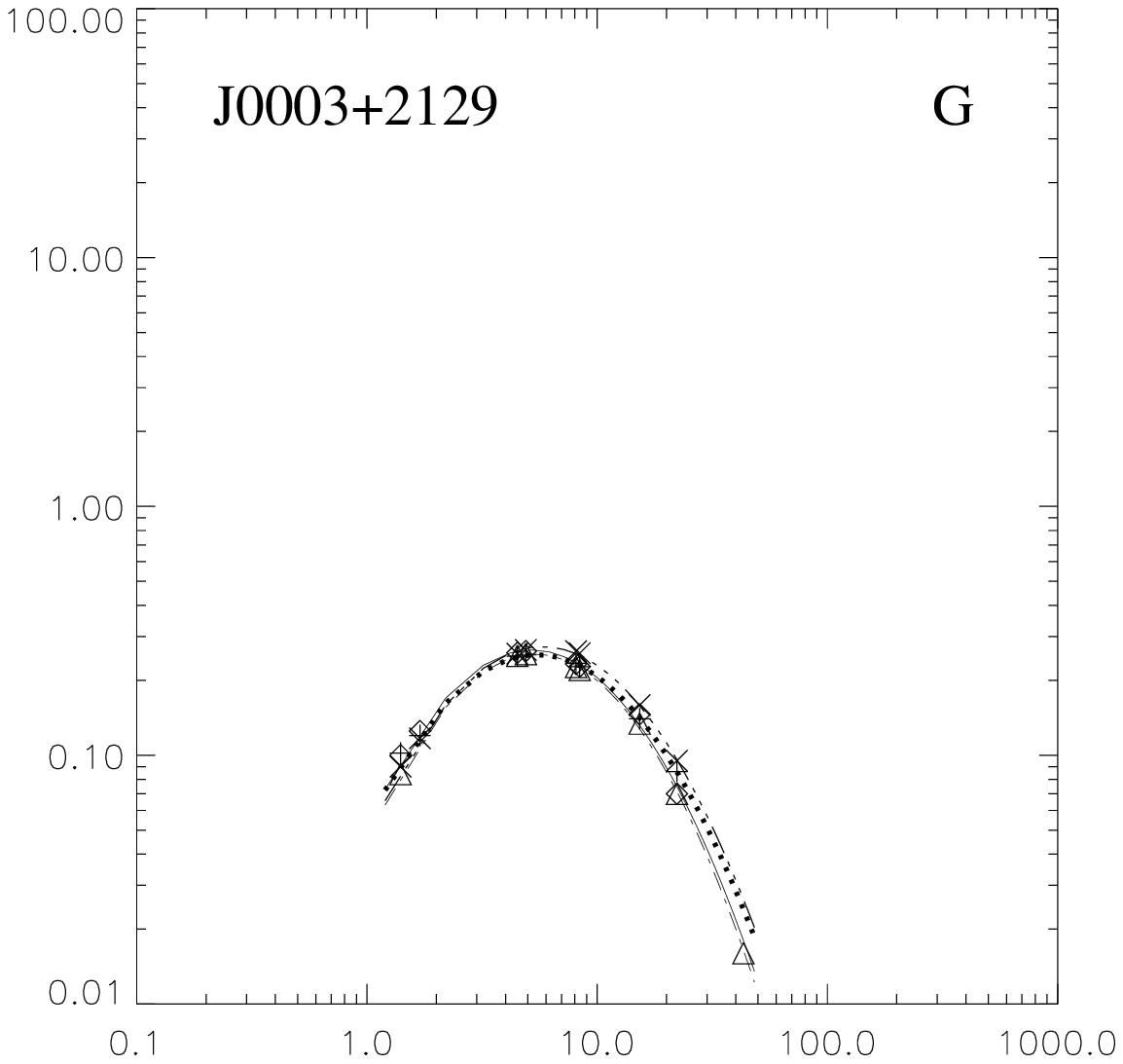}
\includegraphics{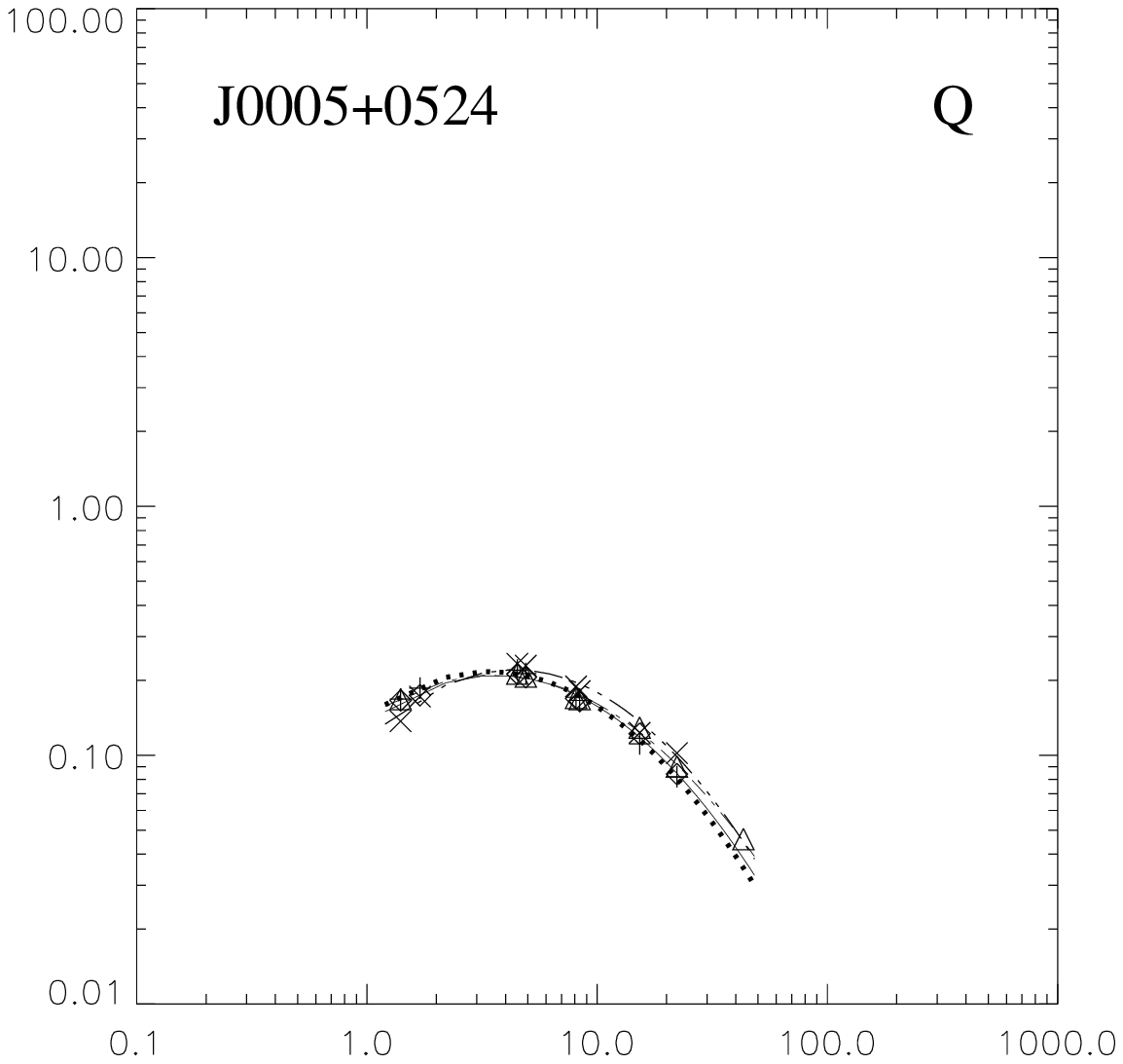}
\includegraphics{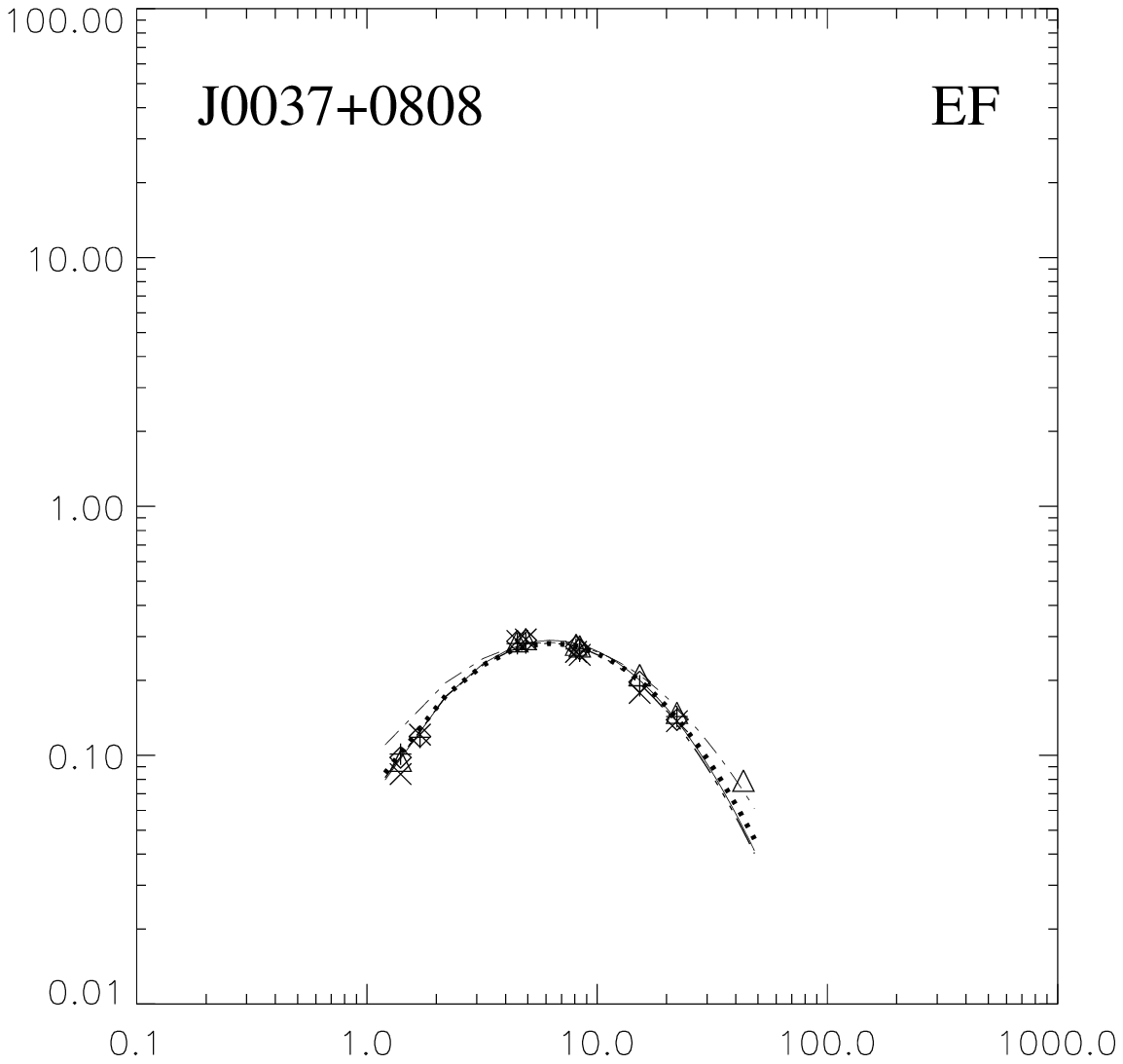}
\includegraphics{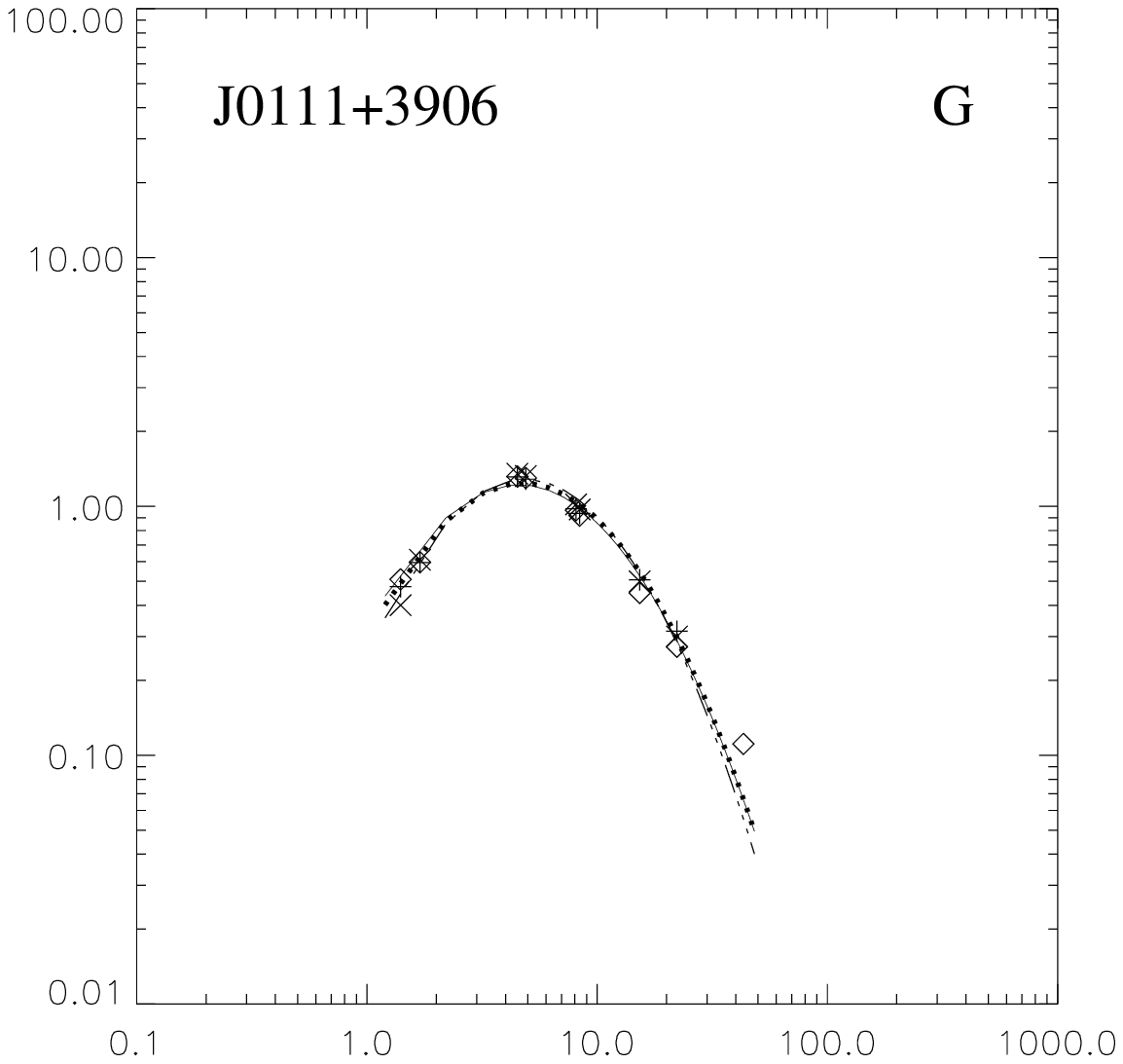}
\includegraphics{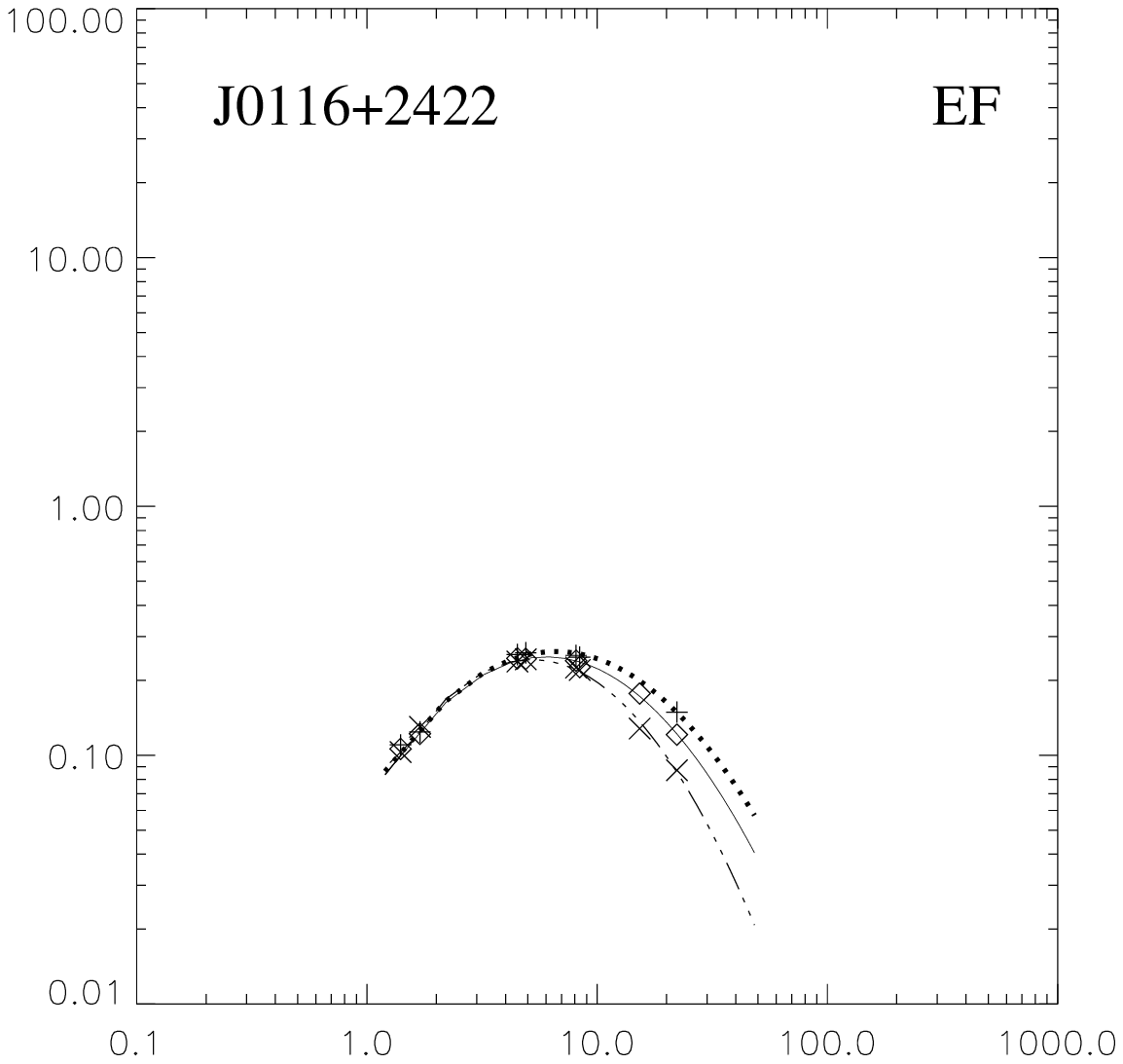}
\includegraphics{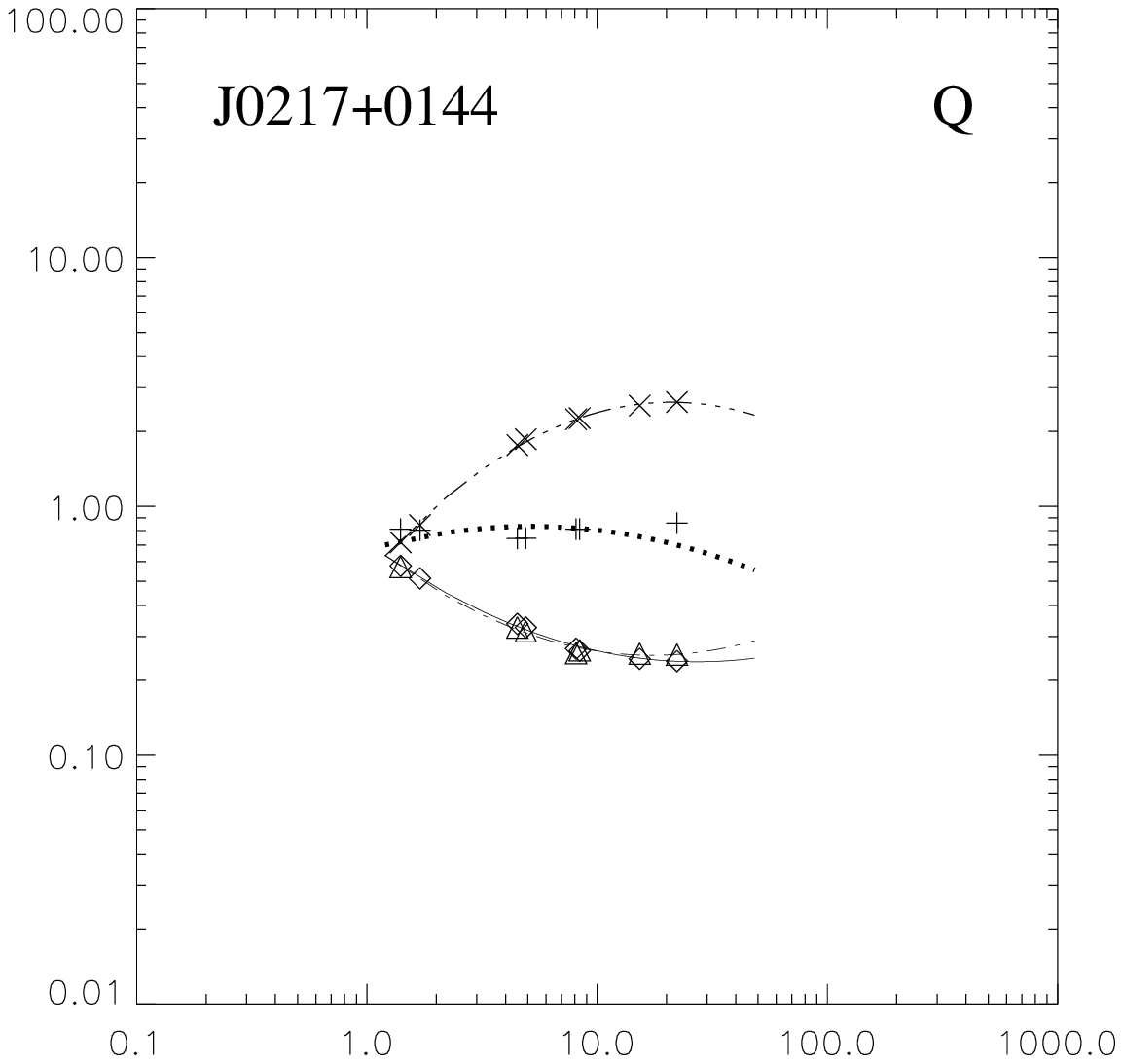}
\includegraphics{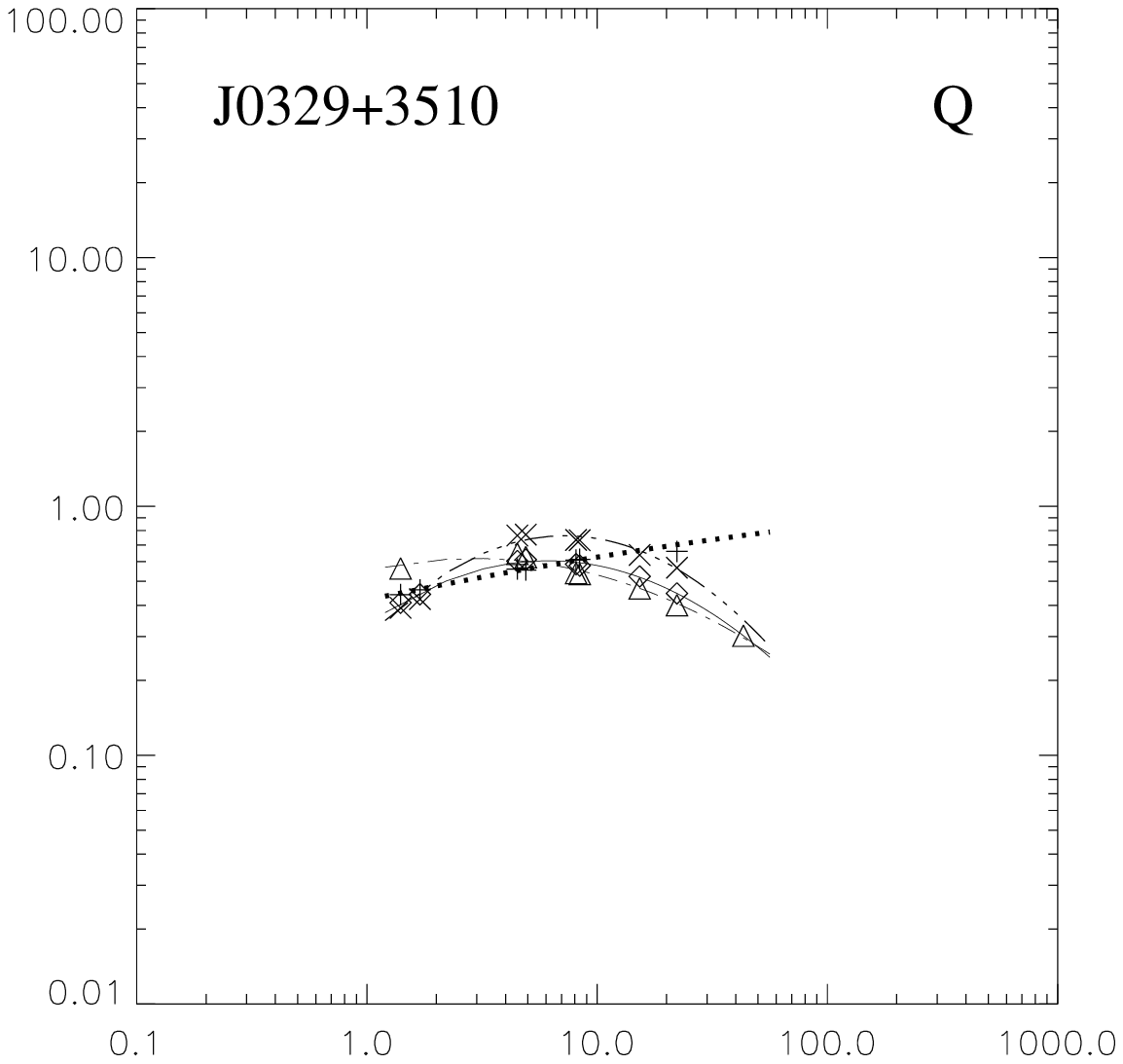}
\includegraphics{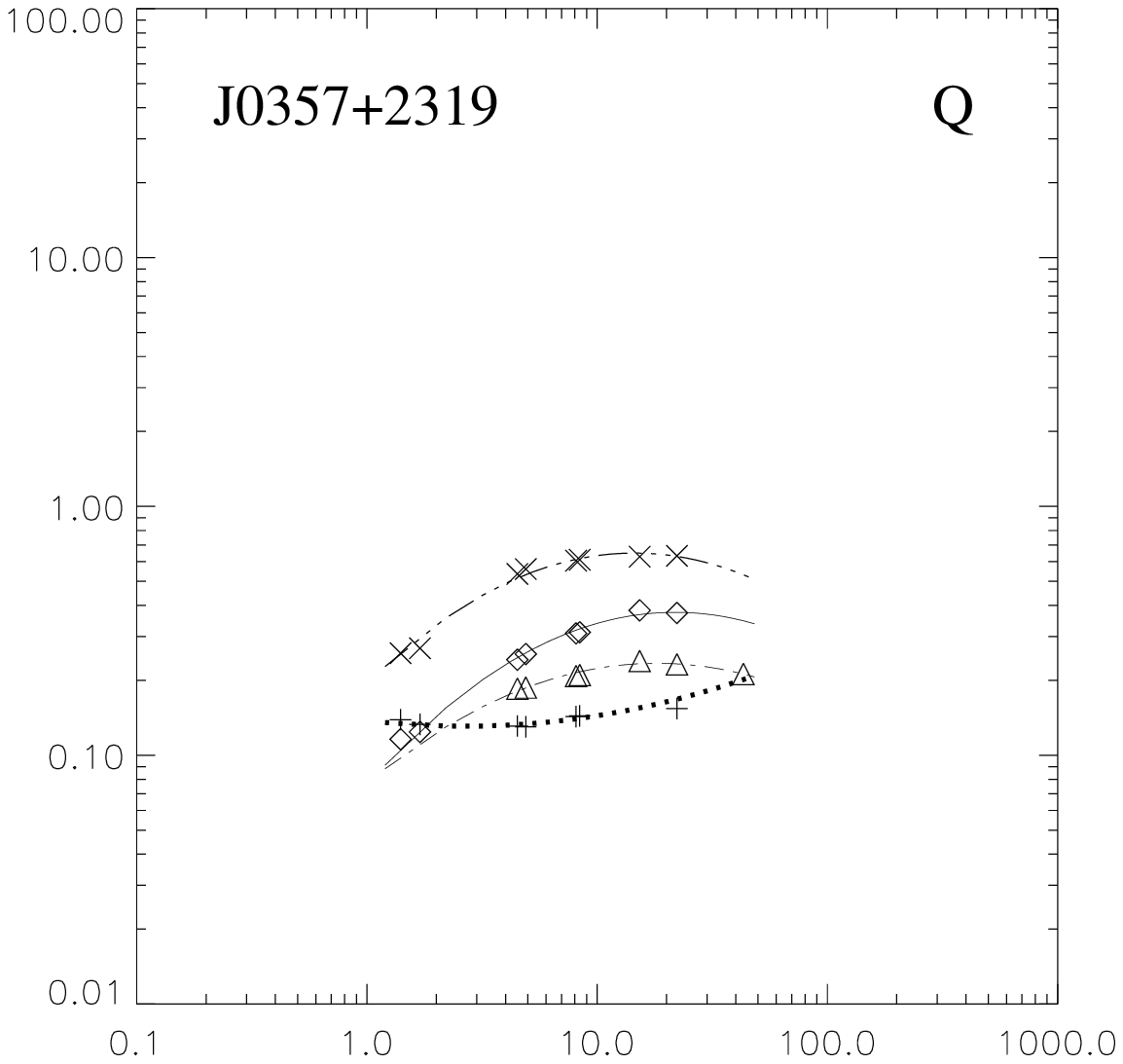}
\includegraphics{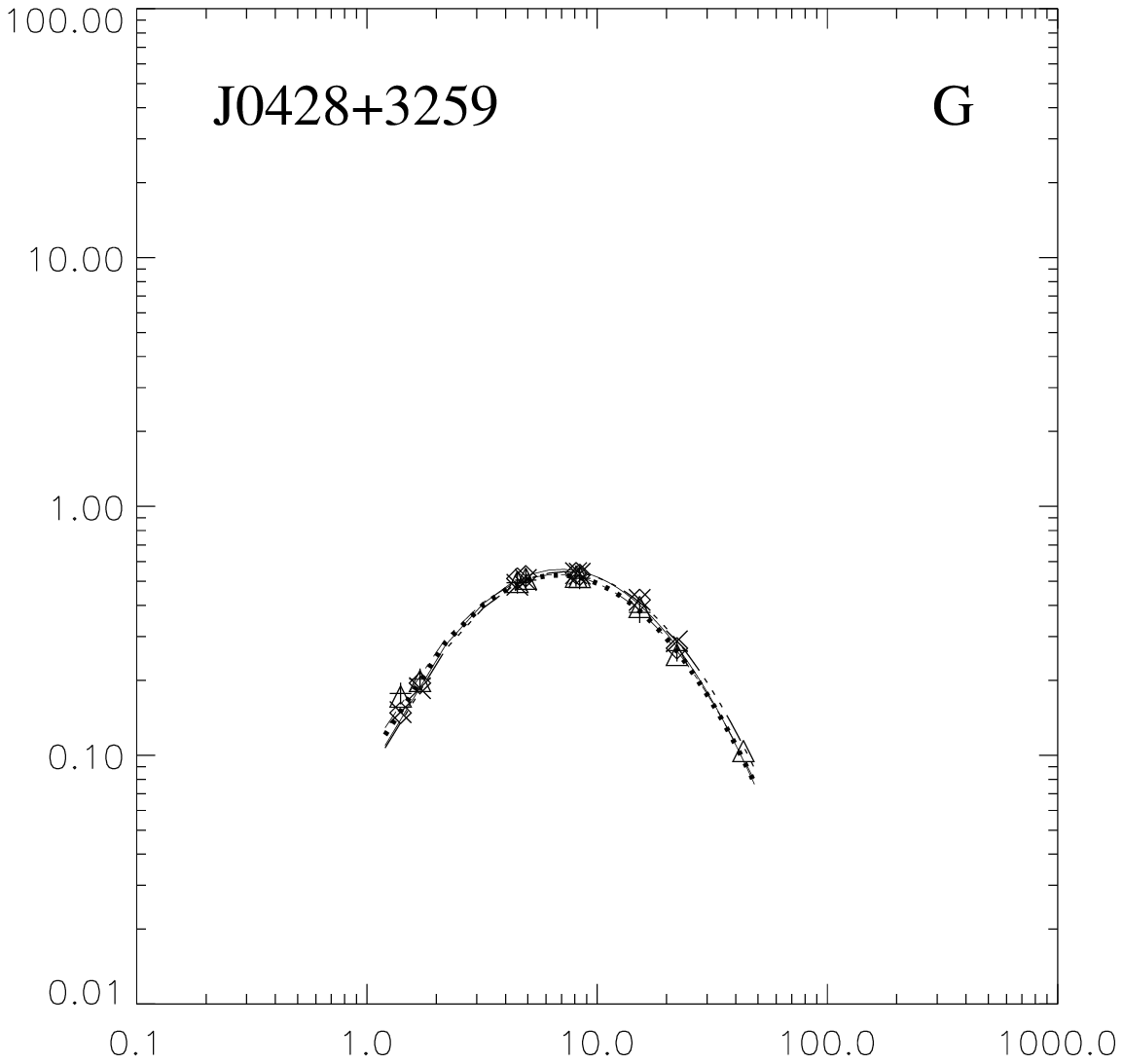}
\includegraphics{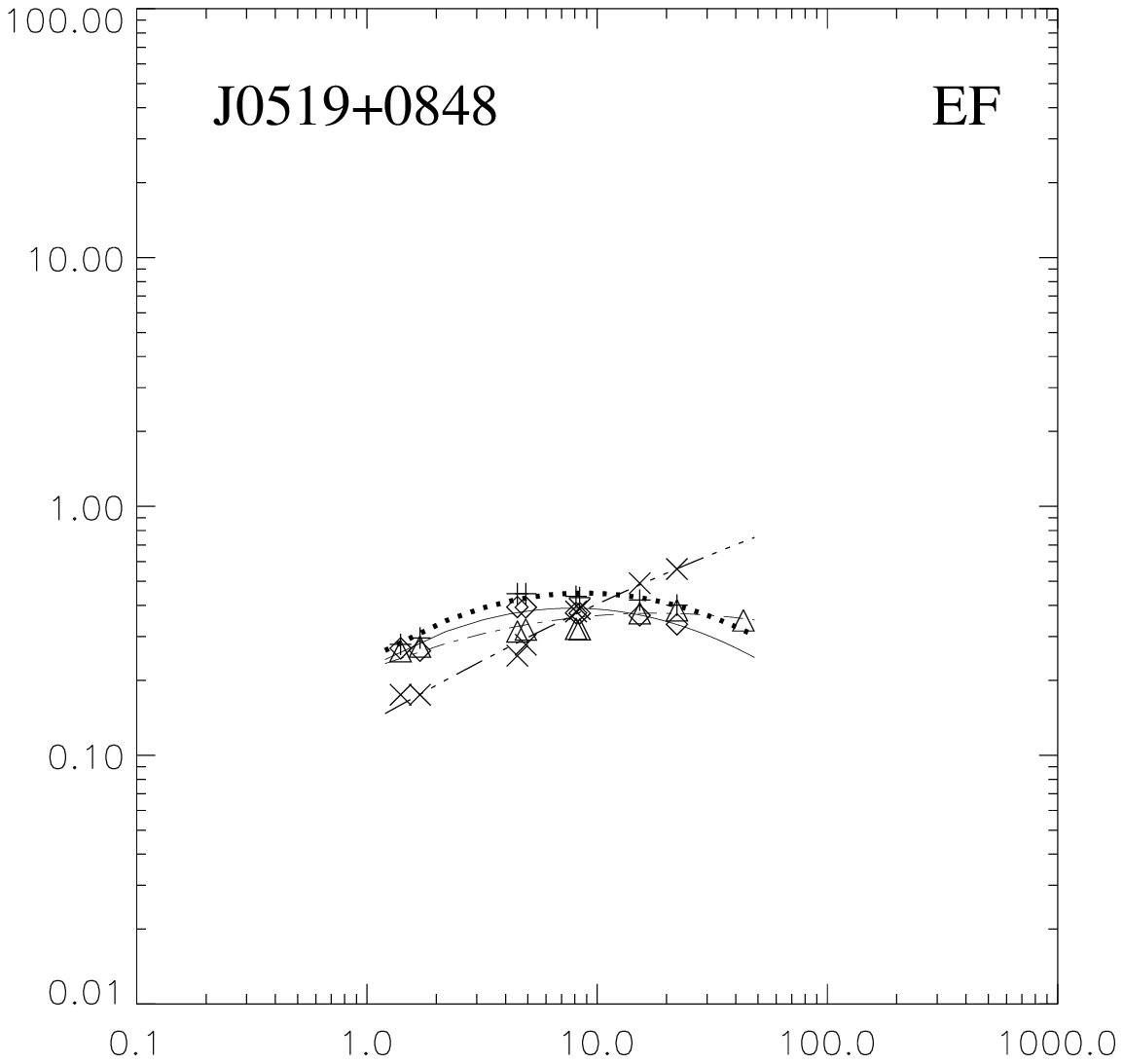}
\includegraphics{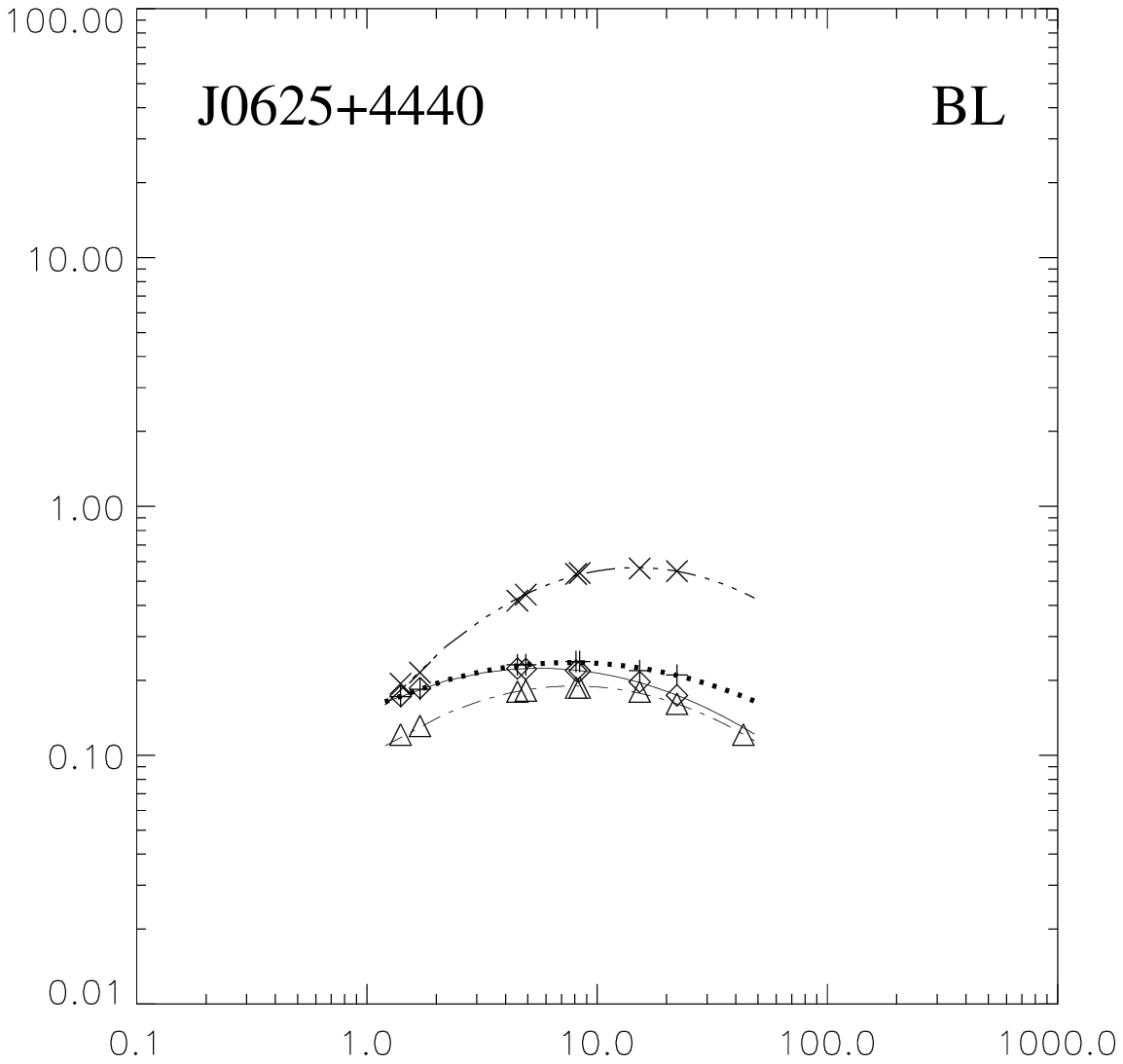}
\includegraphics{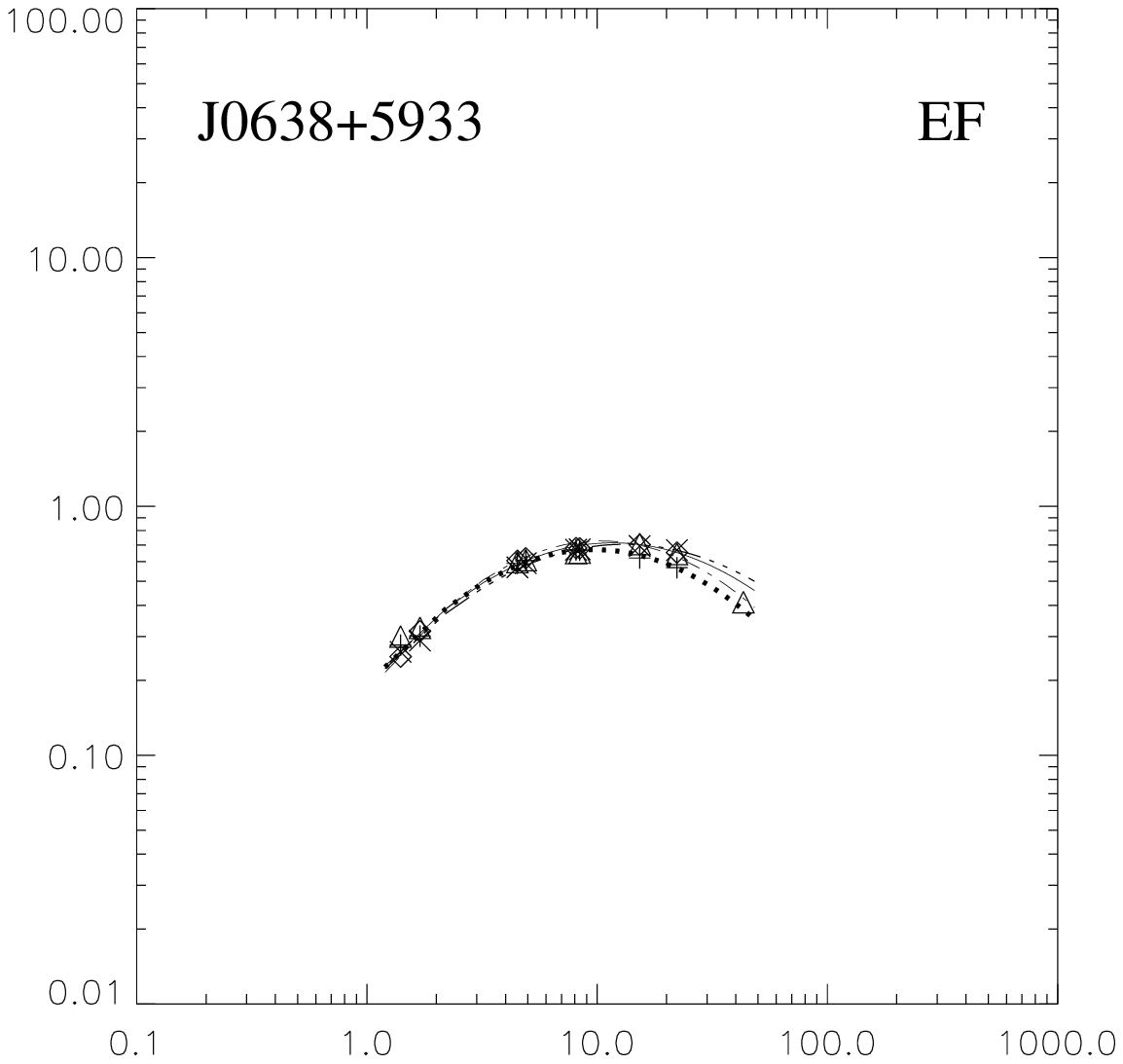}
\includegraphics{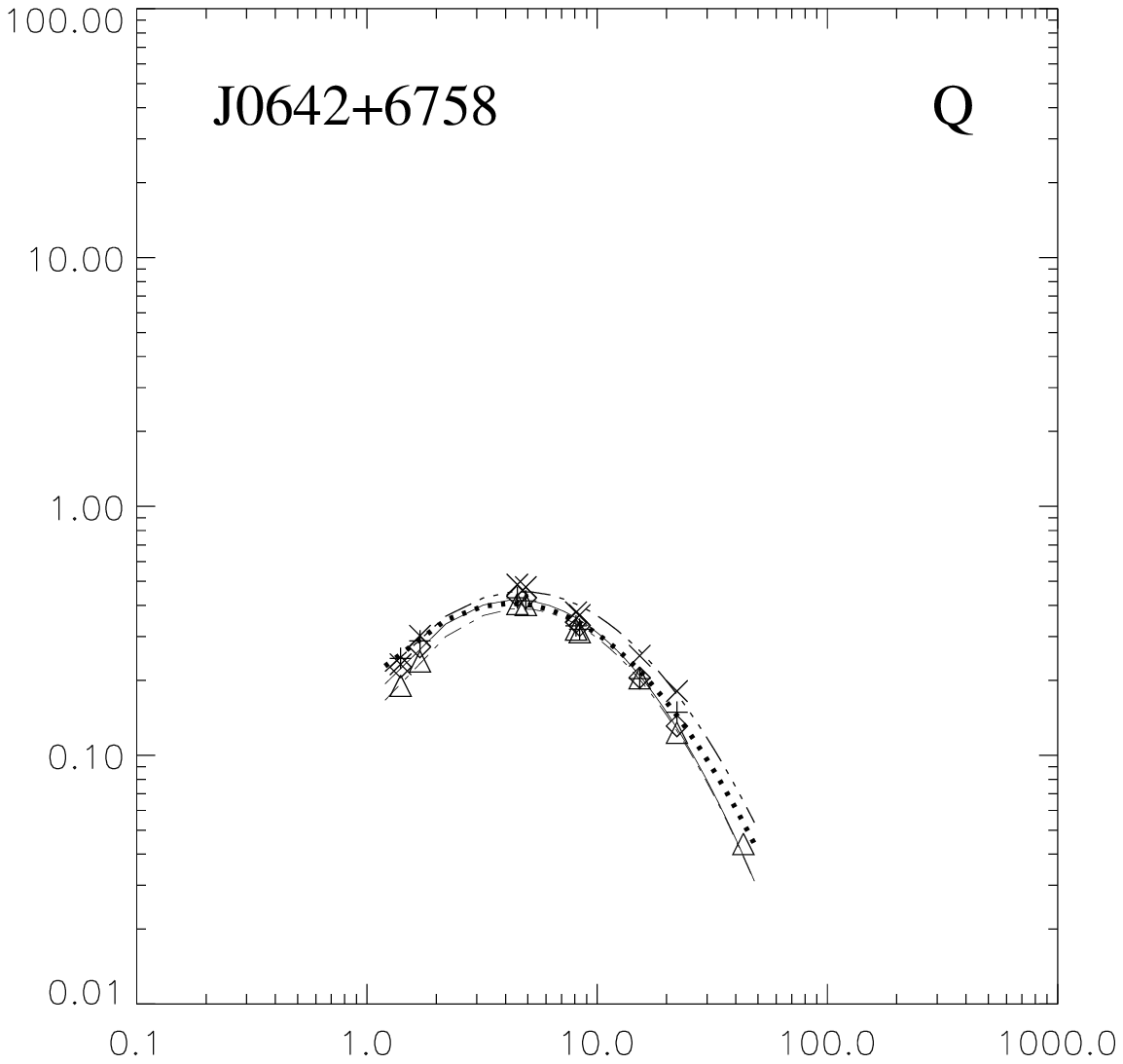}
\includegraphics{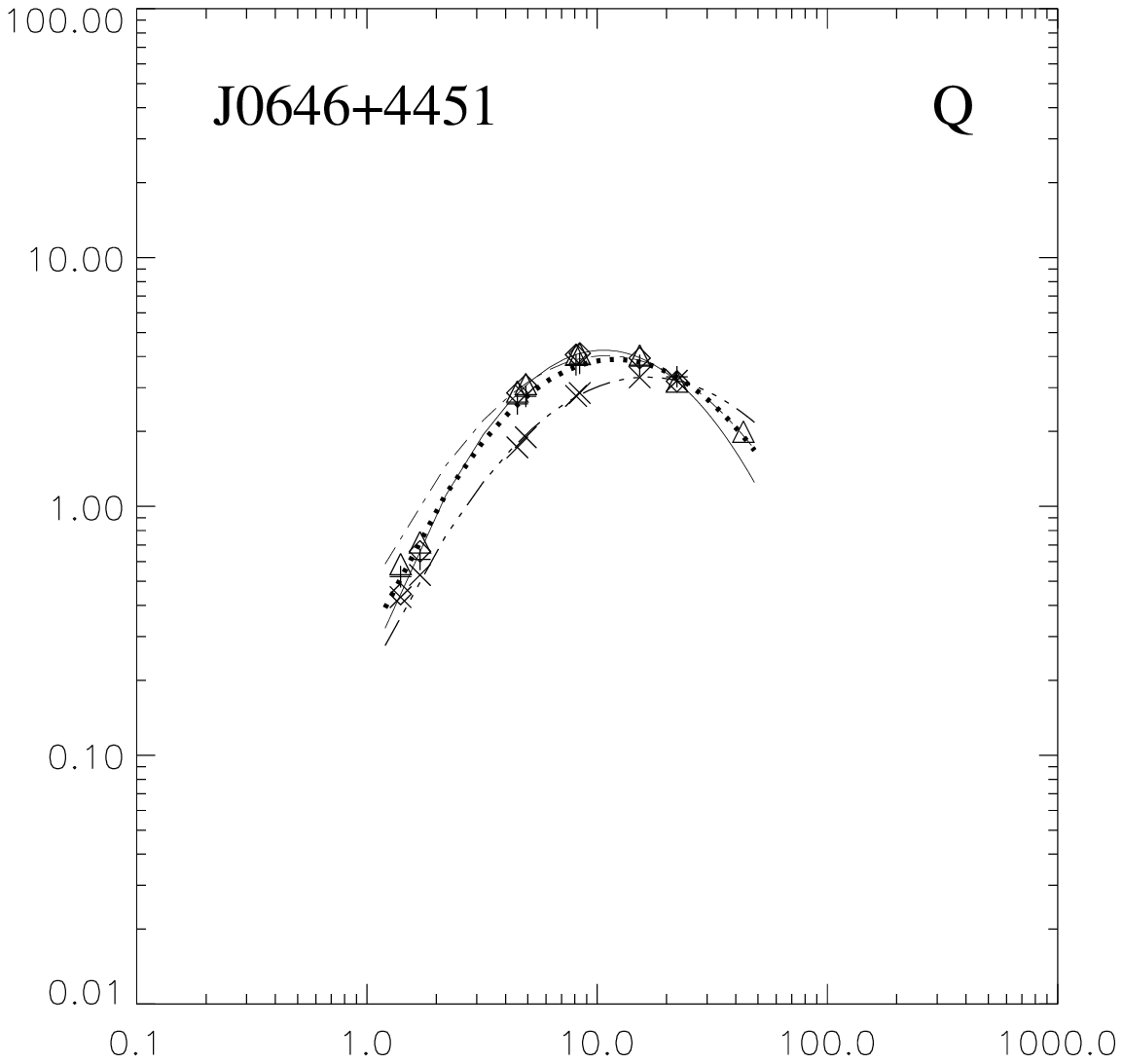}
\includegraphics{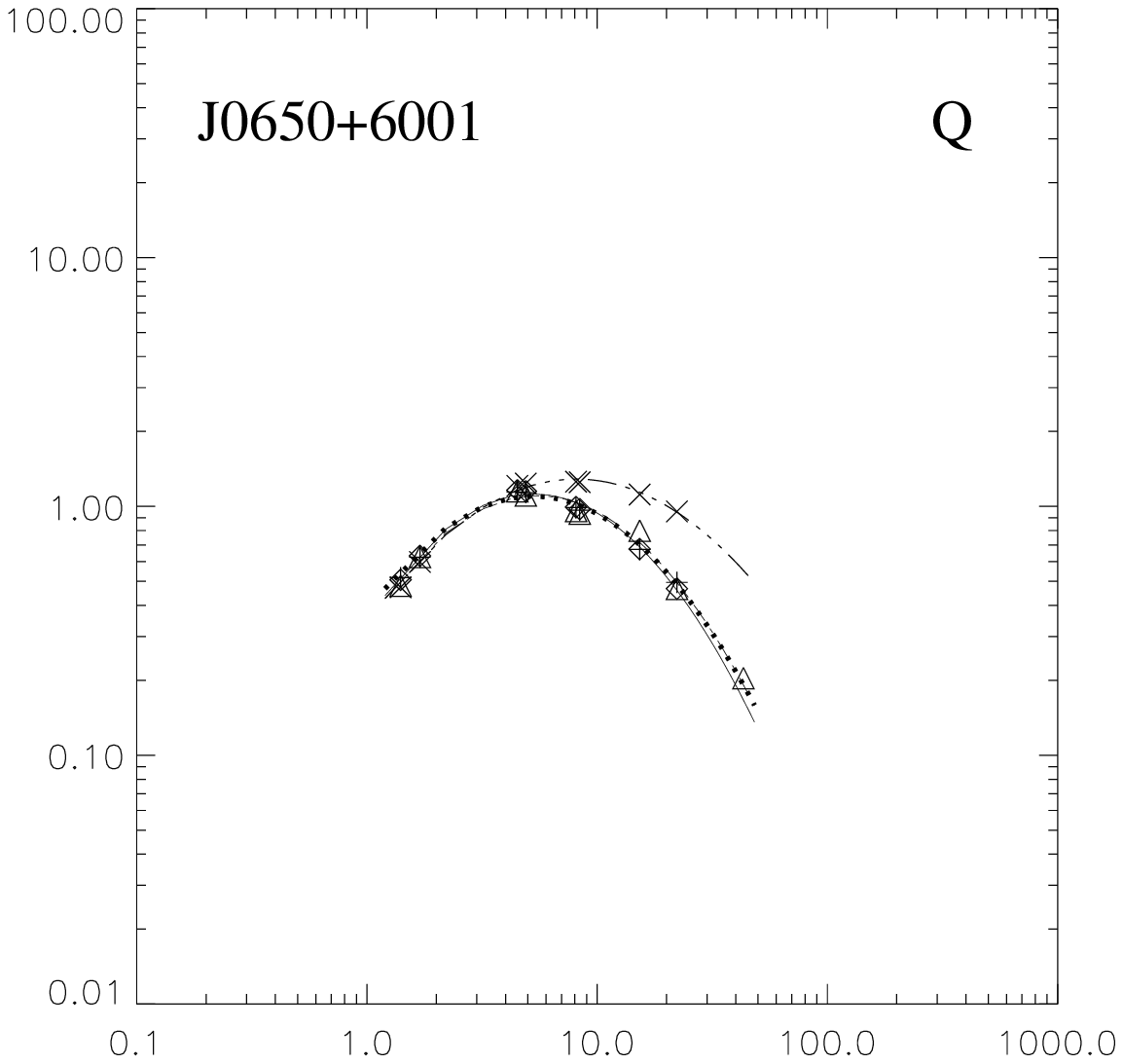}
\includegraphics{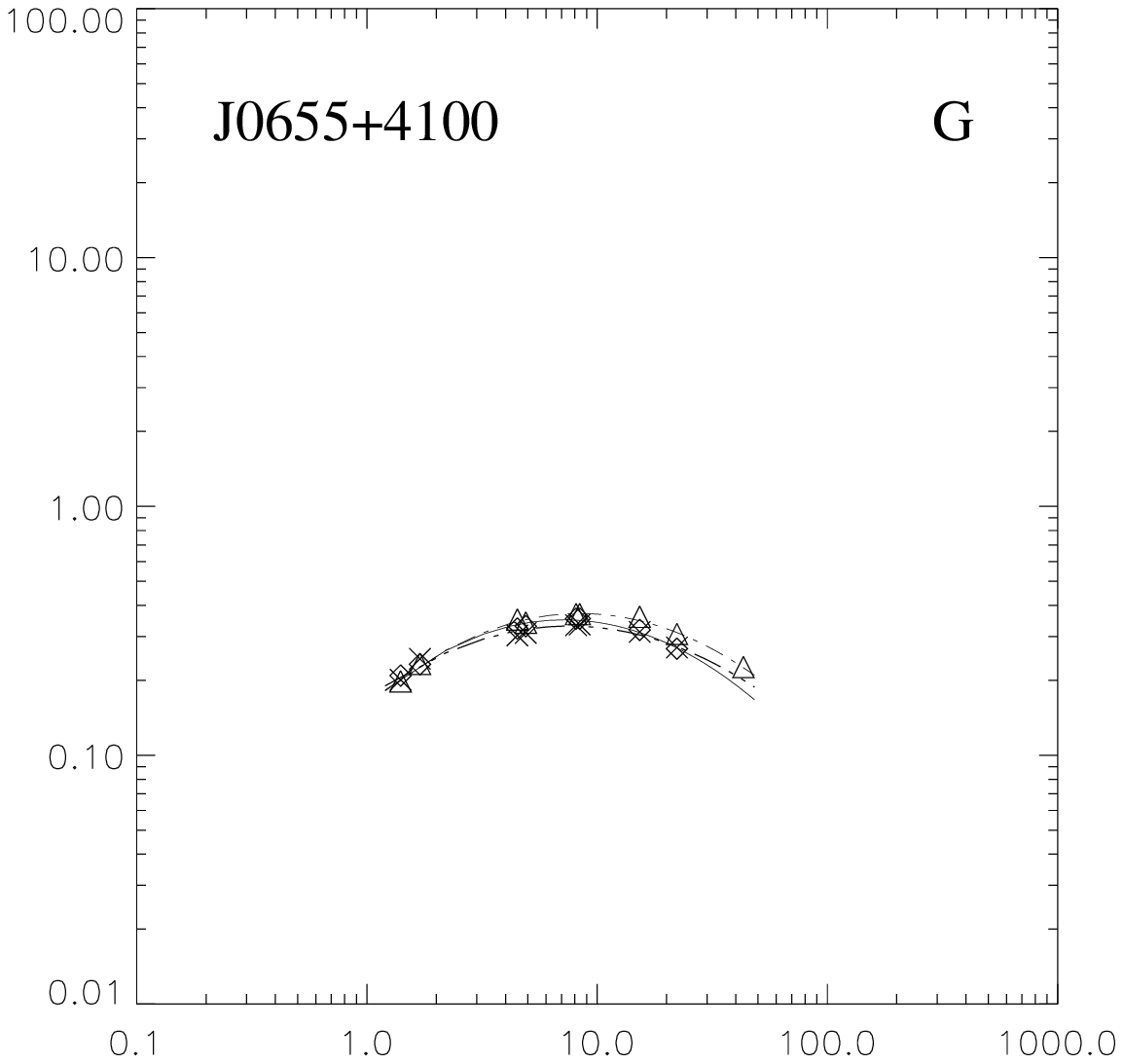}
\includegraphics{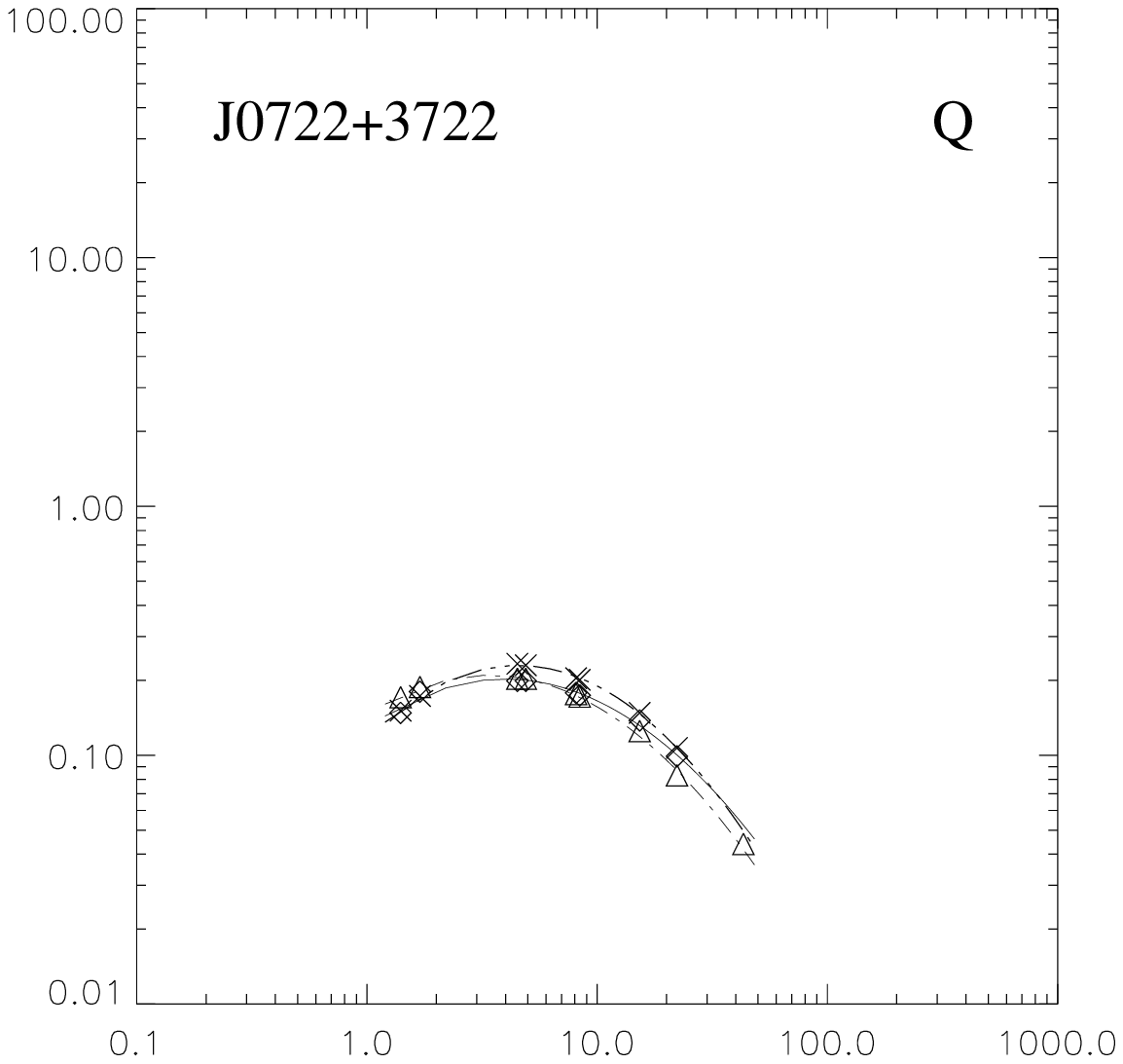}
\includegraphics{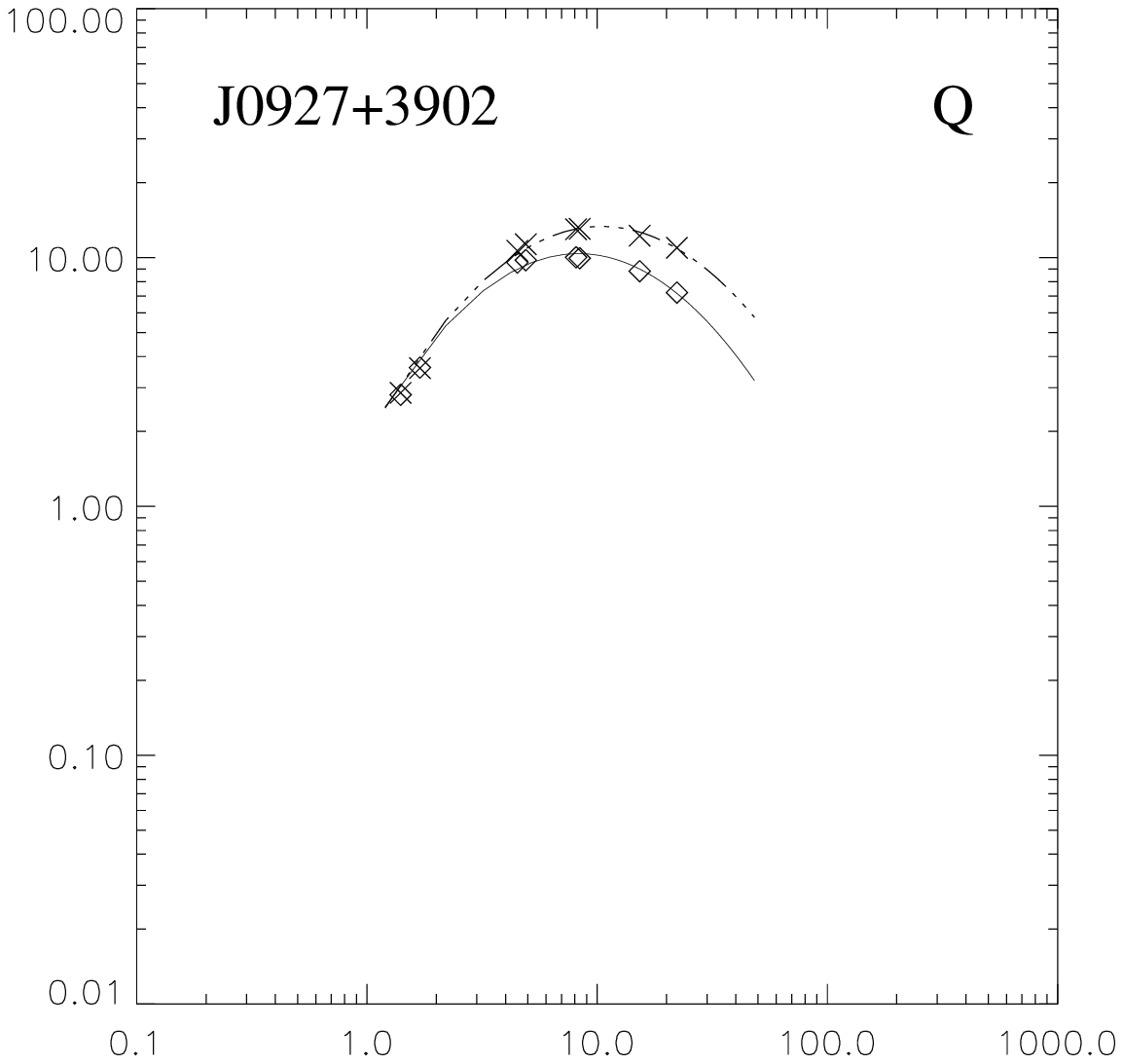}
\includegraphics{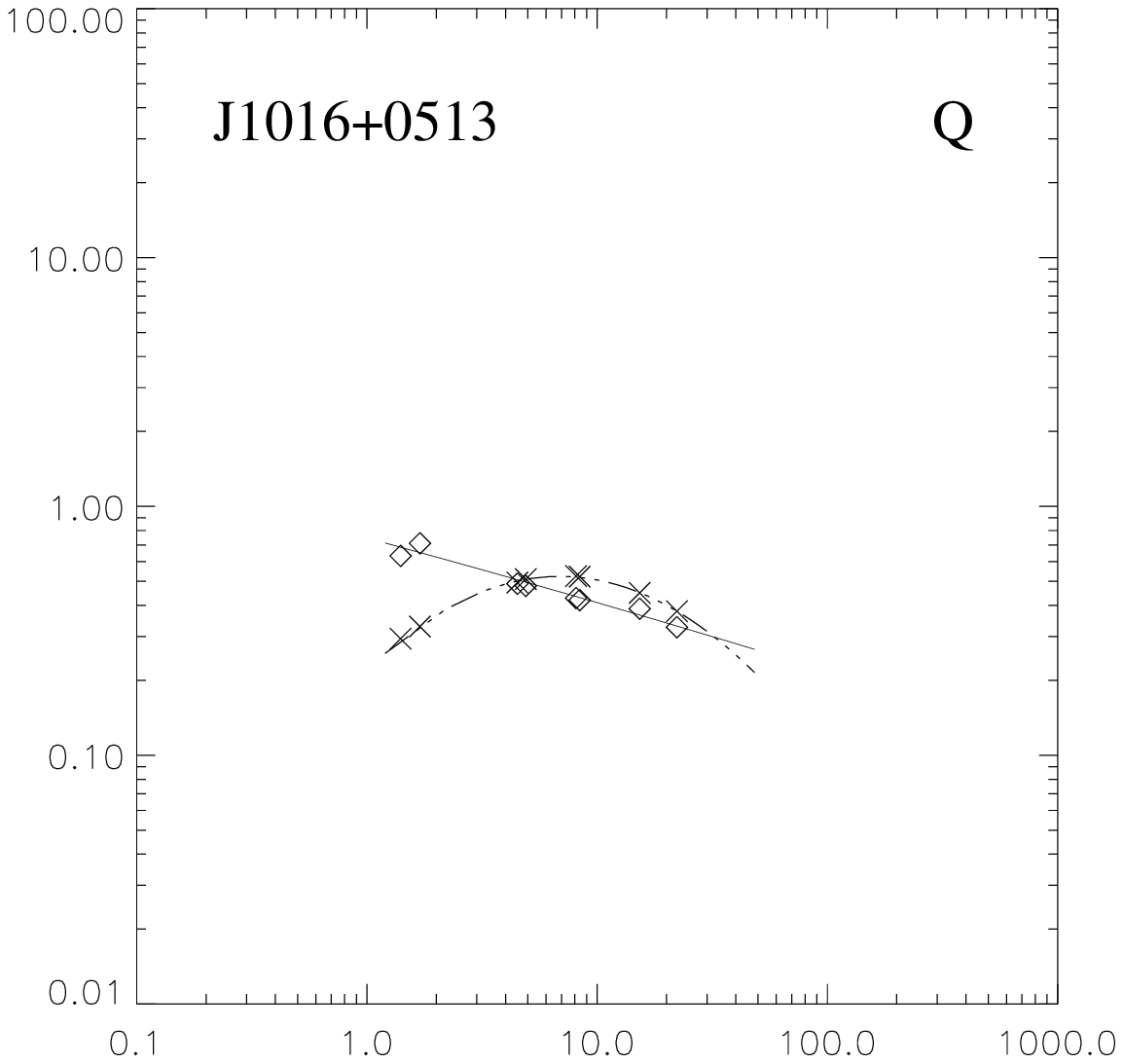}
\includegraphics{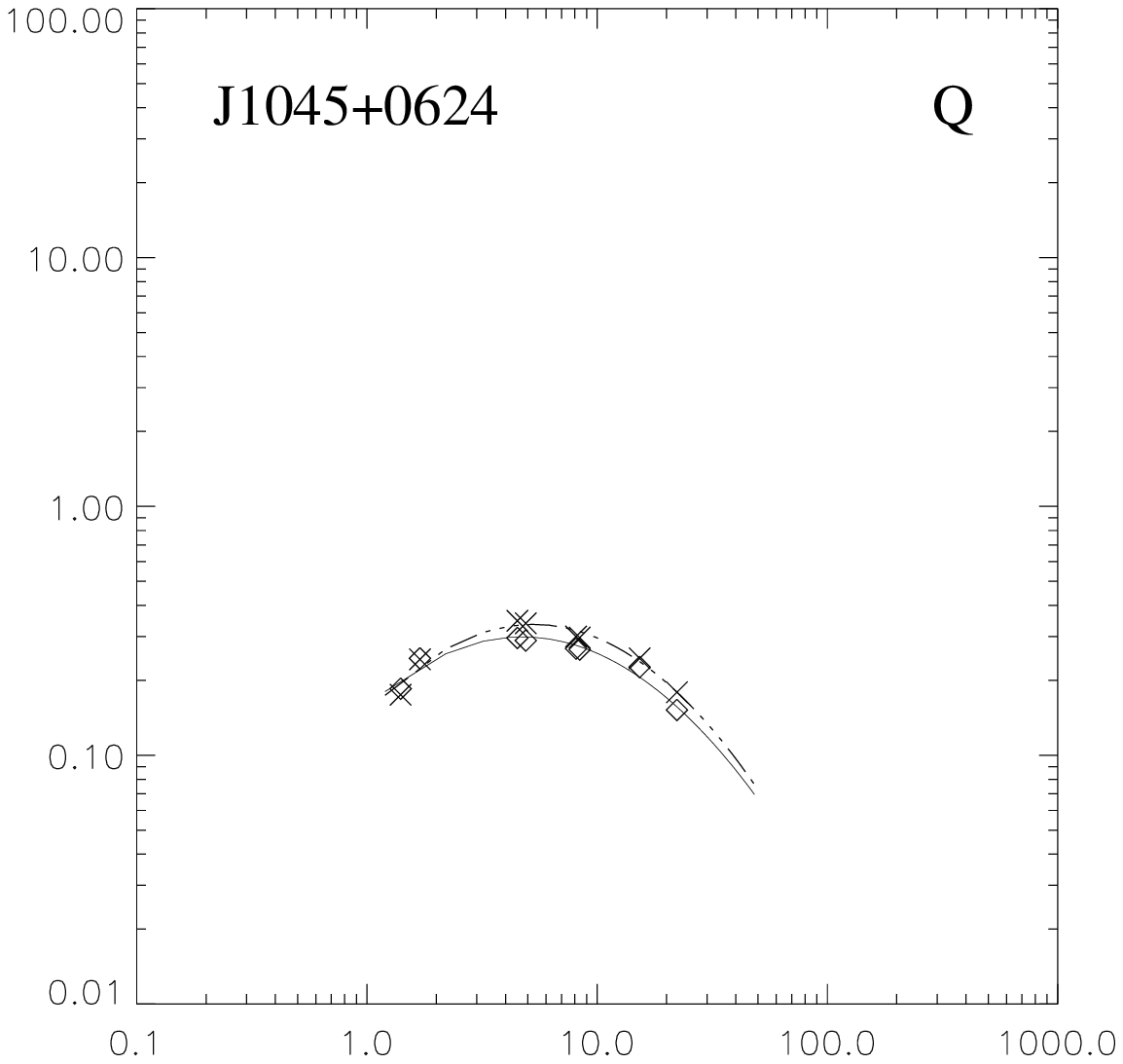}
\vspace{21.5cm}
\caption{Radio spectra of the 51 candidate HFPs observed with the VLA
  during the observing runs presented in this paper [$S_{\nu}$ (Jy) vs
  $\nu$ (GHz)]: crosses, plus signs and diamonds refer
to the first (1998-1999; Dallacasa et al. \cite{dd00}), second
(2002; Tinti et
al. \cite{st05}), and our third epoch (2003-2004;
  dates are reported in Table $\ref{obslog}$)
of simultaneous VLA
observations, while the dashed alternated by three dots,
the dotted and the solid lines indicate the corresponding fits. When a
fourth epoch was available (2004; dates are
  reported in Table $\ref{obslog}$), it is represented by triangles and a
dash-dot line.}
\label{plot}
\end{center}
\end{figure*}

\addtocounter{figure}{-1}
\begin{figure*}[h!]
\begin{center}
\includegraphics{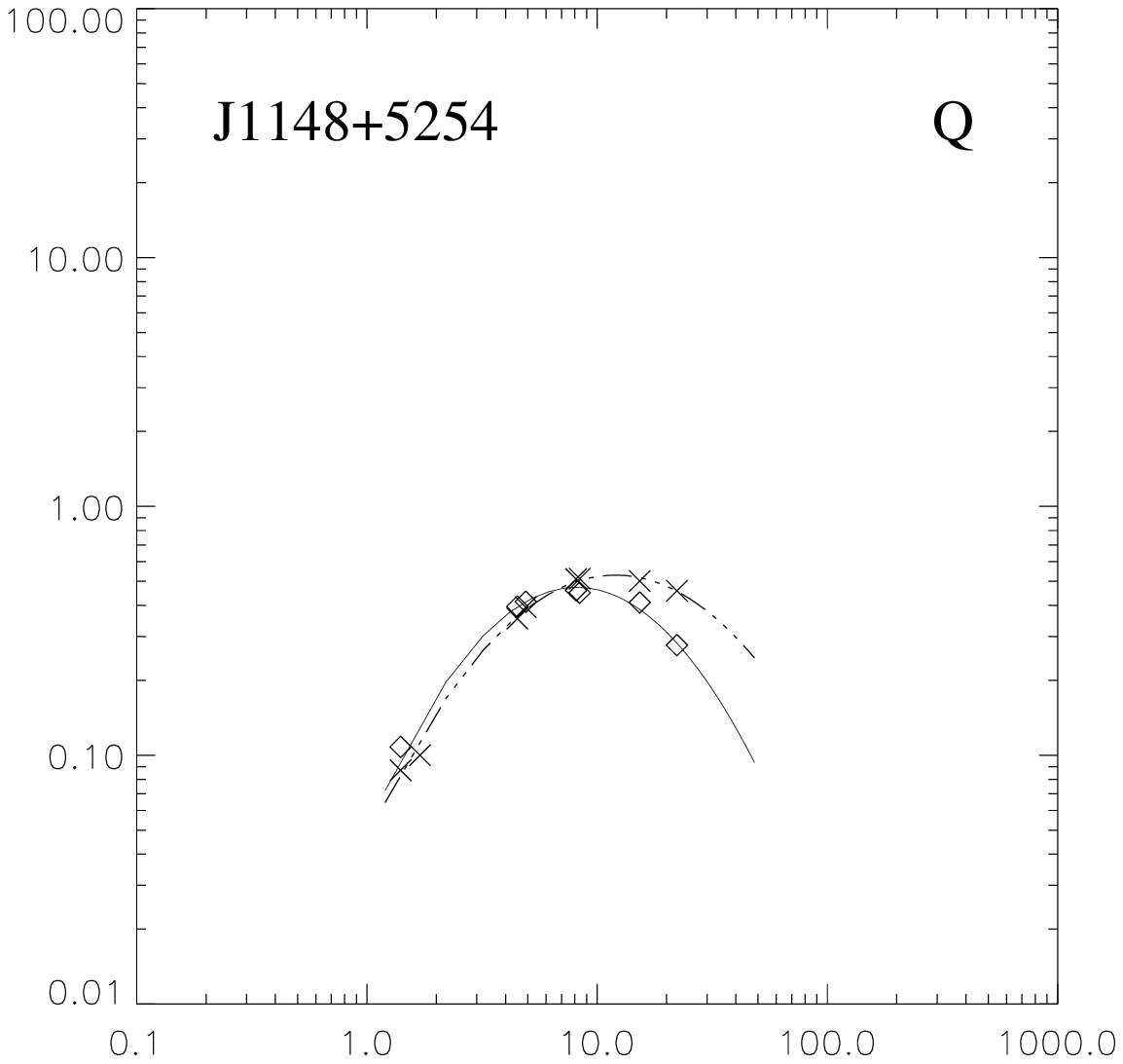}
\includegraphics{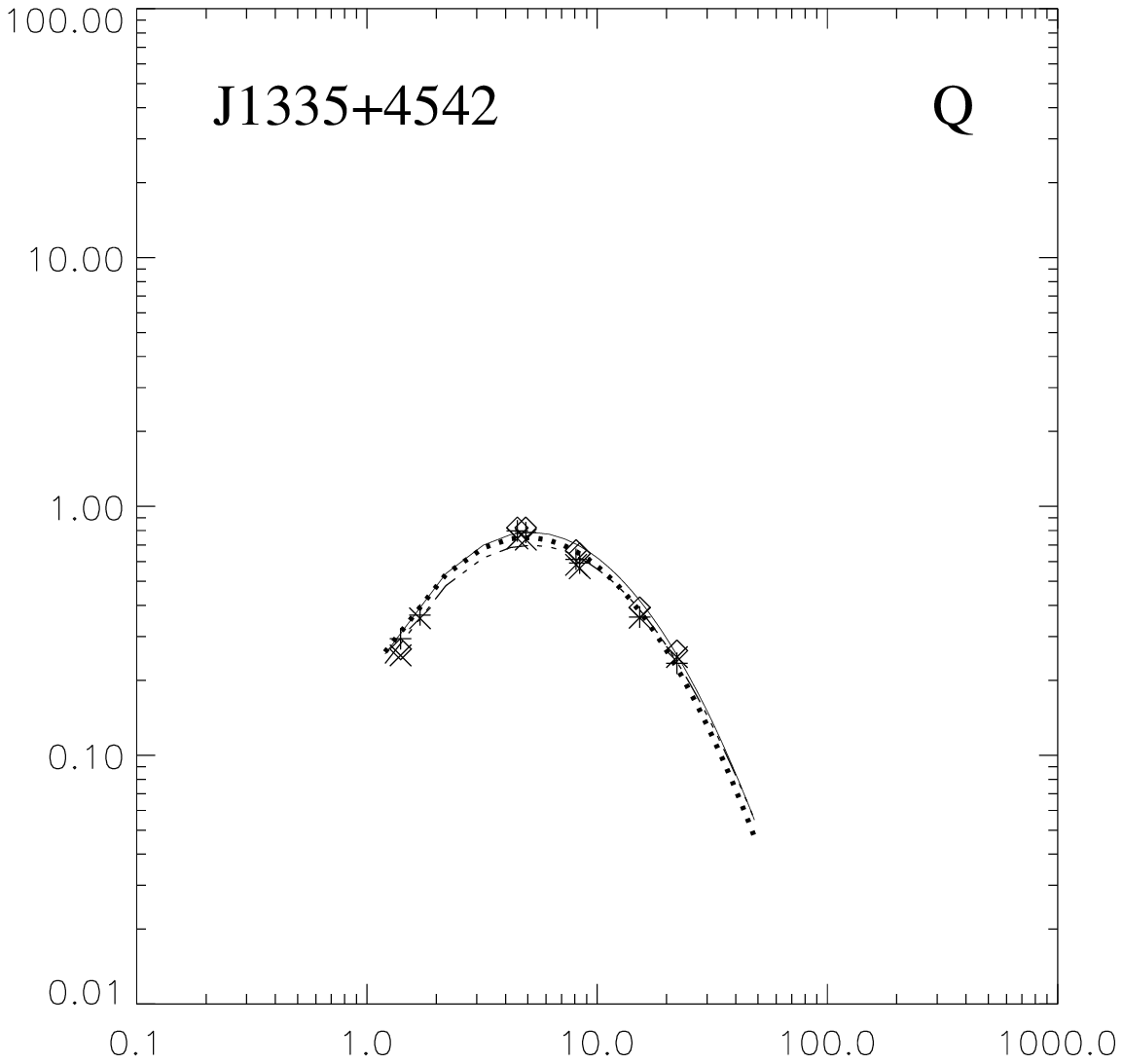}
\includegraphics{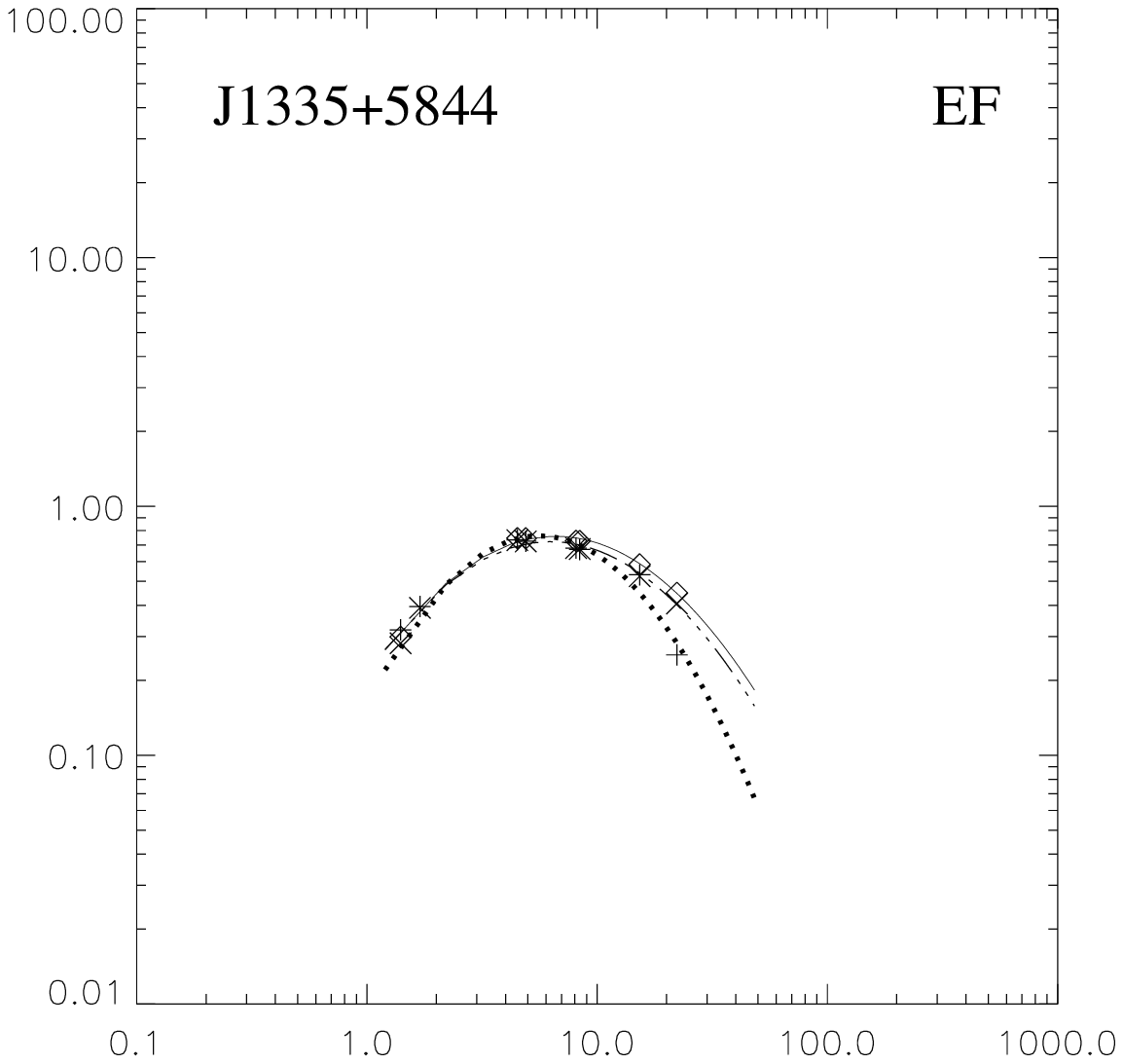}
\includegraphics{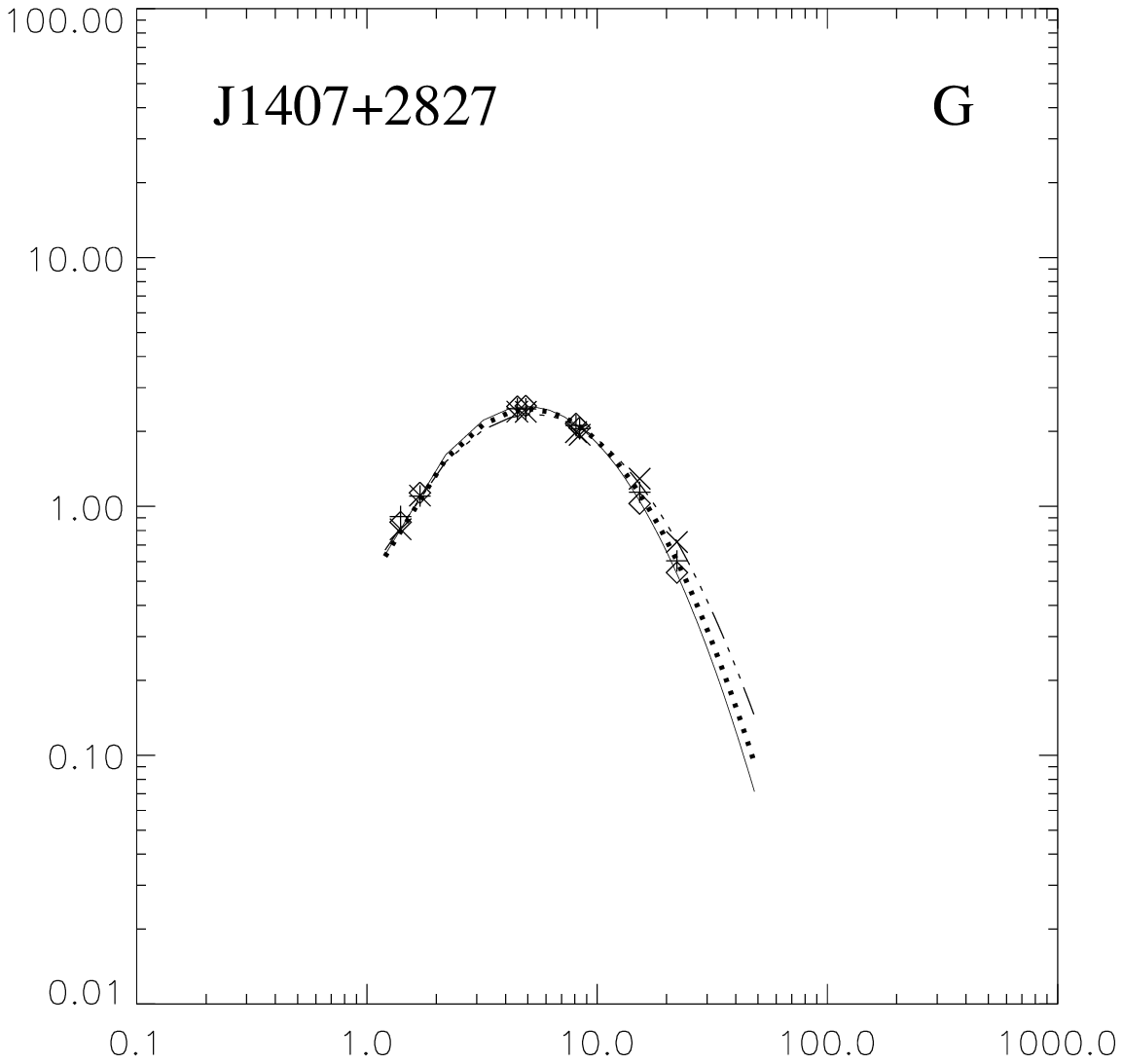}
\includegraphics{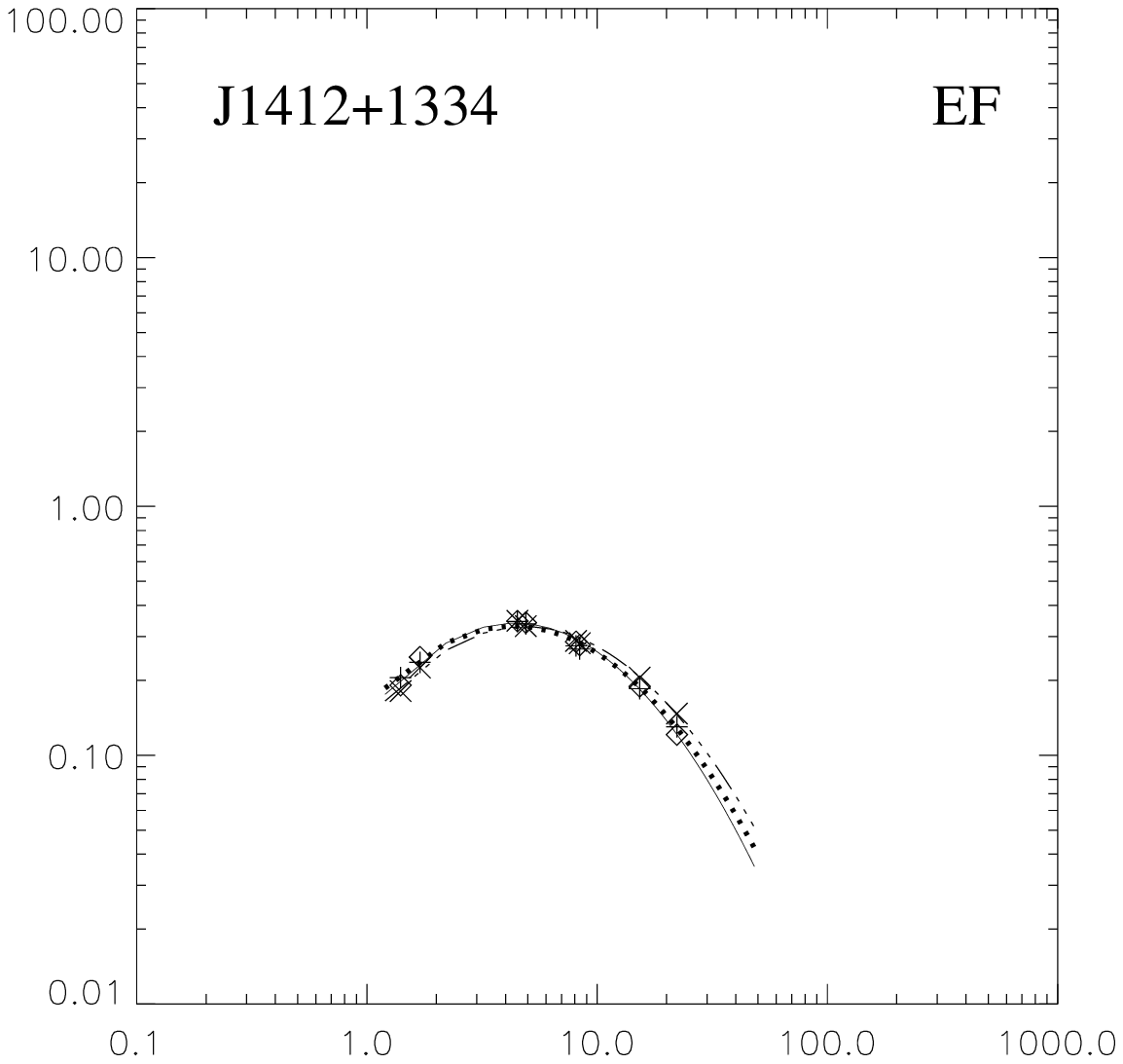}
\includegraphics{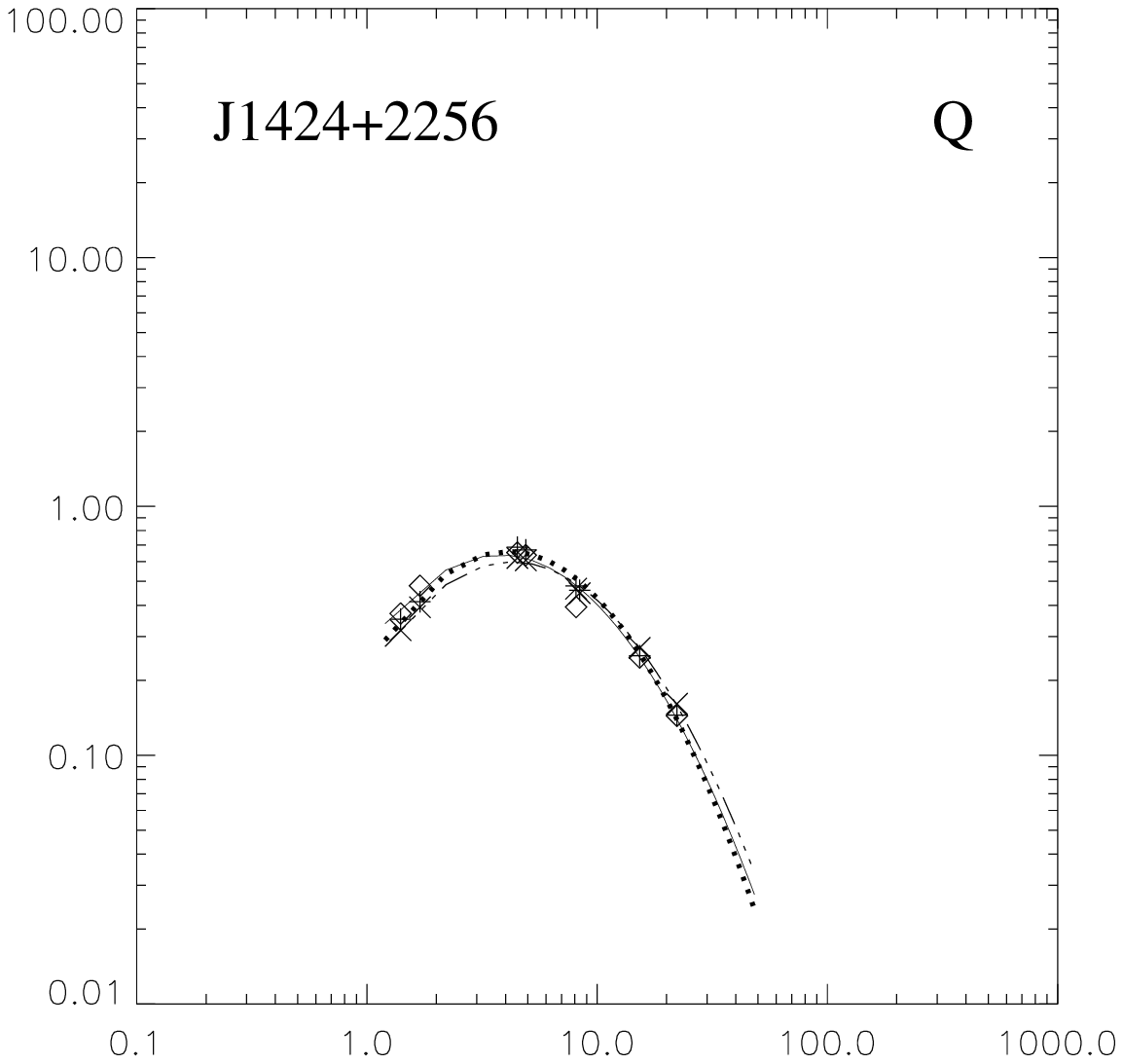}
\includegraphics{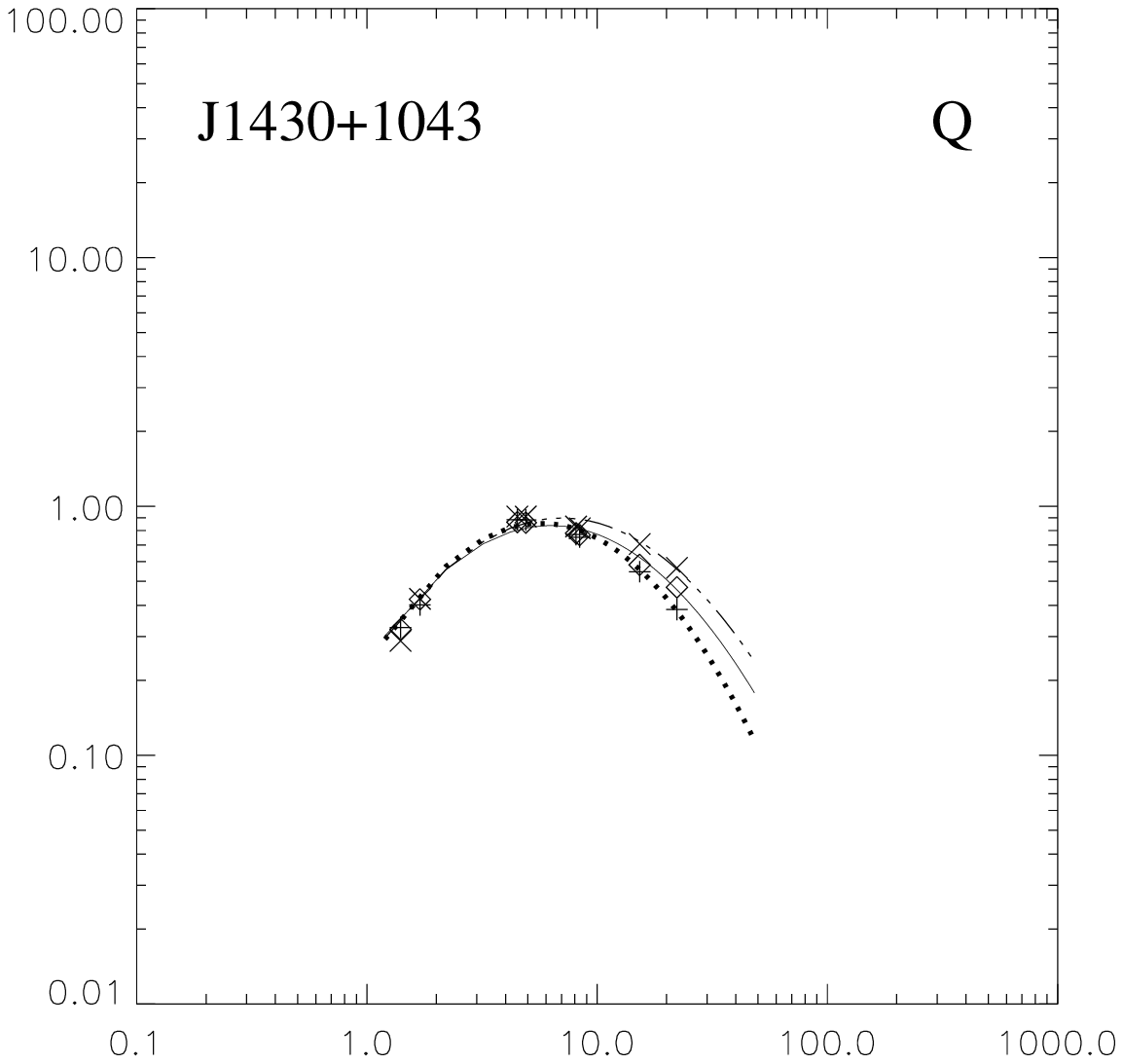}
\includegraphics{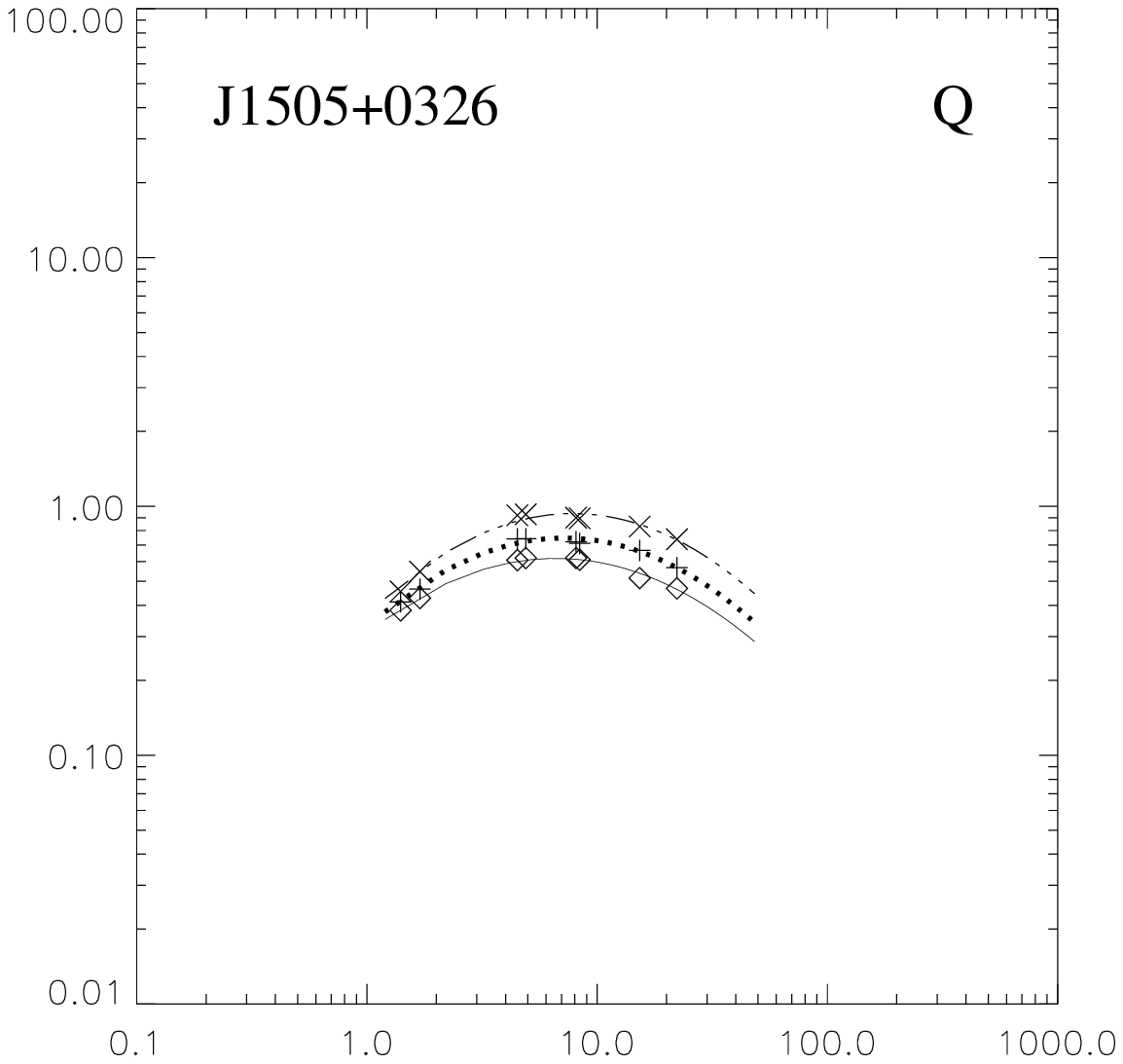}
\includegraphics{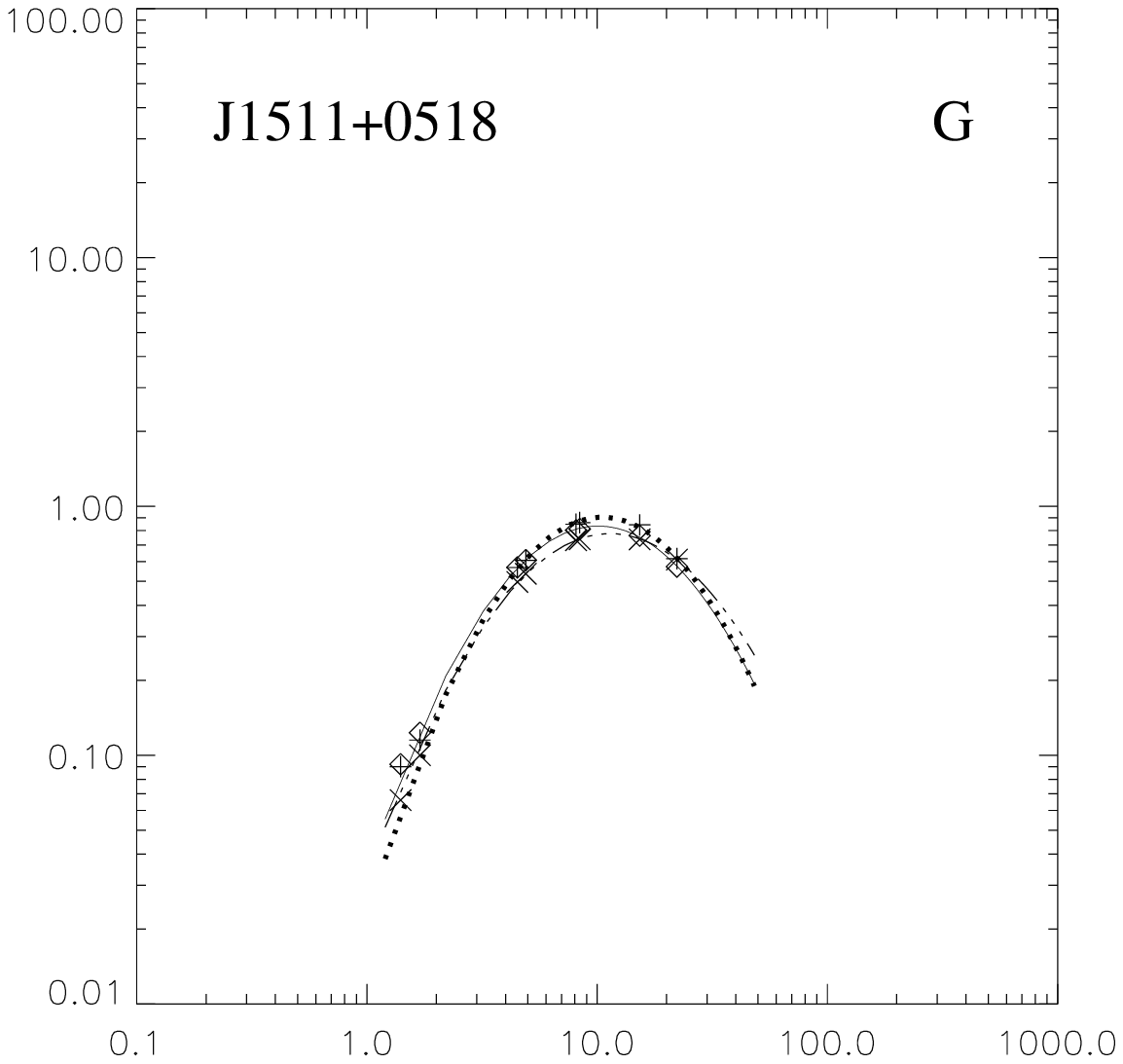}
\includegraphics{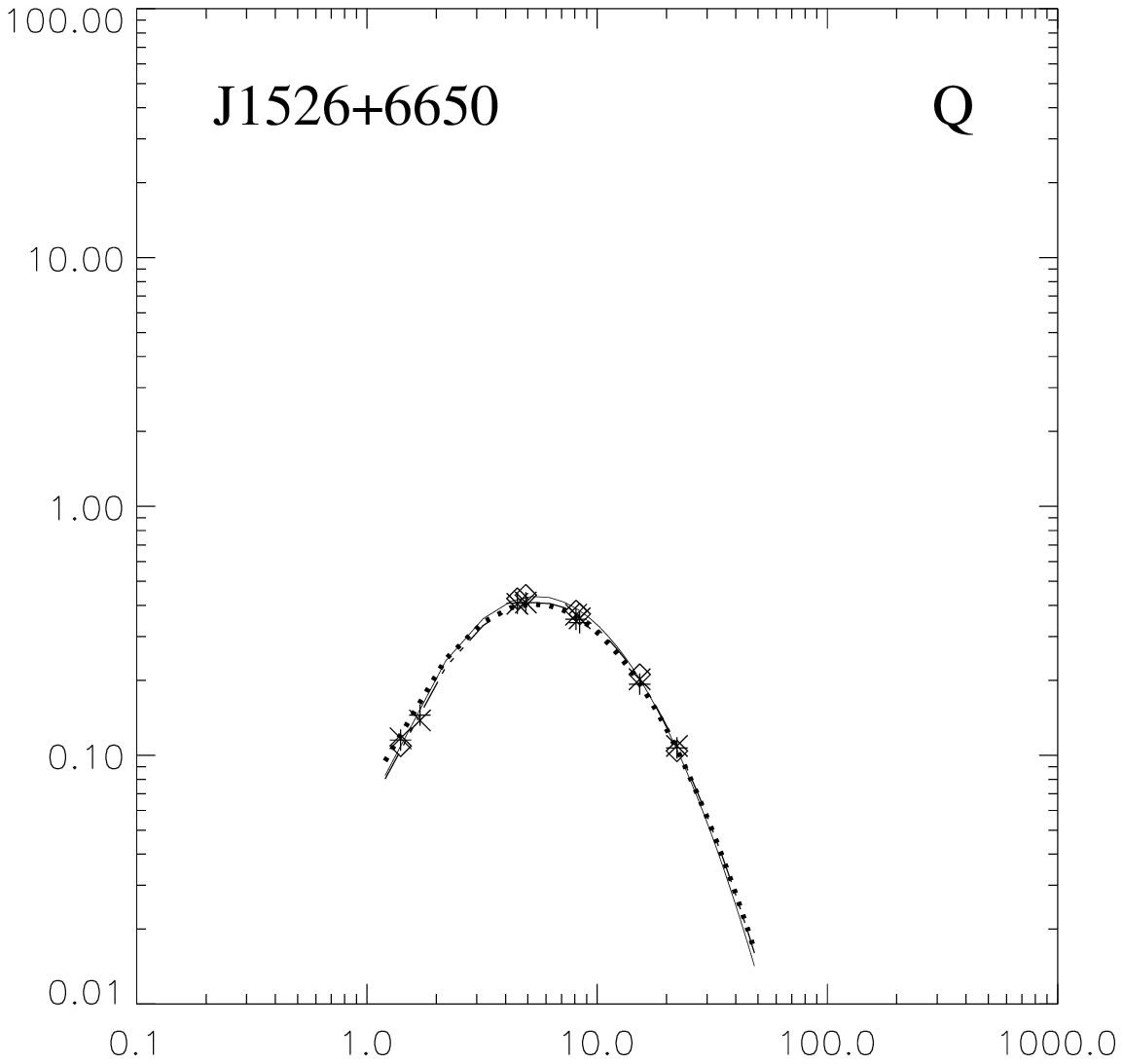}
\includegraphics{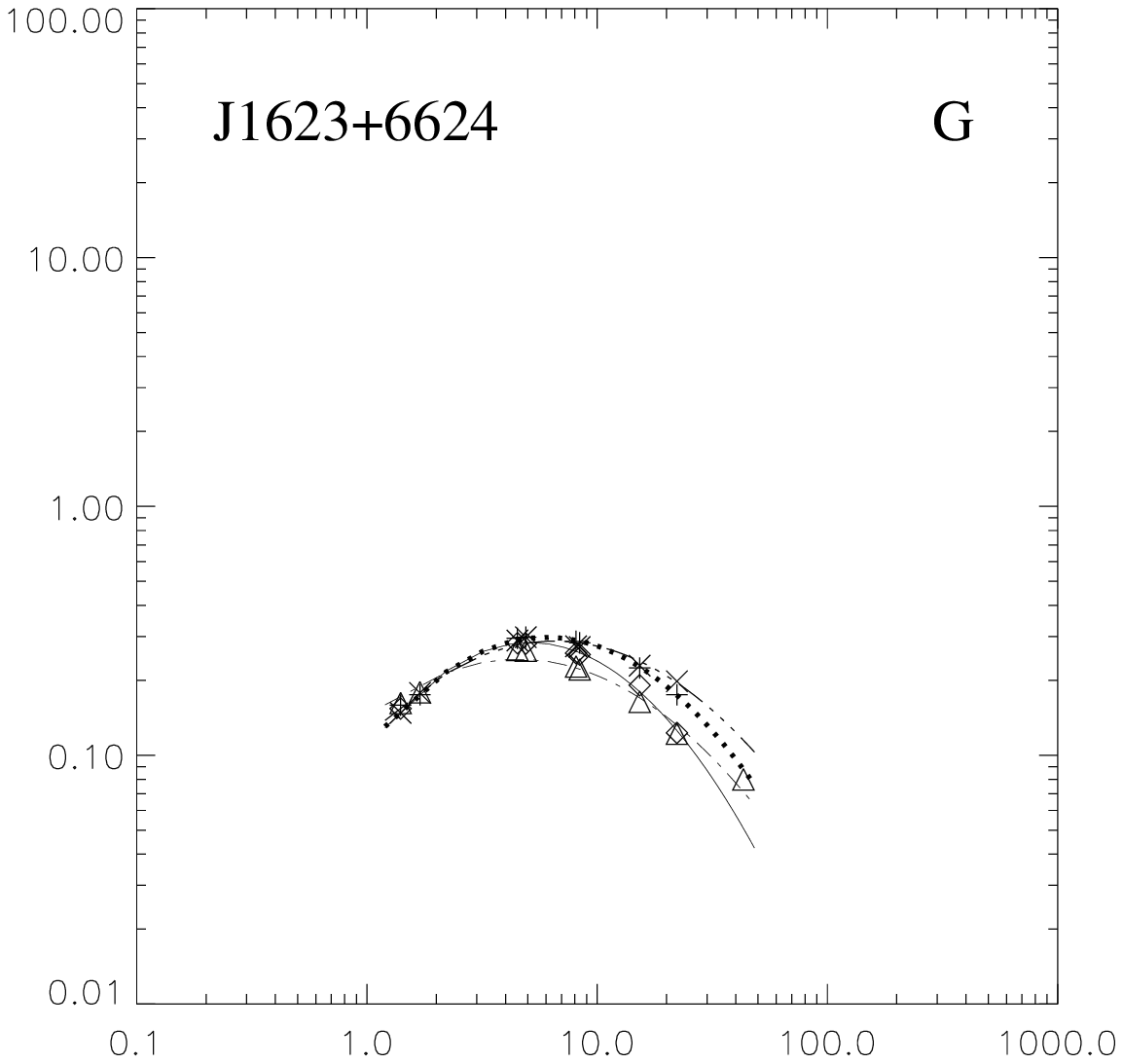}
\includegraphics{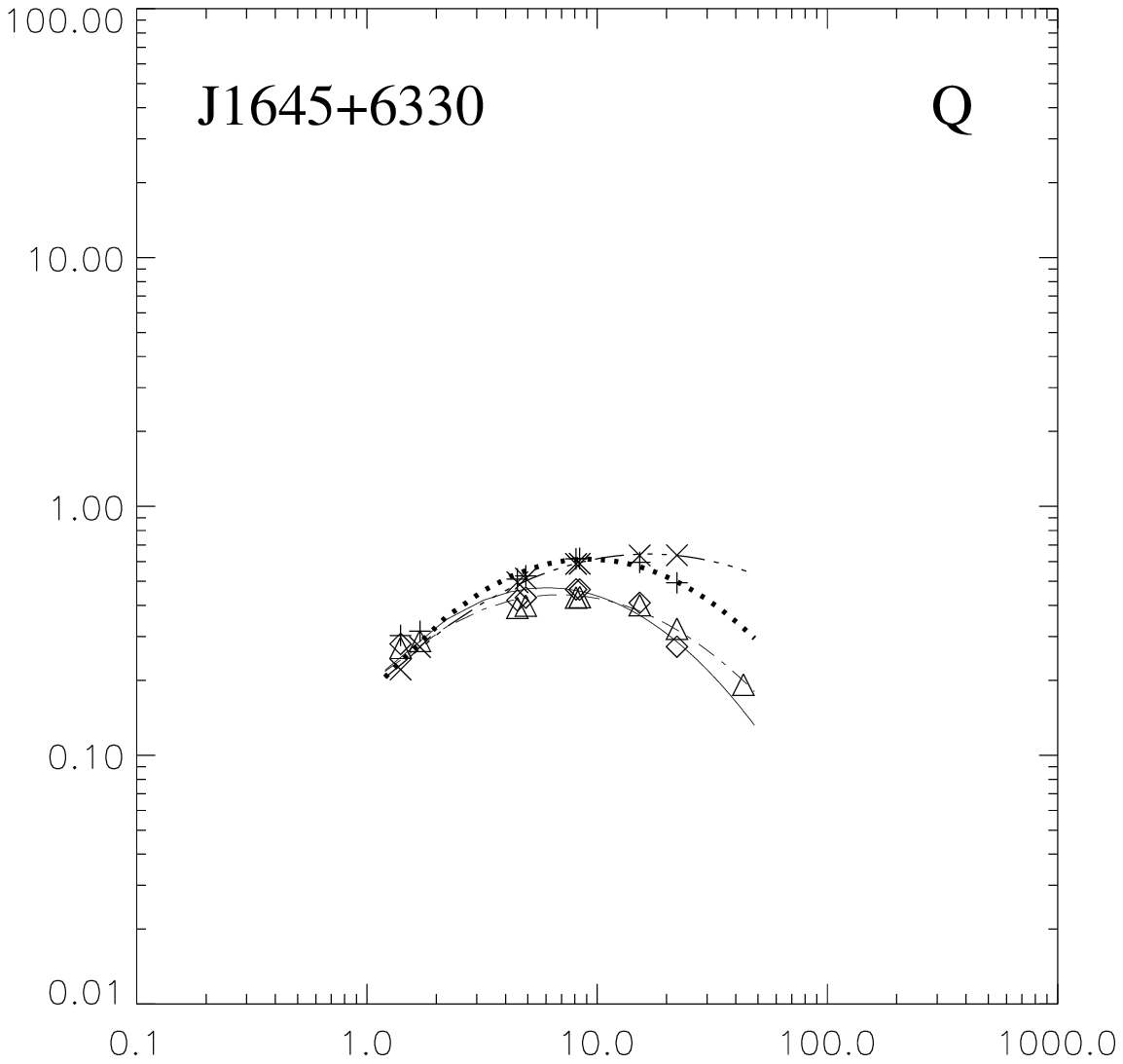}
\includegraphics{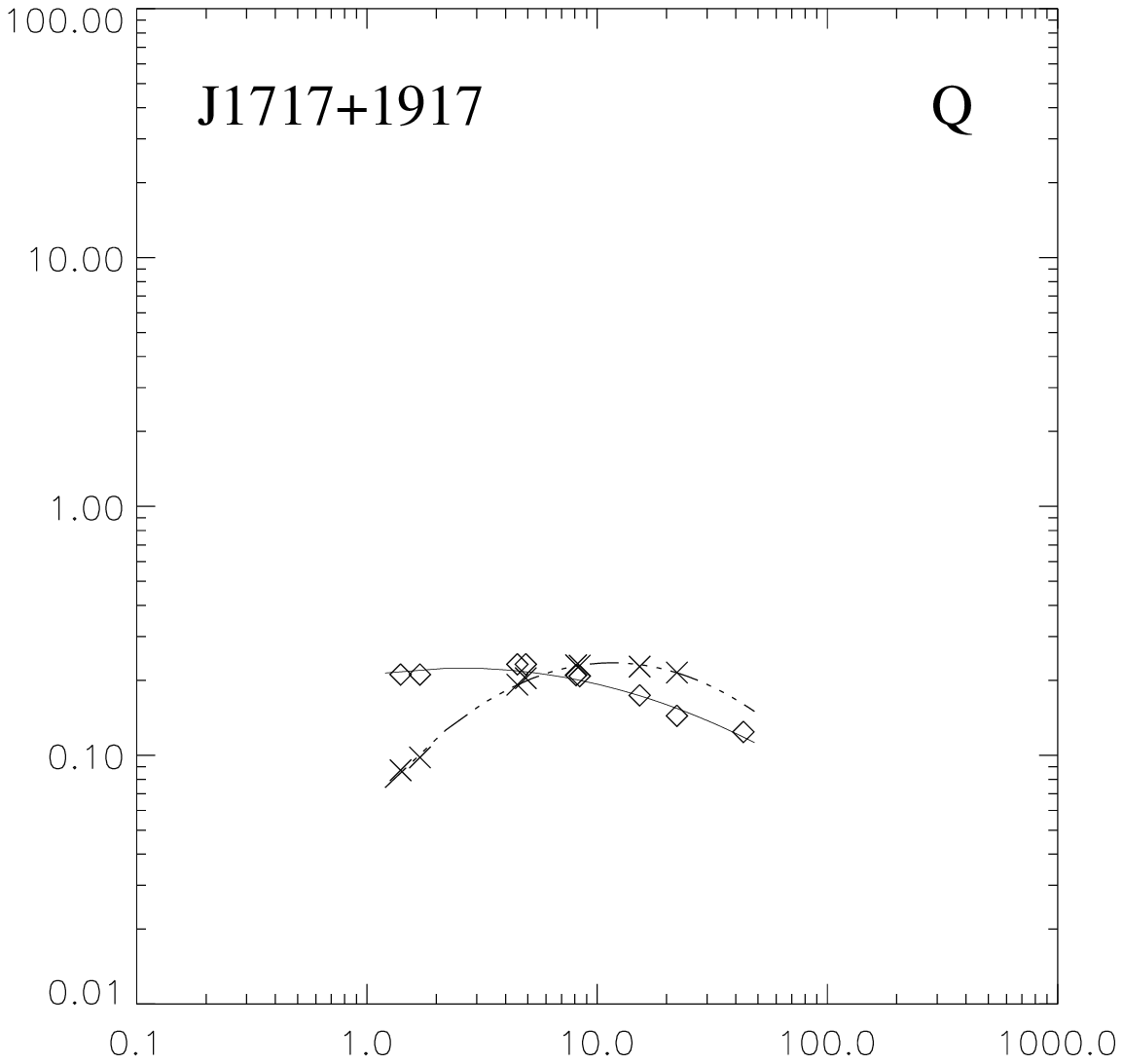}
\includegraphics{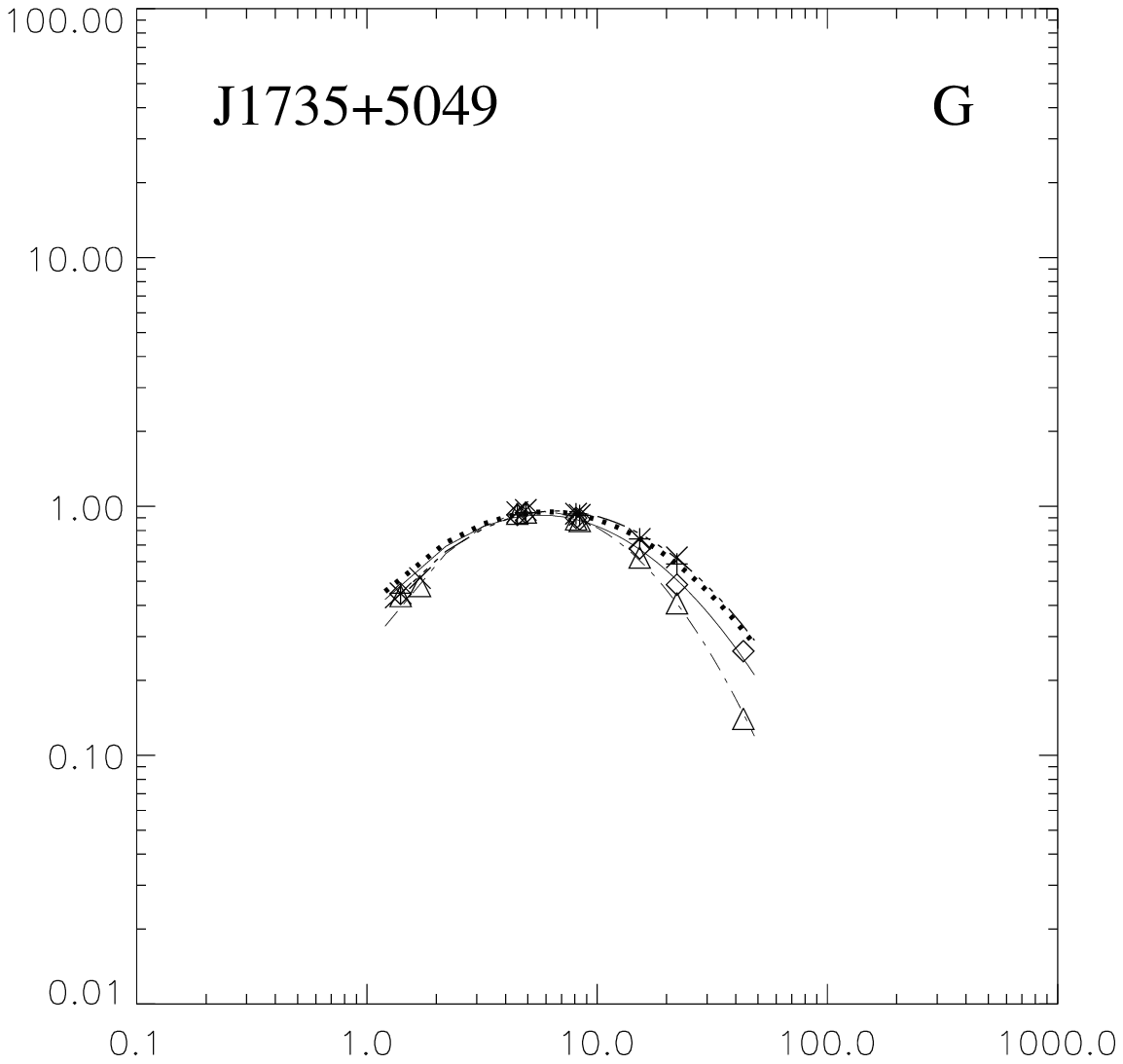}
\includegraphics{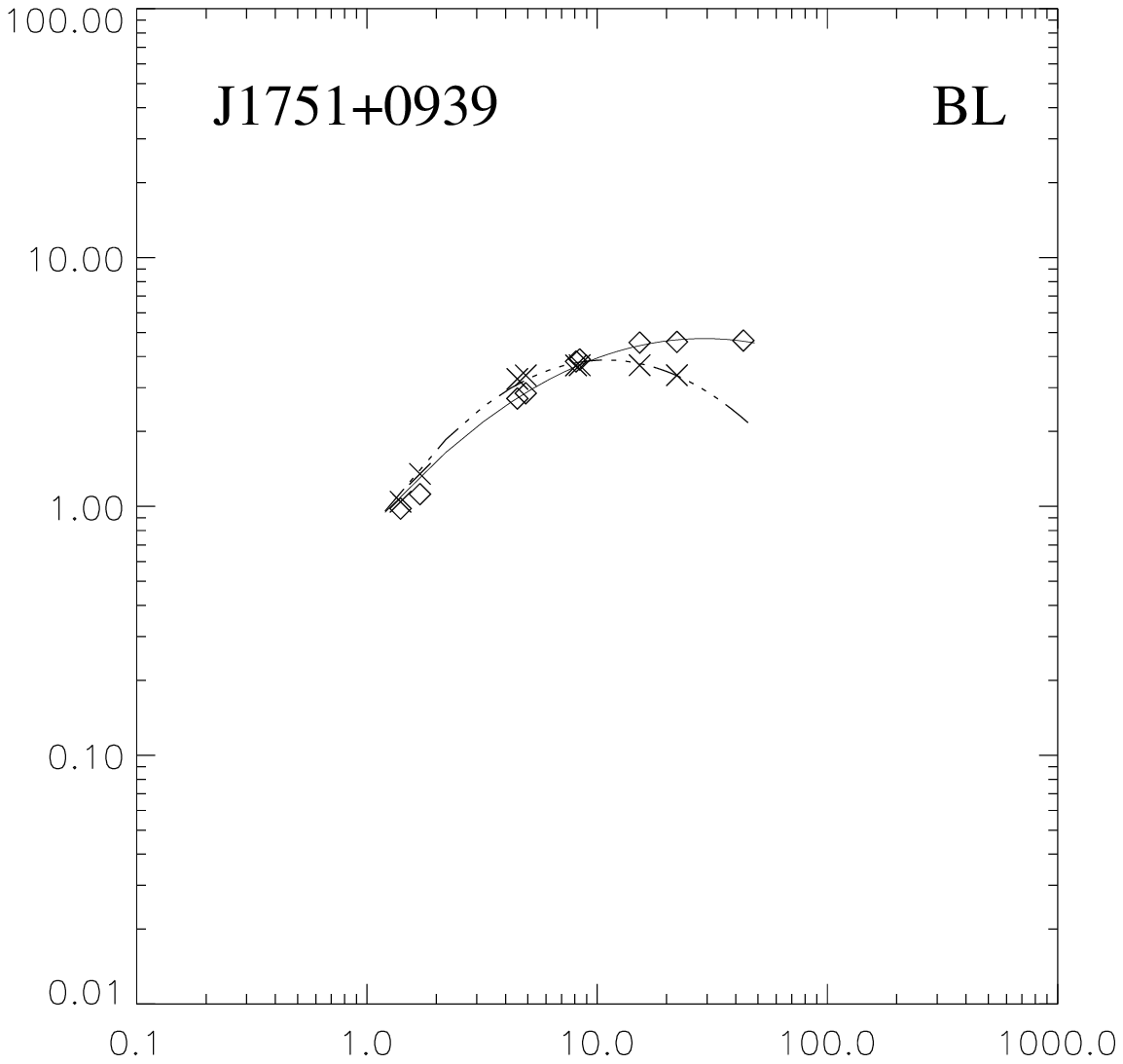}
\includegraphics{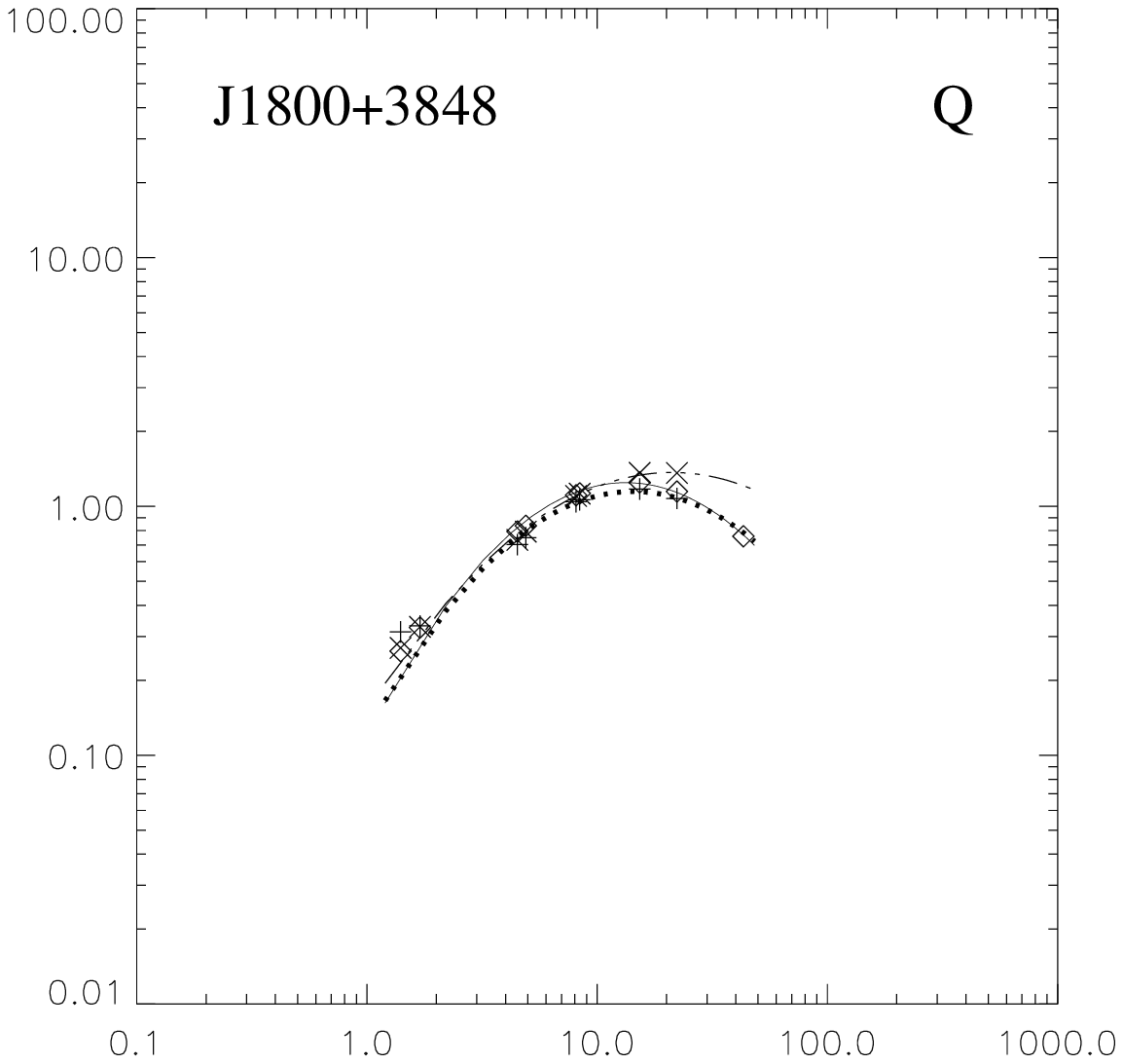}
\includegraphics{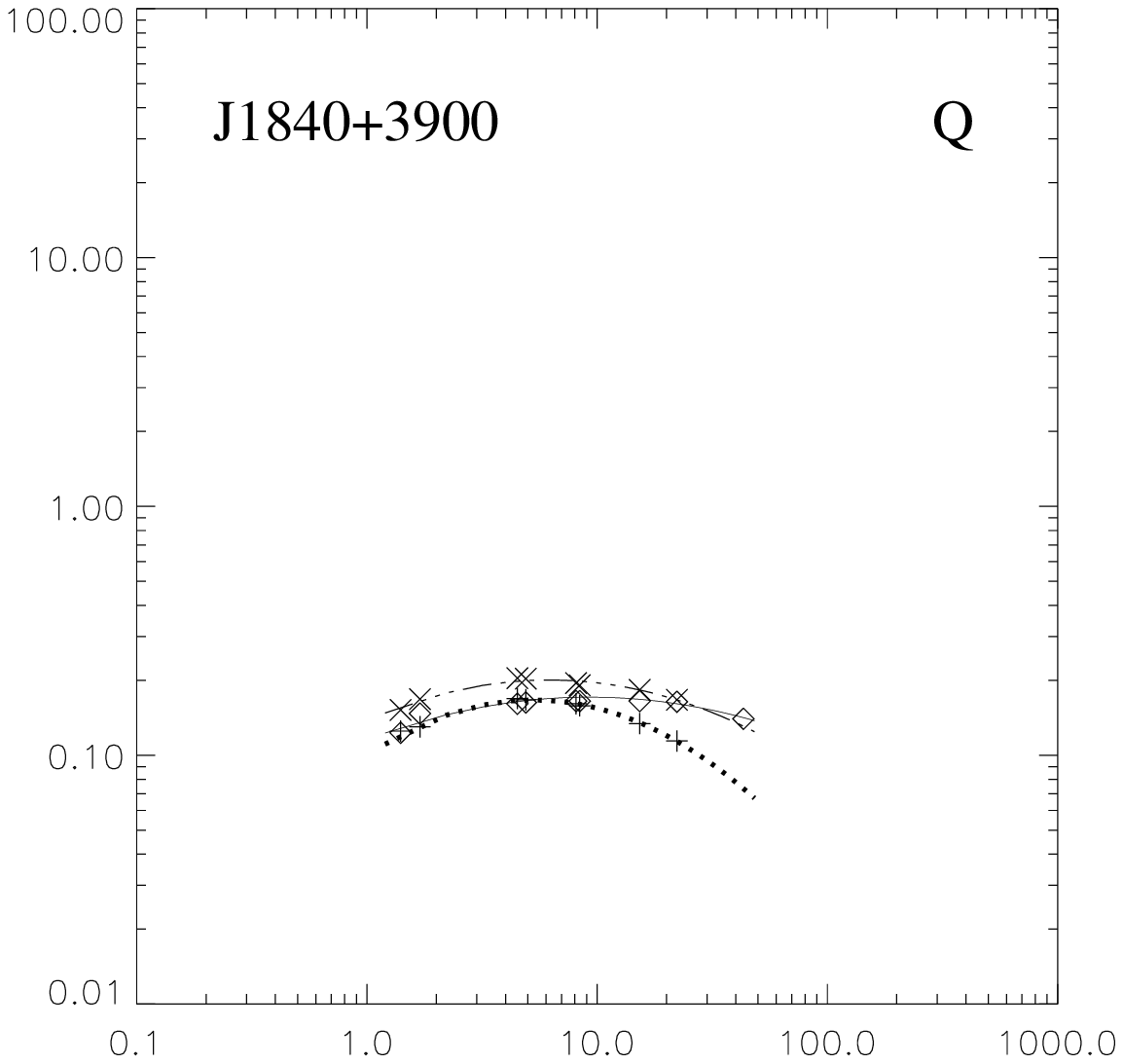}
\includegraphics{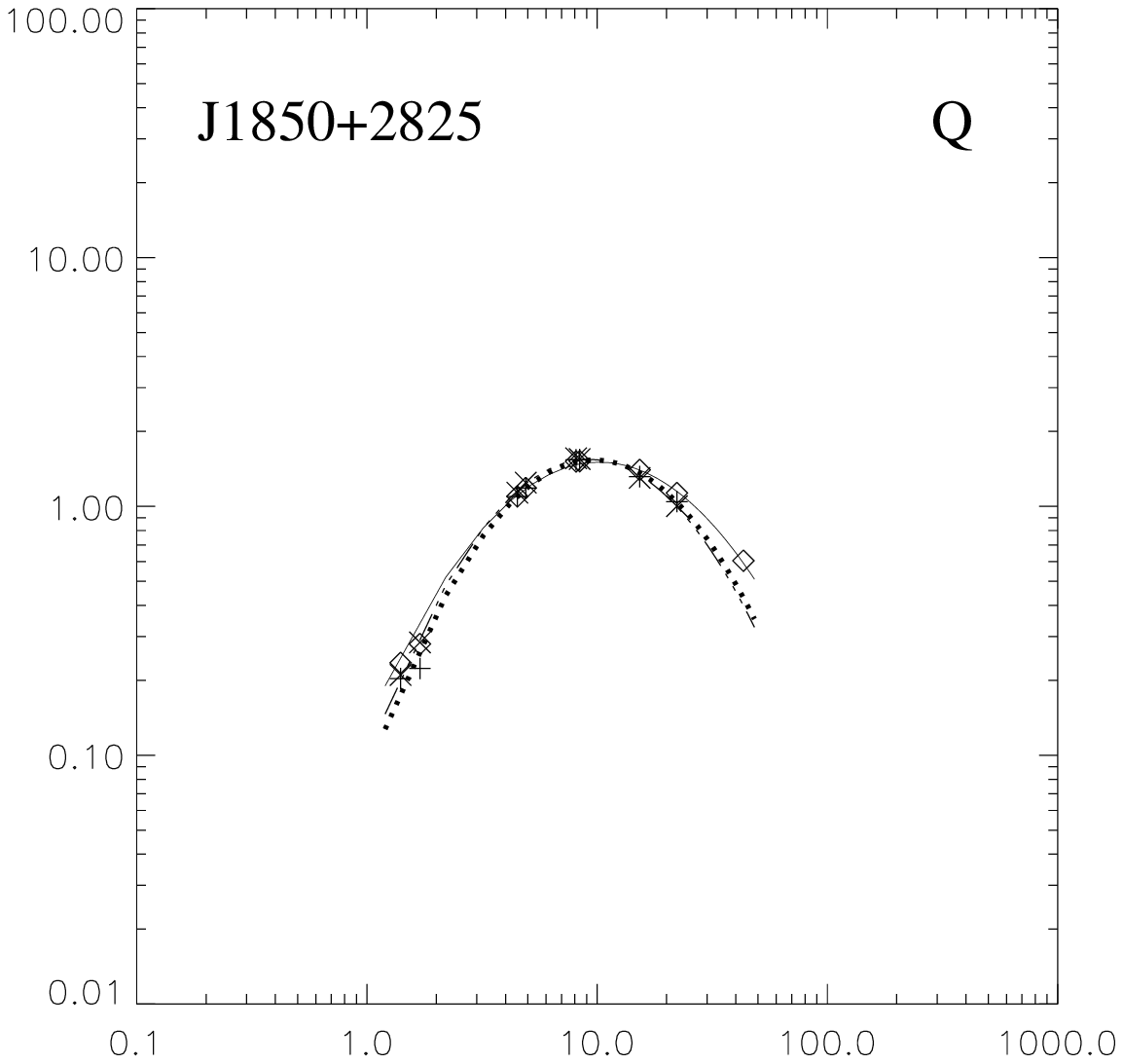}
\includegraphics{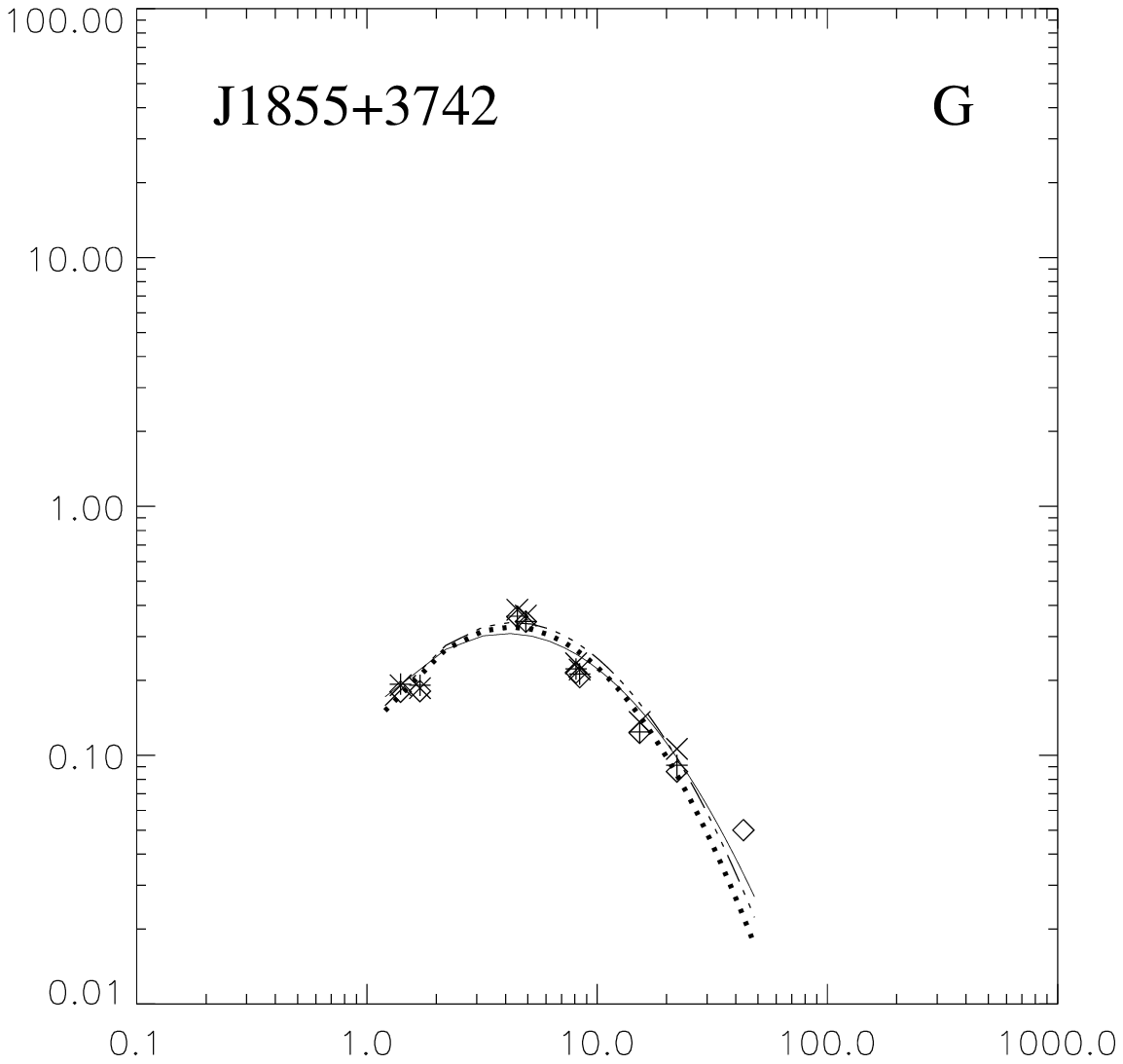}
\includegraphics{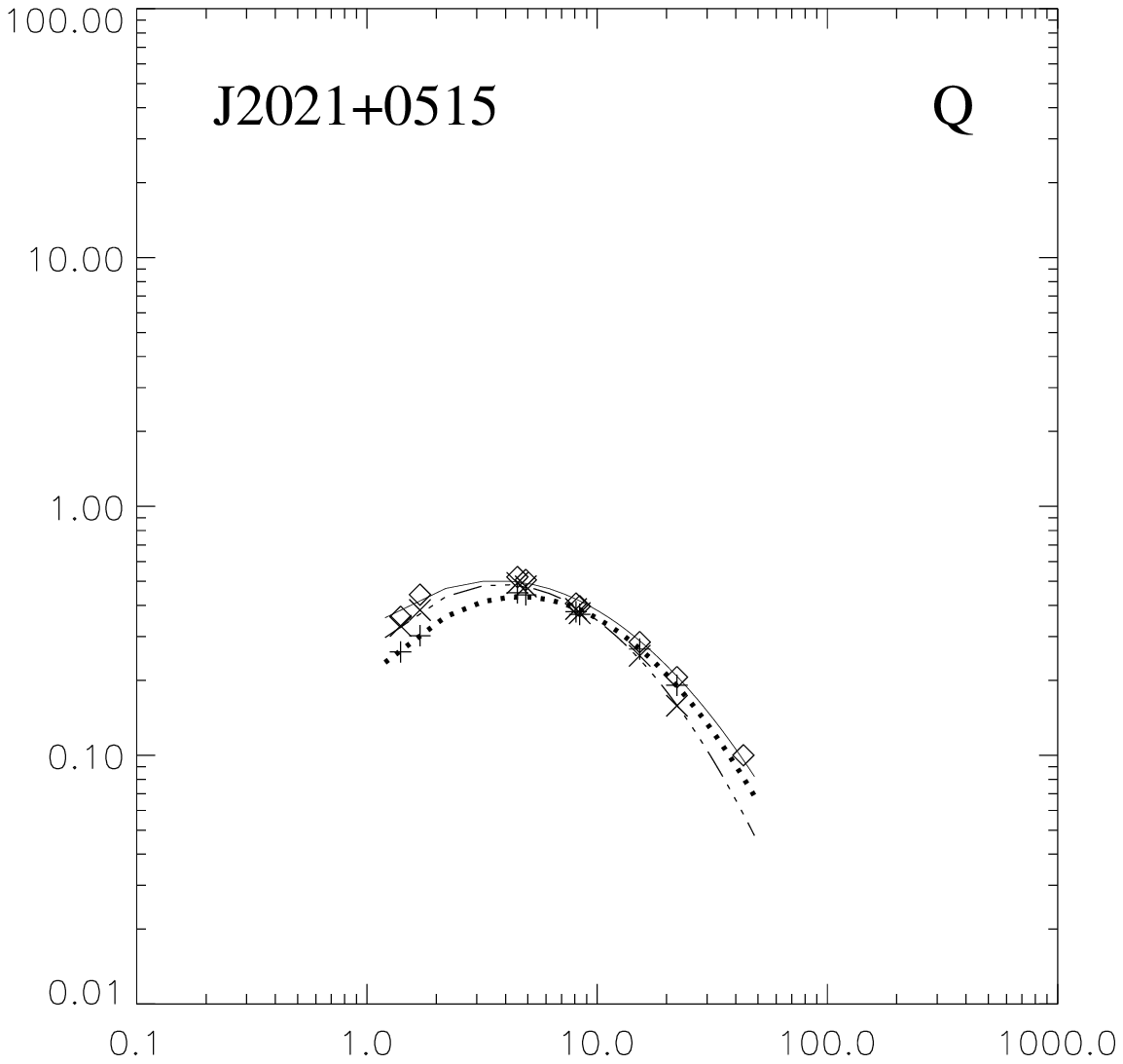}
\vspace{21.5cm}
\caption{Continued.}
\end{center}
\end{figure*}

\addtocounter{figure}{-1}
\begin{figure*}[h!]
\begin{center}
\includegraphics{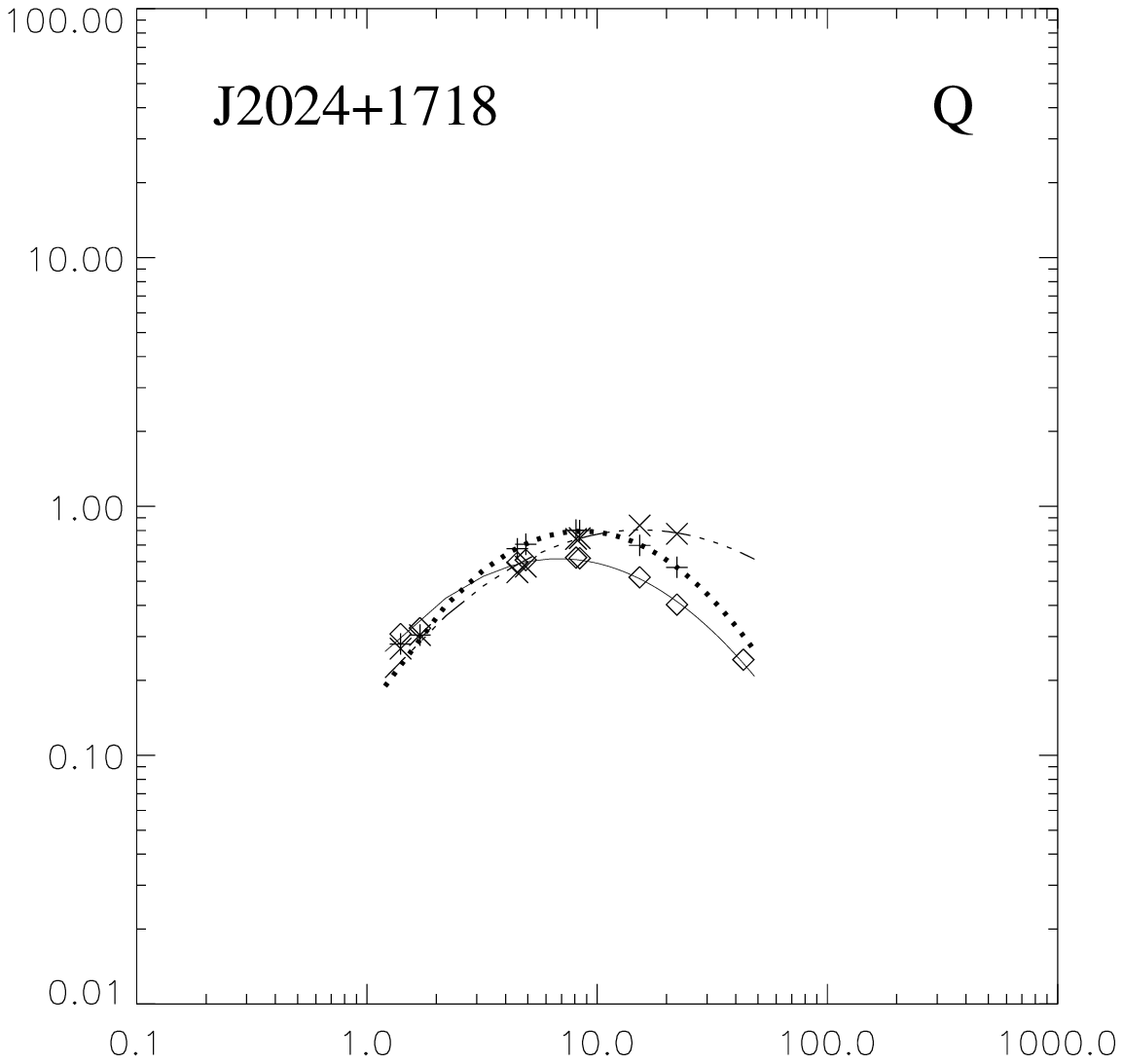}
\includegraphics{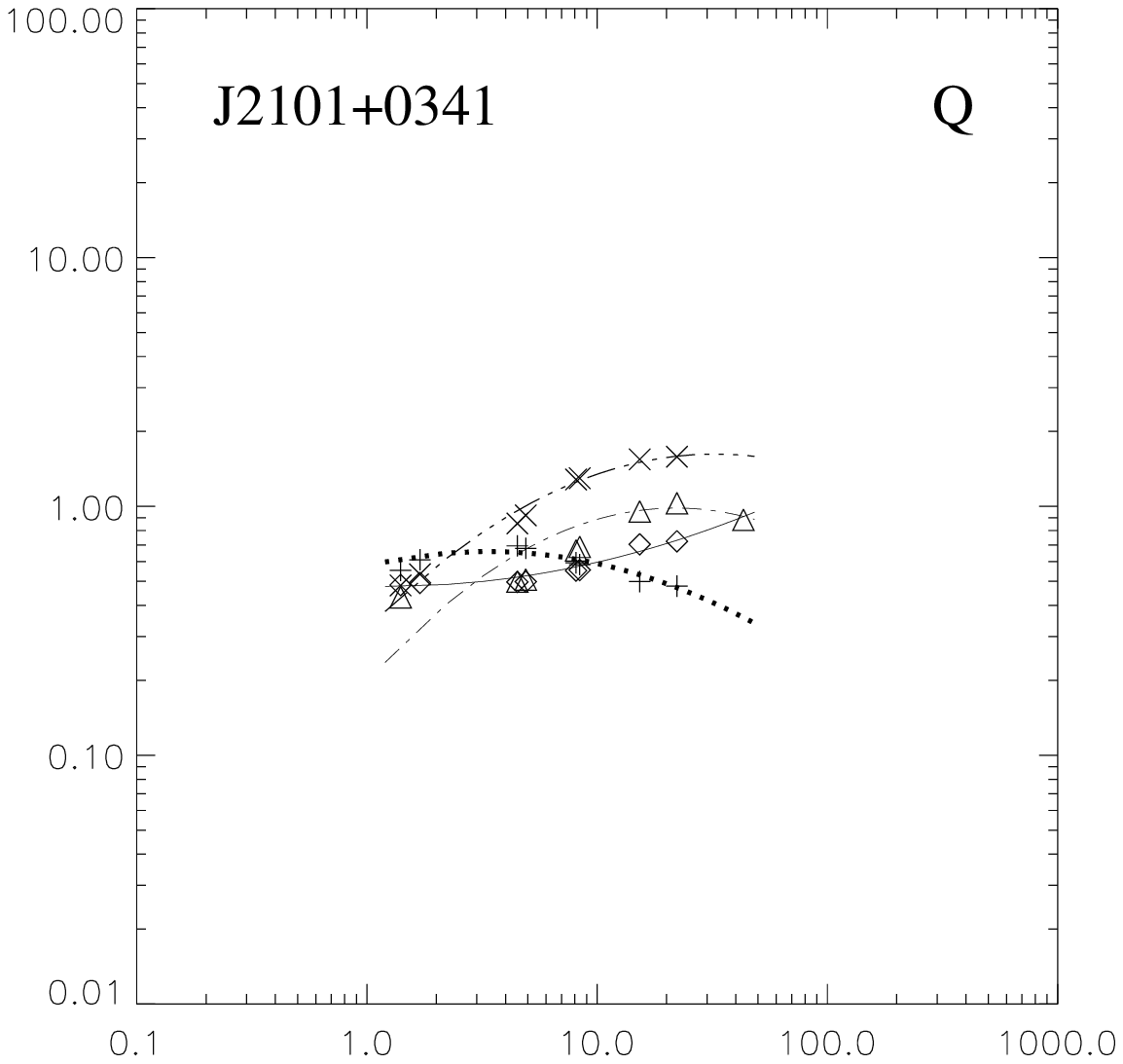}
\includegraphics{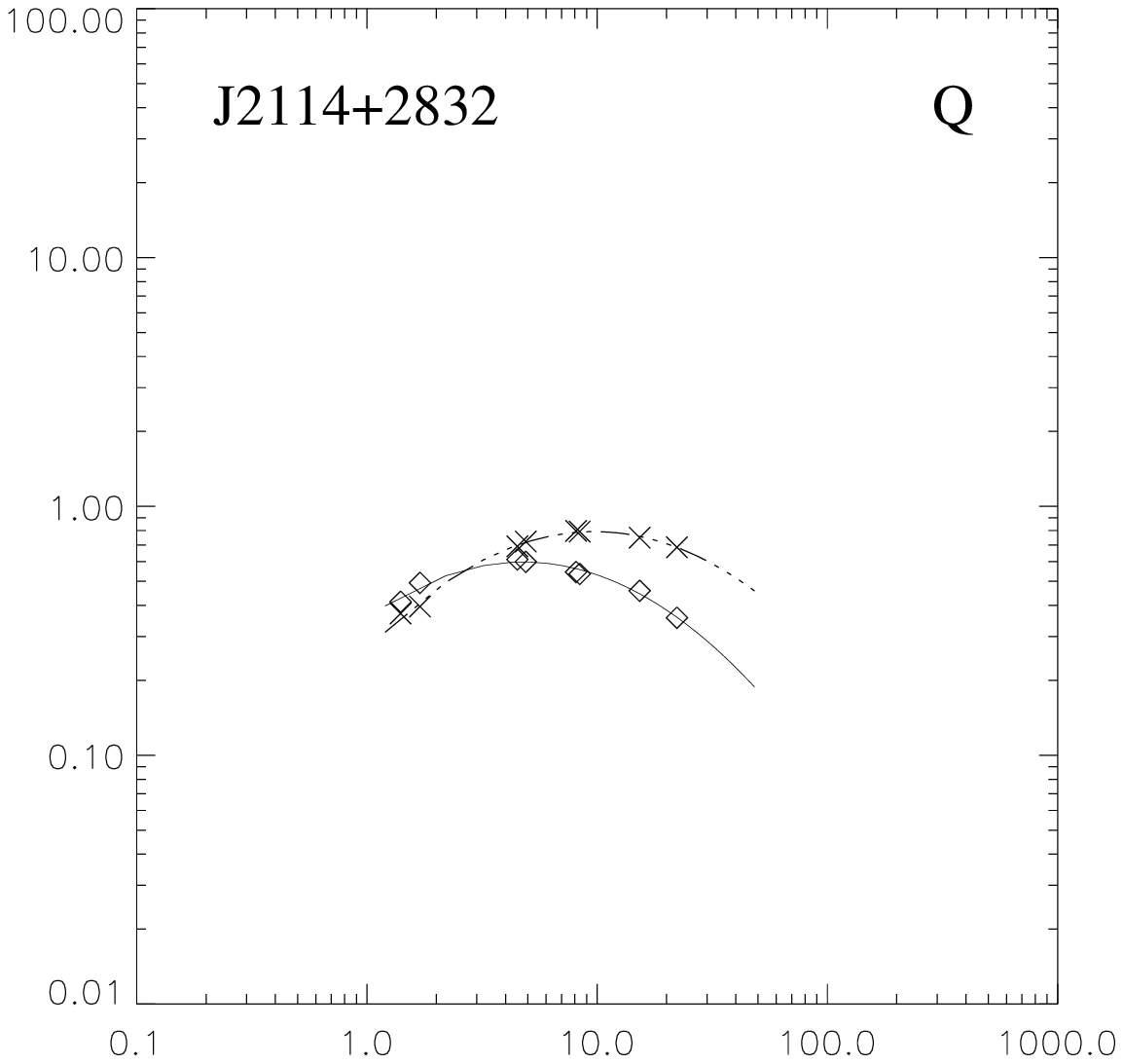}
\includegraphics{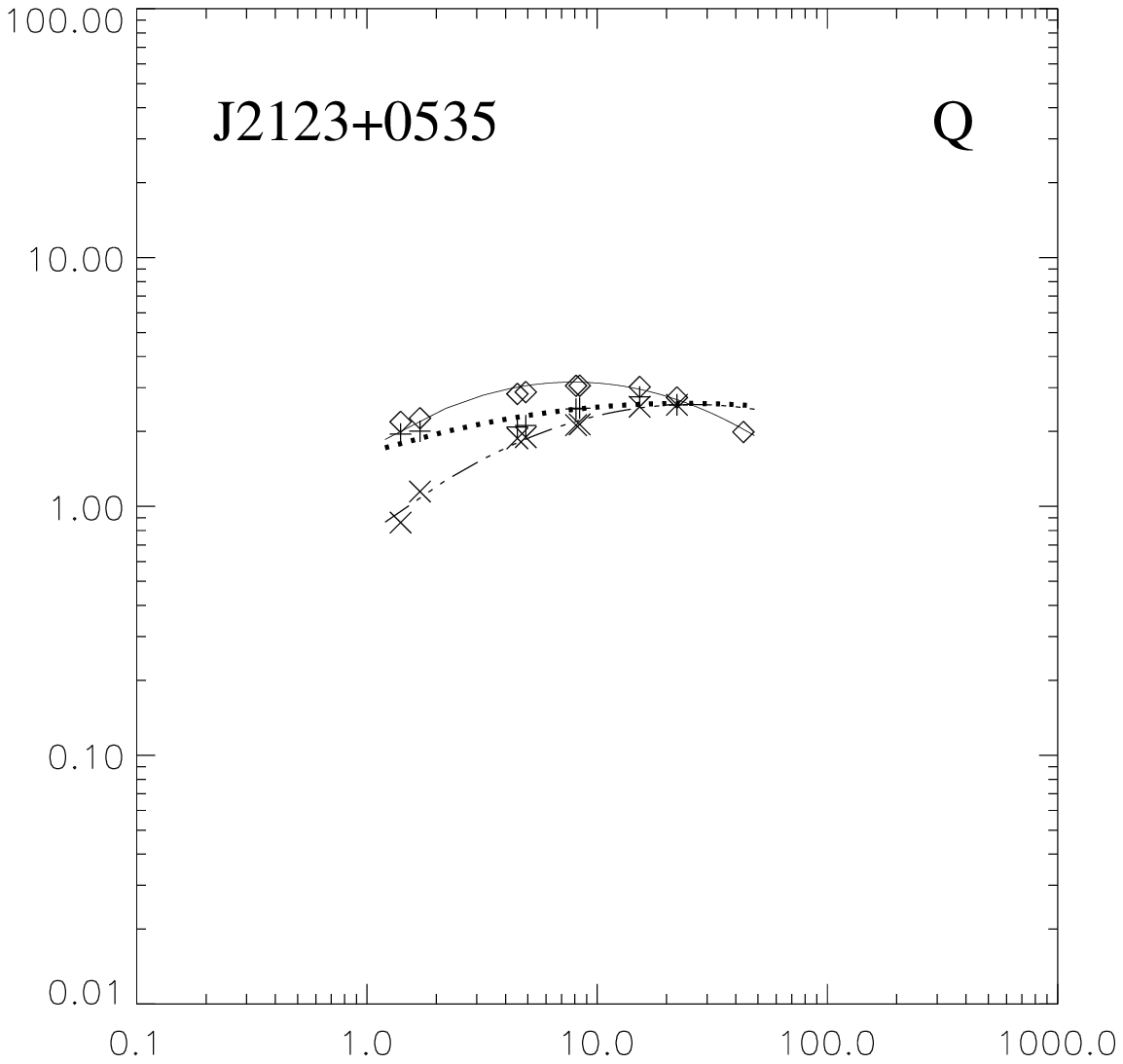}
\includegraphics{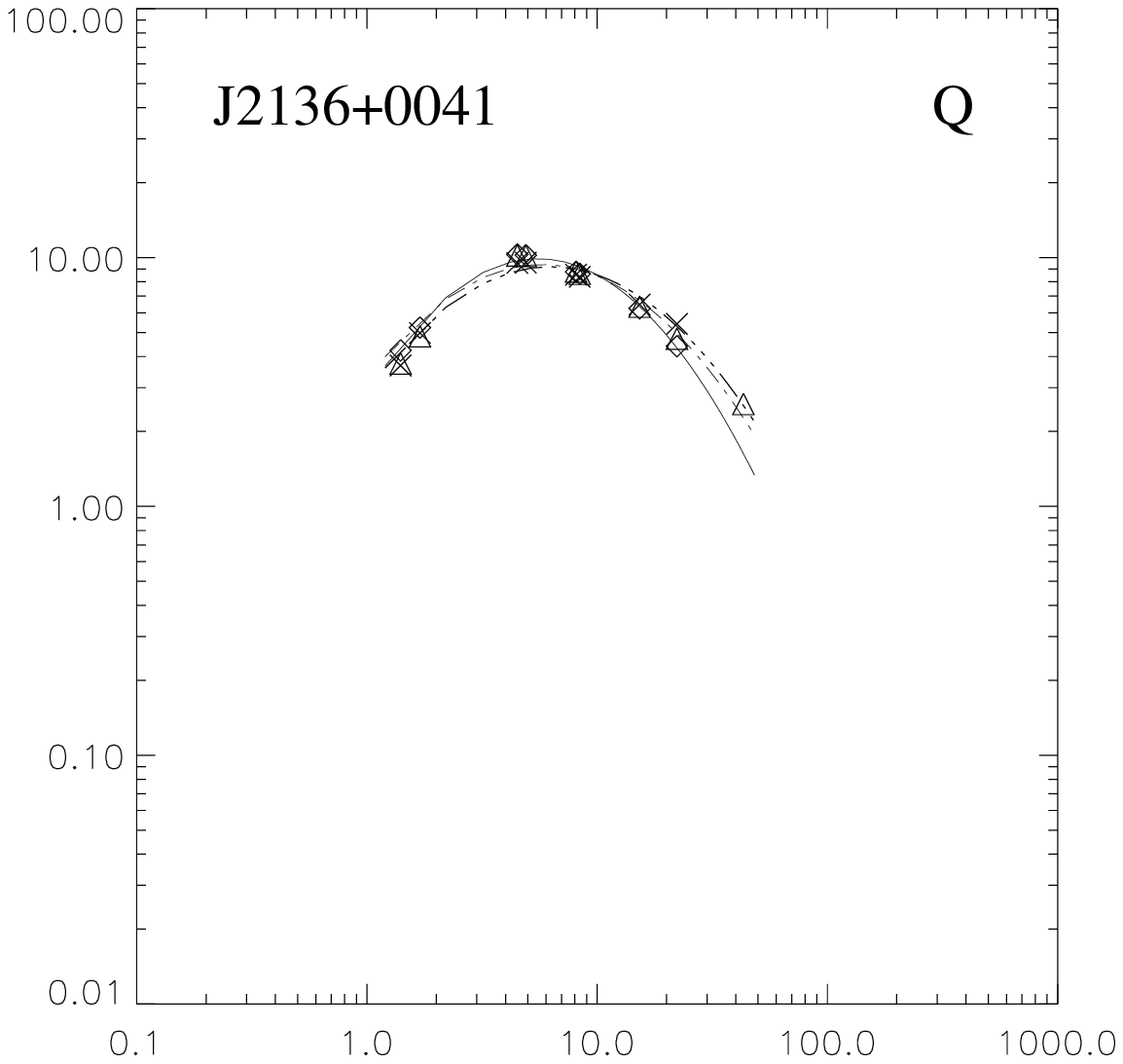}
\includegraphics{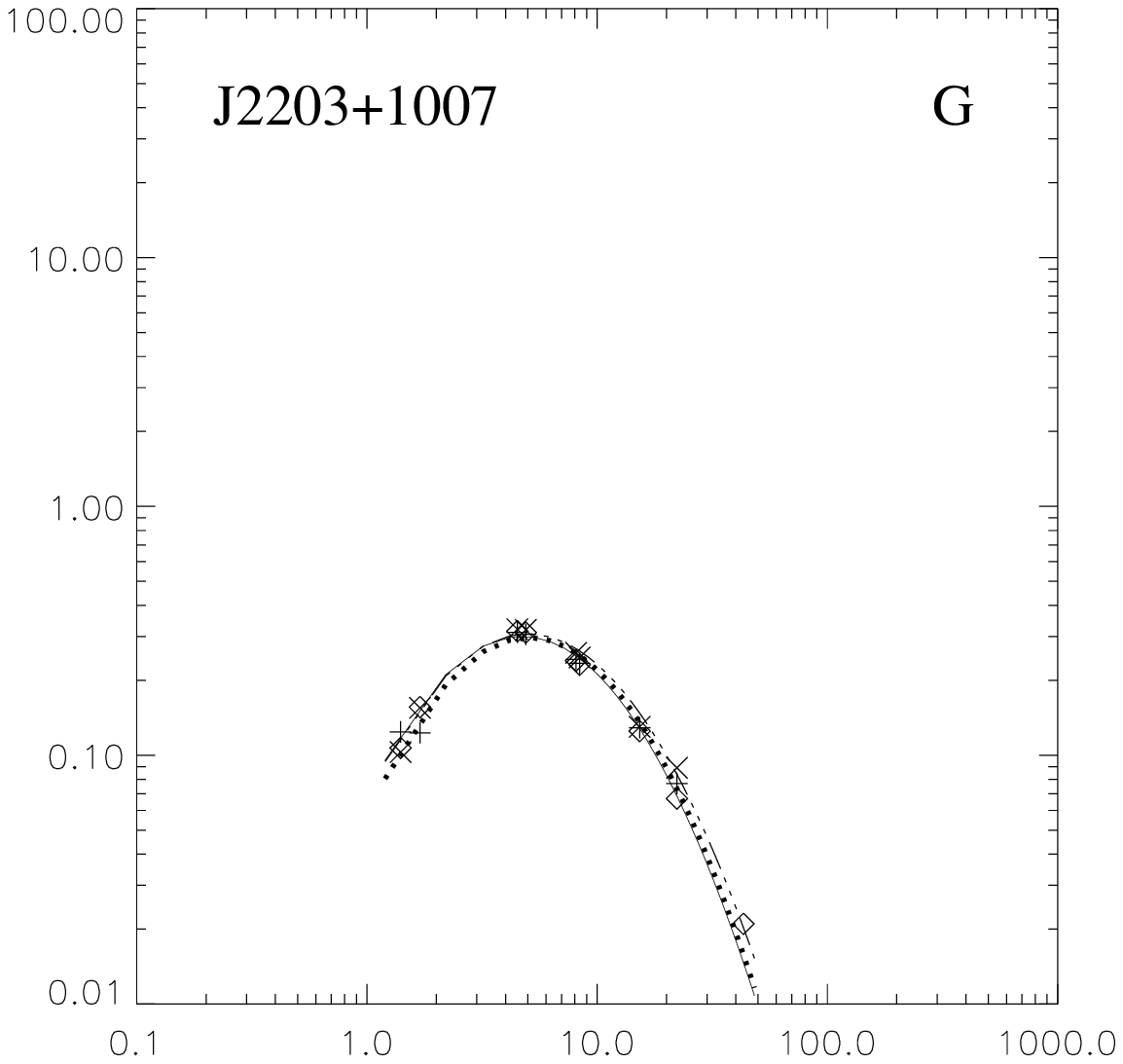}
\includegraphics{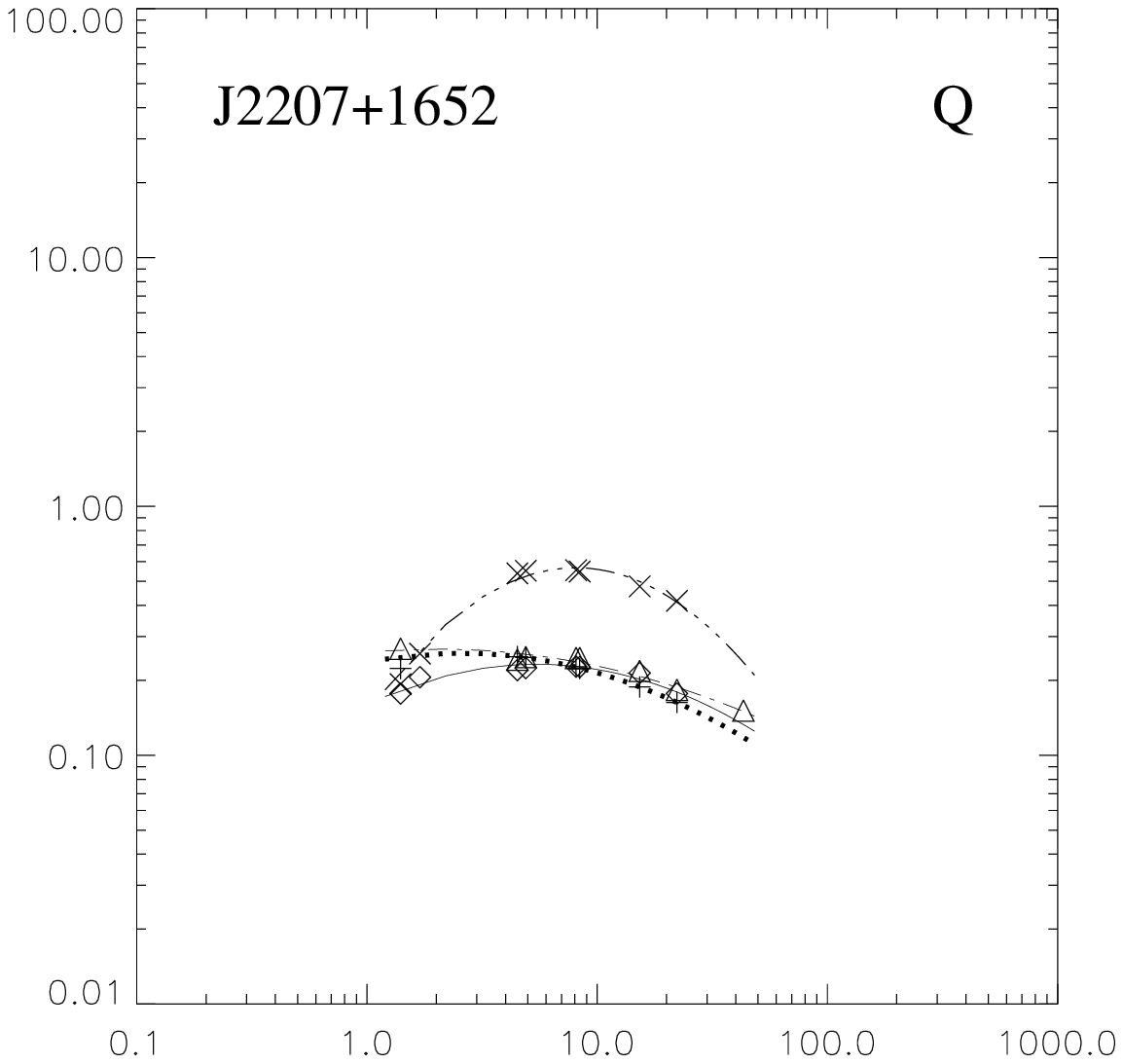}
\includegraphics{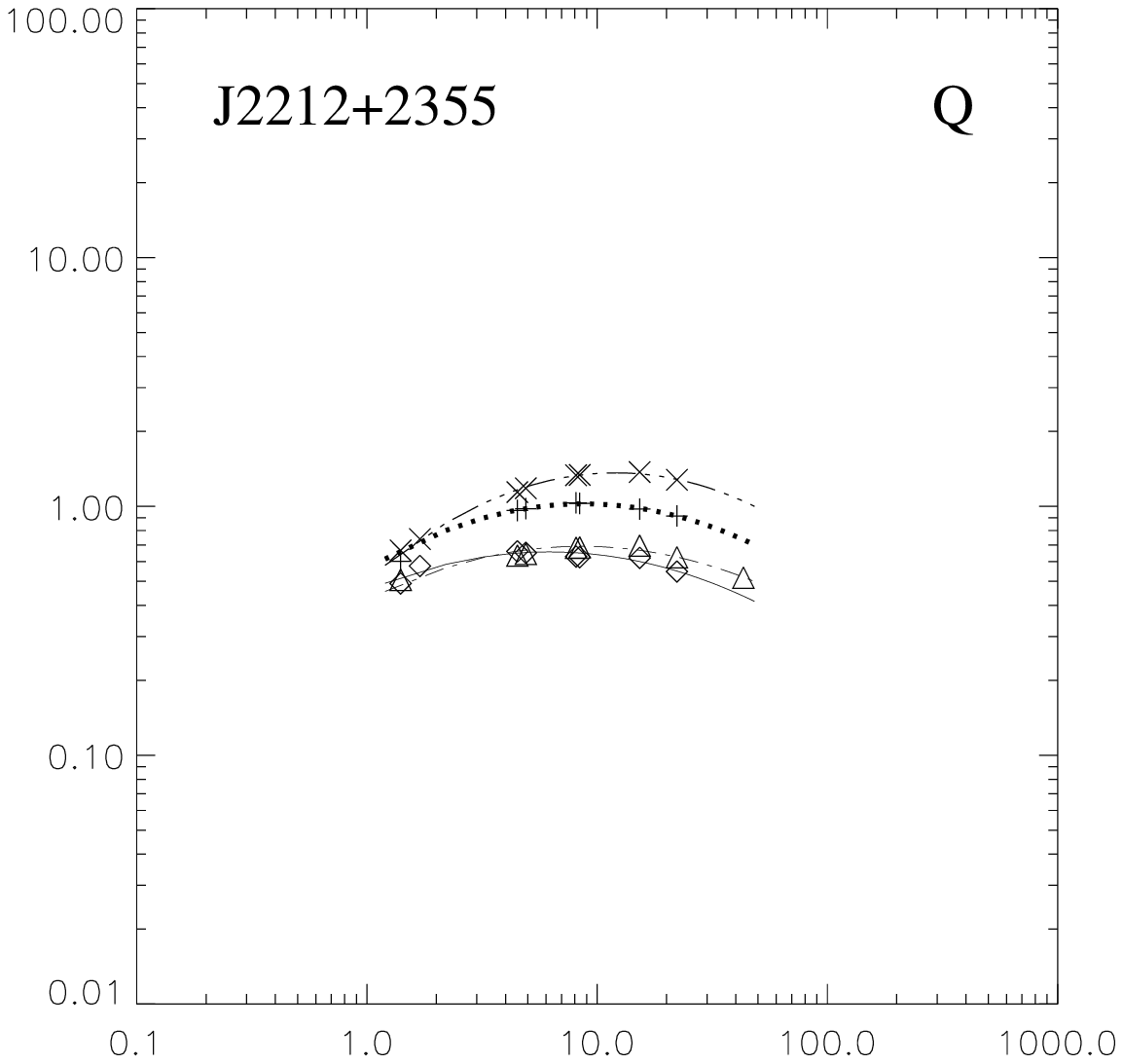}
\includegraphics{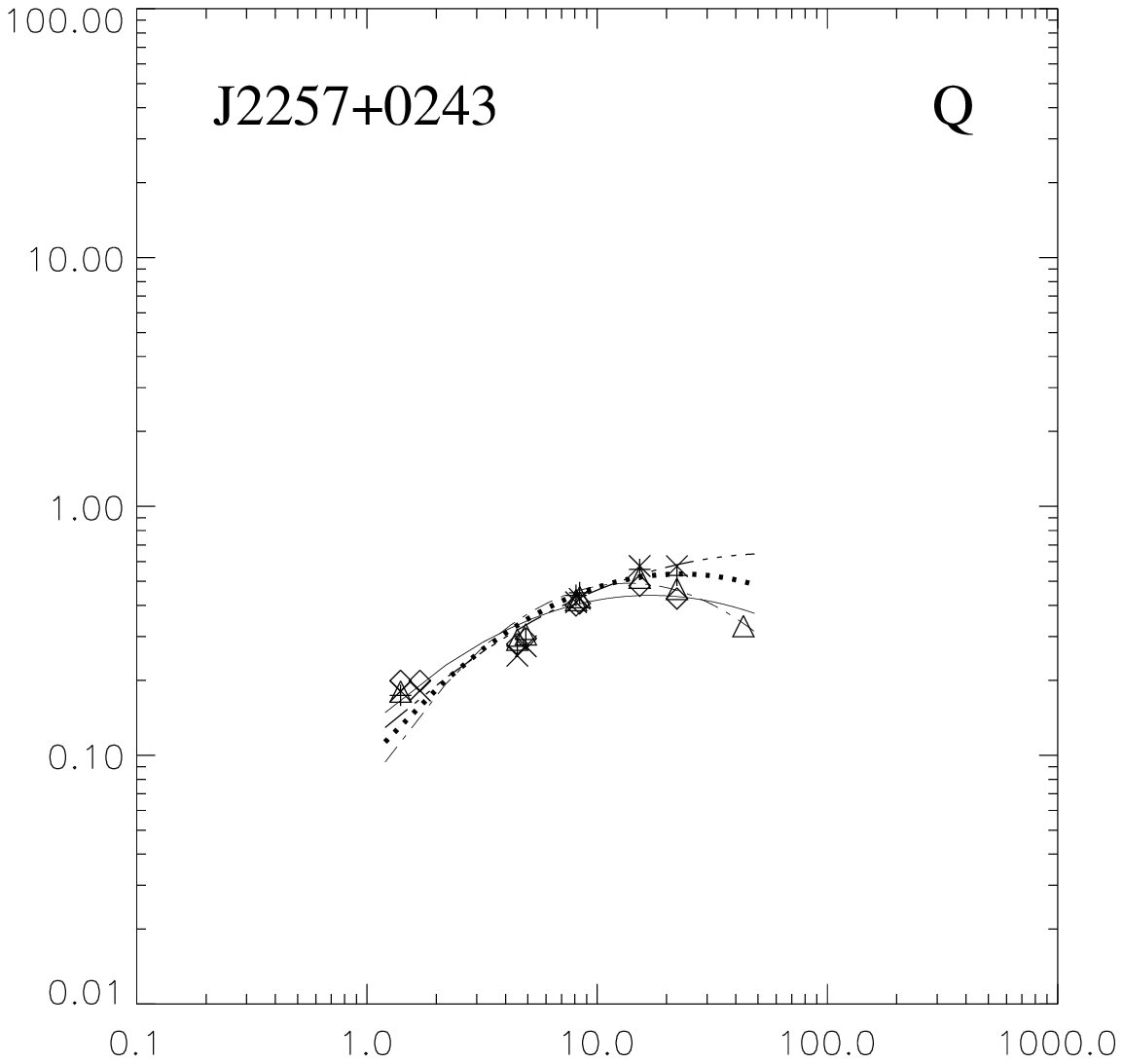}
\includegraphics{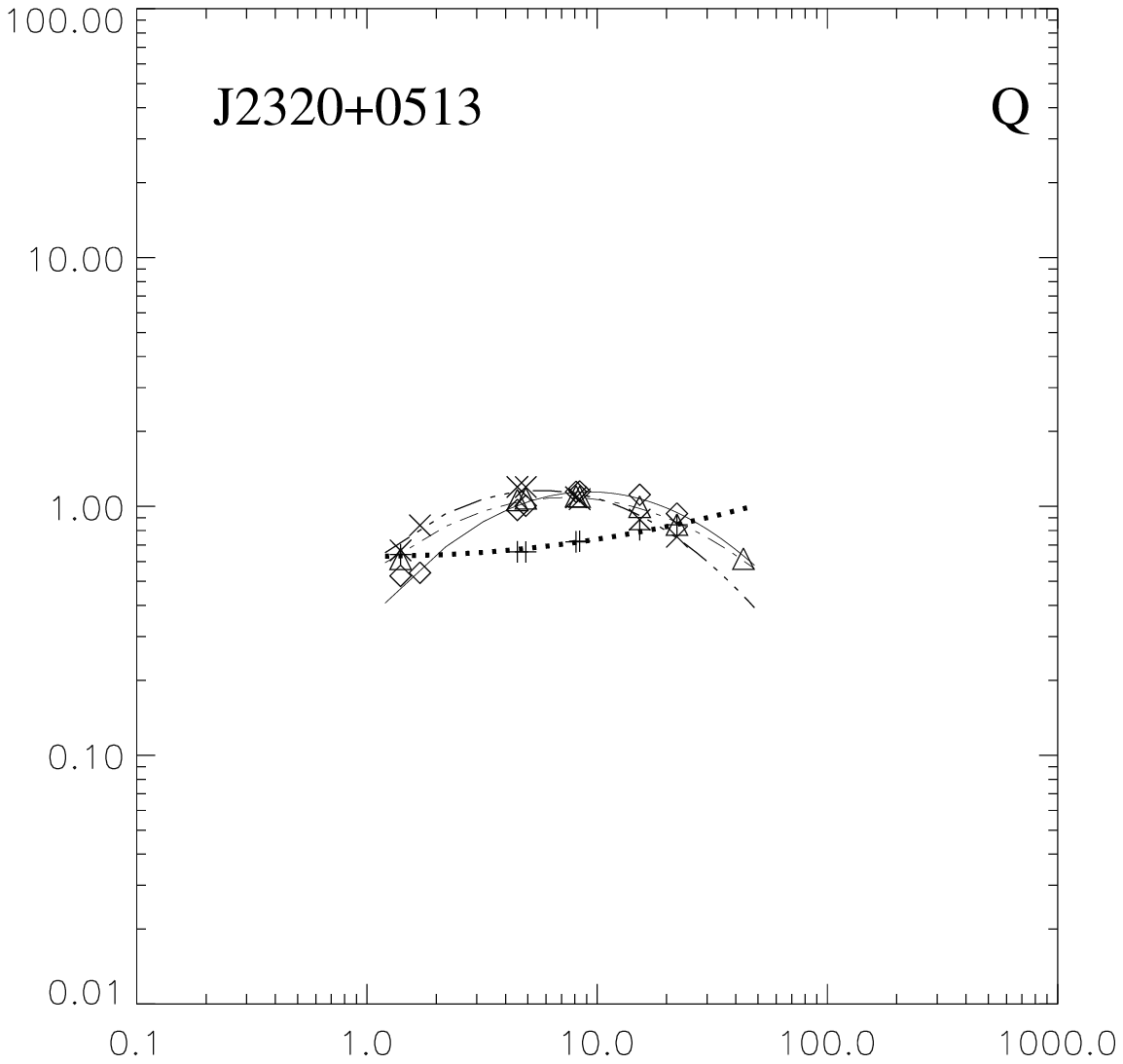}
\includegraphics{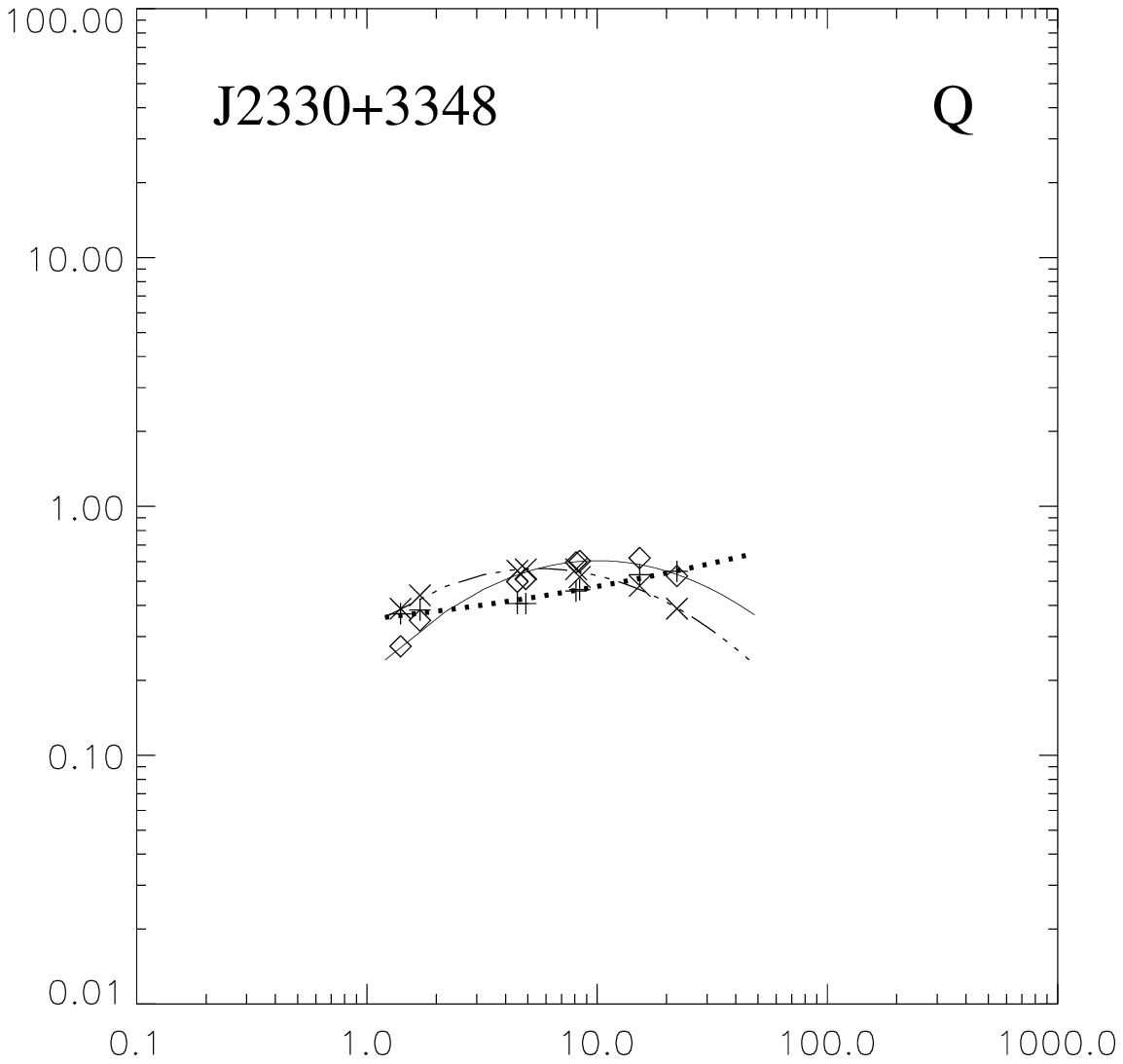}
\vspace{12.9cm}
\caption{Continued.}
\end{center}
\end{figure*}

\section{Multi-frequency VLA observations and data reduction}

Multi-frequency VLA observations of 51 of the 55 candidate
HFPs were carried out in different runs
from September 2003 to March 2004, using filler time. 
About 25 objects were observed twice, in order to identify short-period
variability.\\
The observing bandwidth was chosen to be 50 MHz per IF. 
Separate analysis for each IF in L, C and X bands was carried out to
improve the spectral coverage of the data, as done in previous works
(Dallacasa et al. \cite{dd00}; Tinti et al. \cite{st05}).\\
We obtained the flux density measurements 
in L band (IFs at 1.465 and 1.665 GHz), C band (4.565 and
4.935 GHz), X band (8.085 and 8.465 GHz), U band (14.940 GHz), K band
(22.460 GHz), and for a few datasets in Q band (43.340 GHz).\\
Each source was typically observed for 50 seconds at each band,
cycling through frequencies. Therefore, the flux density measurements
can be considered simultaneous.\\
For each observing run and at each frequency, 
about 3 min were spent on the primary flux
density calibrator, either 3C\,286 or 3C\,48.
Secondary calibrators, chosen to minimize the
telescope slewing time, were observed for 1.5 min at
each frequency every $\sim$20 min. 
This implies that the observations do not have astrometric accuracy.
Accurate positions of the sources can be found in 
the JVAS catalogue (Patnaik et al. \cite{patn92}; Browne et
al. \cite{browne98}; Wilkinson et al. \cite{wil98}).  
Information on the date and duration of the observing runs is
summarized in Table \ref{obslog}. The total observing time also takes into
account 
scans on candidate HFPs from the ``faint'' HFP sample
(Stanghellini et al., in preparation), in order to optimize the observing
schedule.\\
The data reduction was carried out following the standard procedures
for the VLA, implemented in the NRAO AIPS package. Images for each
frequency were produced. 
In order to obtain accurate flux density measurements in the L band,
it was necessary to image several confusing sources falling within
the primary beam, and often accounting for most of the flux density
within the field of view.\\
During some observing runs, strong RFI affected the 1.665 GHz data,
precluding the flux density measurements 
for a few sources at such frequency.\\
The final images were
produced after a few phase-only self-calibration iterations, and source
parameters were measured by means of the task JMFIT, which
performs a Gaussian fit. The integrated flux density was checked with
TVSTAT. 
The flux density measurements at
each frequency and epoch are reported in Table \ref{flux_tab}.\\
Apart from a few sources that were already known to possess an extended
emission (Tinti et al. \cite{st05}), the majority of HFPs are unresolved
with the VLA, even in the Q band.\\
The r.m.s. noise level on the image plane is not relevant for bright
radio sources as our targets. In this case, the main uncertainty comes
from the amplitude calibration errors, which is within (1$\sigma$)
3\% in L, C and X band, 5\% in U band, and 10\% in K and Q band.\\
With these new-multi-frequency observing runs, 48 out of the 55
sources from the bright HFP sample have at least three epochs of VLA
data. The comparison between each epoch allows us to determine the
variability properties of each single source.\\

\section{Results}
The lack of any 
spectral variability is a key element for the selection of genuine
young radio sources. The new epochs of simultaneous
multi-frequency VLA observations, carried out 5-6 years after the first
observing run (Dallacasa et al. \cite{dd00}), provide further information
on the spectral shape and variability of the candidate HFPs, allowing
us to better discriminate between genuine young radio sources and contaminant
objects.\\ 

\begin{table*}
\begin{center}
\begin{tabular}{ccccccccccccccc}
\hline
Source& Id.&z&code&S$_{1.4}$&S$_{1.7}$&S$_{4.5}$&S$_{4.9}$&S$_{8.1}$&S$_{8.4}$&S$_{15.3}$&S$_{22.2}$&S$_{43.2}$&$\alpha_{\rm
below}$&$\alpha_{\rm above}$\\
(1)&(2)&(3)&(4)&(5)&(6)&(7)&(8)&(9)&(10)&(11)&(12)&(13)&(14)&(15)\\
\hline
%& & & & & & & & & & & & & &\\
&&&&&&&&&&&&&&\\
J0003+2129&G &  0.455&d& 100&  125&  258&  262&  233&  227&  146&  70&  -&-0.8&0.9\\
          &  &       &e&  84&  -  &  251&  255&  226&  220&  134&  70&  16&-0.9&1.3\\
J0005+0524&Q &  1.887&c& 166&  174&  212&  206&  169&  165&  122&  84&  -&-&0.6\\
          &  &       &e& 168&  -  &  213&  207&  170&  168&  129&  90&  46&-&0.6\\
J0037+0808&G &       &c&  98&  119&  287&  290&  273&  267&  197& 139&  -&-0.9&0.5\\ 
          &  &       &e&  95&  -  &  287&  293&  278&  274&  210& 148&  79&-0.9&0.6\\
J0111+3906&G &  0.668&c& 509&  597& 1319& 1301&  965&  918&  449& 273& 111&-0.8&1.1\\ 
J0116+2422&EF&       &c& 106&  122&  245&  244&  240&  226&  177& 121&  -&-0.7&0.4\\
          &  &       &e&  99&  -  &  258&  238&  240&  225&  177& 129&  -&-&0.4\\
J0217+0144&Q &  1.715&c& 577&  514&  337&  326&  268&  263&  244& 239&  -&-&0.3\\
          &  &       &e& 571&  -  &  326&  315&  257&  264&  256& 254&  -&-&0.3\\
J0329+3510&Q &  0.5  &d& 410&  443&  603&  611&  586&  578&  523& 447& -&-0.3&0.2\\
          &  &       &g& 563&  -  &  644&  625&  540&  535&  468& 401& 302&-&0.2\\
J0357+2319&Q &       &d& 116&  124&  242&  255&  309&  312&  382& 373& -&-0.5&-\\
          &  &       &g& -  &  -  &  185&  187&  208&  210&  239& 232& 212&-0.2&0.1\\
J0428+3259&G &  0.479&d& 148&  195&  513&  524&  539&  531&  421& 269& -&-0.8&0.6\\
          &  &       &g& 172&  199&  497&  512&  522&  520&  394& 253& 104&-0.7&0.9\\
J0519+0848&EF&       &c& 269&  263&  394&  393&  373&  372&  365& 336& -&-0.4&0.1\\
          &  &       &g& 262&  273&  314&  320&  322&  321&  372& 382& 348&-0.1&0.1\\
J0625+4440&BL&       &c& 173&  186&  223&  222&  219&  218&  197& 174& -&-0.2&0.1\\
          &  &       &g& 121&  131&  180&  182&  187&  187&  180& 161& 121&-0.2&0.2\\
J0638+5933&EF&       &c& 249&  315&  605&  619&  680&  678&  701& 653& -&-0.5&-\\
          &  &       &g& 300&  325&  597&  608&  649&  651&  706& 643&  412&-0.4&0.5\\
J0642+6758&Q &  3.180&c& 226&  272&  436&  429&  343&  332&  205& 131& -&-0.5&0.7\\
          &  &       &g& 190&  238&  405&  402&  320&  312&  204& 123&  44&-0.6&0.9\\
J0646+4451&Q &  3.396&c& 444&  660& 2860& 3071& 4068& 4124& 3944&3184& -&-1.3&0.3\\
          &  &       &g& 586&  717& 2894& 3103& 4064& 4094& 4039&3175& 1996&-1.1&0.5\\
J0650+6001&Q &  0.455&c& 507&  633& 1161& 1150&  994&  975&  674& 467& -&-0.7&0.5\\
          &  &       &g& 480&  626& 1150& 1106&  958&  935&  798& 466&  204&-0.7&0.7\\
J0655+4100&G &0.02156&c& 209&  232&  323&  330&  354&  354&  319& 268& -&-0.3&0.3\\
          &  &       &g& 198&  233&  351&  341&  369&  369&  360& 307& 226&-0.3&0.3\\
J0722+3722&Q &1.63   &d& 148&  180&  199&  199&  178&  174&  138&  99&  -&-&0.4\\
          &  &       &g& 171&  188&  203&  203&  176&  172&  125&  83&  44&-&0.7\\
J0927+3902&Q &0.6948 &d&2810& 3615& 9545& 9760&10027& 9937& 8813&7237& -&-0.8&0.3\\
J1016+0513&Q &       &d& 633&  711&  489&  478&  428&  420&  388& 327& -&-&0.3\\
J1045+0624&Q &1.507  &d& 185&  245&  296&  289&  268&  266&  226& 152& -&-&0.4\\
J1148+5254&Q &1.632  &a& 108&  -  &  396&  414&  460&  450&  411& 277&  -&-0.9&0.4\\
J1335+4542&Q &2.449  &a& 267&  -  &  821&  821&  666&  646&  392& 263&  -&-0.9&0.7\\
J1335+5844&EF&       &a& 299&  -  &  745&  744&  727&  726&  585& 449&  -&-0.8&0.3\\
J1407+2827&G &0.0769 &b& 865& 1133& 2519& 2532& 2147& 2071& 1027& 542&  -&-0.8&1.0\\
J1412+1334&EF&       &b& 191&  248&  346&  341&  286&  277&  189& 121&
-& -0.5& 0.6\\
J1424+2256&Q &3.626  &b& 371&  480&  652&  637&   - &  394&  247& 144&  -&-0.4&0.9\\
J1430+1043&Q &1.710  &b& 321&  423&  865&  861&  780&  767&  582& 473&  -&-0.8&0.4\\
J1505+0326&Q &0.411  &b& 382&  428&  608&  620&  620&  610&  515& 468&  -&-0.3&0.2\\
J1511+0518&G &0.084  &b&  92&  123&  569&  608&  801&  811&  763& 573&  -&-1.3&0.3\\
J1526+6650&Q &3.02   &d& 109&  -  &  426&  440&  380&  369&  211& 104&  -&-1.1&1.0\\  
J1623+6624&G &0.203  &d& 154&  -  &  282&  281&  257&  254&  191& 123&  -&-0.5&0.5\\
          &  &       &i& 162&  179&  265&  263&  227&  221&  164& 122&  80&-0.4&0.5\\
J1645+6330&Q &2.379  &d& 280&  -  &  419&  430&  464&  463&  410& 273&  - &-0.3&0.5\\
          &  &       &i& 270&  288&  391&  399&  430&  432&  401& 322& 192&-0.3&0.5\\
J1717+1917&Q &1.81   &h& 211&  211&  232&  232&  210&  208&  174& 144&  124&-0.1&0.3\\
J1735+5049&G &       &h& 445&  -  &  925&  943&  898&  888&  678& 485&  262&-0.6&0.6\\ 
          &  &       &i& 436&  479&  935&  945&  886&  873&  623& 408&  140&-0.6&0.9\\
J1751+0939&BL&0.322  &h& 980& 1120& 2714& 2848& 3825& 3897& 4556&4583& 4642&-0.5&-\\
J1800+3848&Q &2.092  &f& 262&  326&  793&  836& 1109& 1128& 1249&1147&  758&-0.7&0.5\\
&&&&&&&&&&&&&&\\
\hline
\end{tabular} 
\vspace{0.5cm}
\end{center}   
\caption{The VLA flux density of the 51 candidates HFPs observed
  with the VLA during the observing 
  runs presented in this paper. Col. 1: source name
  (J2000); Col. 2: optical identification; Col. 3: redshift; Col. 4:
  the observation code, from Table \ref{obslog};
  Col. 5, 6, 7, 8, 9, 10, 11, 12, and 13: VLA flux density in mJy at 1.4,
  1.7, 4.5, 4.9, 8.1, 8.4, 15.3, 22.2 and 43.2 GHz respectively;
  Col. 14: spectral index below the peak
  frequency; Col. 15: spectral index above the peak frequency.} 
\label{flux_tab}
\end{table*}  

\addtocounter{table}{-1}
\begin{table*}
\begin{center}
\begin{tabular}{ccccccccccccccc}
\hline
Source&
Id.&z&code&S$_{1.4}$&S$_{1.7}$&S$_{4.5}$&S$_{4.9}$&S$_{8.1}$&S$_{8.4}$&S$_{15.3}$&S$_{22.2}$&S$_{43.2}$&$\alpha_{\rm
below}$&$\alpha_{\rm above}$\\
(1)&(2)&(3)&(4)&(5)&(6)&(7)&(8)&(9)&(10)&(11)&(12)&(13)&(14)&(15)\\
\hline
&&&&&&&&&&&&&&\\
J1840+3900&Q &3.095  &f& 123&  147&  161&  163&  165&  165&  165& 164&  140&-0.1&0.1\\
J1850+2825&Q &2.560  &f& 235&  280& 1097& 1185& 1515& 1520& 1402&1132&  605&-1.1&0.6\\
J1855+3742&G &       &f& 180&  181&  360&  344&  215&  206&  123&  86&   50&-0.6&0.9\\
J2021+0515&Q &       &f& 360&  442&  520&  506&  407&  397&  285& 206&  100&-0.3&0.7\\
J2024+1718&Q &1.05   &f& 307&  324&  594&  609&  623&  619&  519& 403&  242&-0.5&0.6\\
J2101+0341&Q &1.013  &c& 483&  493&  497&  498&  552&  555&  704& 724&  -&-0.2&-\\
          &  &       &f& 431&  -  &  498&  508&  665&  687&  954&1031& 883&-0.3&0.2\\
J2114+2832&Q &2.345  &d& 414&  493&  612&  599&  545&  535&  458& 357& -&-0.3&0.3\\
J2123+0535&Q &1.878  &f&2185& 2260& 2831& 2879& 3054& 3057& 3021&2740& 1992&-0.1&0.2\\
J2136+0041&Q &1.932  &c&4234& 5231&10284&10225& 8763& 8601& 6252&4401& -&-0.7&0.5\\
          &  &       &f&3752& 4823&10193&10150& 8779& 8599& 6343&4720& 2578&-0.8&0.6\\
J2203+1007&G &       &c& 107&  156&  315&  311&  240&  231&  125&  67&   21&-0.9&1.2\\
J2207+1652&Q &1.64   &d& 177&  206&  220&  224&  227&  226&  214& 177&  -&-0.1&0.2\\
          &  &       &e& 268&  -  &  242&  248&  246&  246&  218& 183&  151&-&0.2\\
J2212+2355&Q &1.125  &d& 490&  576&  659&  651&  629&  625&  631& 547&  -&-0.2&0.1\\
          &  &       &e& 507&  -  &  634&  644&  682&  684&  695& 624&  517&-0.1&0.3\\
J2257+0243&Q &2.081  &c& 199&  199&  280&  295&  400&  408&  480& 426&  -&-0.4&0.3\\
          &  &       &e& 180&  -  &  291&  306&  420&  432&  517& 466&  330&-0.5&0.5\\
J2320+0513&Q &0.622  &c& 526&  541&  968& 1006& 1144& 1148& 1114& 936&  -&-0.5&0.2\\
          &  &       &e& 615&  -  & 1060& 1076& 1098& 1094&  996& 844&  616&-0.4&0.3\\
J2330+3348&Q &1.809  &d& 274&  -  &  498&  511&  596&  603&  620& 525&  -&-0.3&0.4\\
&&&&&&&&&&&&&&\\
\hline
\end{tabular} 
\vspace{0.5cm}
\end{center}   
\caption{Continued}
\end{table*}

\subsection{Spectral shape}
The shape of the radio spectrum is one of the key identifying
characteristics of young radio sources. The overall shape of their
radio spectra is convex with a peak at high frequencies that is
likely due to synchrotron self-absorption within the small
radio emitting region (Snellen et al. \cite{sn00}), 
although free-free absorption (Bicknell et
al. \cite{bick97}; Kameno et al. \cite{ka00}) may play a role.\\
On the other hand, blazar objects, which usually show flat spectra,
sometimes can display a convex spectrum as a consequence of an
increment of the flux density, when a single, homogeneous and boosted  
component at
the jet base temporarily dominates the radio emission. As the component
adiabatically expands,
the flux density decreases, leading the spectrum back to a flat shape.\\
Following the approach from Dallacasa et al. (\cite{dd00}) and Tinti et
al. (\cite{st05}), we fit the simultaneous radio spectra with a
function that provides the flux density 
and the frequency of the spectral peak, but without any physical content.     
Since most of the sources are optically thin at 43 GHz, the
new flux density measurements at such frequency, which was not
available in the previous epochs, provide very tight
constraints on the fits and a better determination of the peak.\\
In Fig. \ref{plot}, we show the radio spectra of all the sources observed
at the various epochs.\\
Following the approach by Torniainen et al. (\cite{torni05}) we
compute the spectral indices $\alpha_{\rm below}$ and 
$\alpha_{\rm above}$ of the overall spectrum
(Col. 14 and 15 of Table \ref{flux_tab}) fitting a straight line to the parts
below and above the spectral peak respectively. 
We consider ``flat'' those sources with both $\alpha_{\rm below} > -0.5$
and $\alpha < 0.5$ (where $S \propto \nu^{- \alpha}$).
In a few sources, depending on the peak frequency, we could fit either
$\alpha_{\rm below}$ or $\alpha_{above}$, in order to avoid the
flattening near the spectral peak. In this case, sources with the
spectral index in the range of -0.5 and 0.5 are considered
flat-spectrum objects. 
We find that 18 objects, labeled with an ``F'' in Column 9 of 
Table $\ref{variability}$, 
no longer show the convex spectrum. Such sources are labelled
``flat'' in Column 5 and 6 of Table \ref{variability}, and are 
definitely 
classified as ``blazar'', and removed from the sample of genuine
young radio sources. Six of these sources (J0217+0144, J0329+3510,
J0357+2319, J2123+0535, J2320+0513 and J2330+3348) were already
found with a flat spectrum in the second epoch observations, and
already rejected from the sample of candidate HFPs (Tinti et
al. \cite{st05}).\\

\subsection{Variability index}
As already mentioned in Section 3.1, given the large Doppler factors
characterizing the blazar jets, 
the flux density variability can
substantially modify the observed spectrum of blazars on short timescales.\\
On the contrary, young radio sources 
should not display any significant flux density variability and they can be
considered as
the least variable class of 
extragalactic radio sources (O'Dea \cite{odea98}), with a mean variation of
$\sim$ 5\% (Stanghellini et al. \cite{cs05}).
Therefore, genuine young HFP objects, 
considered to be newly born radio sources, should not
display significant variability.\\ 
We analyze the variability of the sources in terms of the quantity:\\

\begin{equation}
V= \frac{1}{m}\sum_{i=1}^m \frac{(S_{i}- \overline S_{i})^2}{\sigma^2_{i}}
\label{eq_var}
\end{equation}
which is a multi-epoch 
generalization of the variability index defined by Tinti et
al. (\cite{st05}).
S$_{i}$ is the flux density at the {\it i}-th frequency
measured at one epoch, while $\overline S_{i}$ is the mean value
computed averaging the flux density at the {\it i}-th frequency
measured at all the available epochs; $\sigma_{i}$ is the error on  
S$_{\rm i}$ $-$ $\overline S_{i}$, and $m$ is the number
of sampled frequencies.\\
Columns 7 and 8 of Table \ref{variability} report the 
variability between each new epoch and the mean value obtained by
averaging all the available epochs.
The variability index V has been computed for each single new epoch,
rather than considering all the epochs together. In this case the
availability of two distinct values better indicates the presence of 
flux-density bursts.\\
Comparing the variability distributions computed from Eq. \ref{eq_var}
for all the sources of each epoch, 
the KS test does not detect any significant difference
($>$99\%). 
This means that the flux density 
of the majority of the observed sources has not changed its behaviour
with time.\\
From the comparison of the multi-epoch radio spectra 
we find:

\begin{itemize}
\item 18 sources with a convex spectrum at the first epoch (Dallacasa
  et al. \cite{dd00}) show a flat spectrum in at least one of the subsequent
  epochs. They have an ``F'' in Column 9 of Table
  \ref{variability} and they are rejected from the sample;
\item Of the aforementioned sources the quasar
  J2320+0513 shows a continuous alternation
  between a flaring-phase with a convex-shape spectrum, and a
  quiescent-phase where the spectrum is flat. In
  particular, during the second-epoch observations, the source
  was characterized by a flat spectrum, while during our
  last observing run its spectrum becomes convex again and with almost
  the same flux density as the first epoch. This example points out
  how important a multi-epoch, multi-frequency flux density
  monitoring is in order to reveal blazar objects.
\item 14 sources maintain a convex spectrum at the various 
  observing epochs,
  although with significant flux density variability ($V$ $>$3), and
  they have a ``V'' in Column 9 of Table \ref{variability}. 
\item 20 sources preserve the convex spectrum and do not show 
  significant flux density variability ($V$ $<$3), and they have
  an ``H'' in Column 9 of Table \ref{variability}.
\end{itemize}

When we compare the variability properties between sources with
different optical identification by means of a KS test, we find that 
there is a  
difference ($>$90\%) 
between the variability of galaxies and quasars, as also found by
  Torniainen et al. (\cite{torni07}), 
supporting the idea that radio sources with different
optical identification represent different radio source populations
(Stanghellini et
al. \cite{cs05}; Orienti et al. \cite{mo06}).
This difference becomes
stronger ($>$99\%) if in the KS test we consider sources with or without a
CSO-like pc-scale morphology (Orienti et al. \cite{mo06}).\\

\subsection{Peak frequency}

So far the anti-correlation (O'Dea \& Baum \cite{odea97}) between the
peak frequency and the projected linear size has been explained mainly
in terms of Synchrotron Self-Absorption (SSA): as the radio source
expands the turnover moves progressively to lower frequencies as the
result of a decreased energy density within the emitting region.\\
In Table \ref{variability} we report the peak frequency measured at the
various epochs. A KS test considering all the observed sources
does not detect any
significant ($>$99\%) difference among the distributions of the peak frequency
at the various epochs. This result is expected since the time
elapsed between the observing runs is too short to detect any modification
in the spectra of the growing sources.\\
However, if we consider individual objects we find that
most of them have a smaller $\nu_{p}$ at the subsequent epochs, consistent
with the evolution models. The median value of the peak frequency of
the whole sample has
continuously decreased: $\nu_{\rm p}$ = 6.7 GHz at the first epoch
(Dallacasa et
al. \cite{dd00}), $\nu_{\rm p}$ = 6.3 GHz at the second epoch
(Tinti et al. \cite{st05}), and $\nu_{\rm p}$ = 6.0 GHz at the
epochs presented here.\\
If we consider young HFP candidates and blazar objects separately,
we find that the former show a decreasing peak frequency, from a median
value of 6.0 GHz
during the first epoch (Dallacasa et al. \cite{dd00}) to 5.5 GHz at
the subsequent epochs. The latter have a median peak frequency which
does not follow a monotonic trend: $\nu_{\rm p}$ = 7.4 GHz at the
first epoch, $\nu_{\rm p}$ = 8.6 GHz at the second epoch, $\nu_{\rm
  p}$ = 6.2 at the third epoch and $\nu_{\rm p}$ = 6.8 GHz
at the last epoch. A KS test does find a difference ($>$99\%)
between the
peak frequency distributions between young HFP candidates and blazars,
although at the first two epochs only.\\

\begin{figure}
\begin{center}
\includegraphics{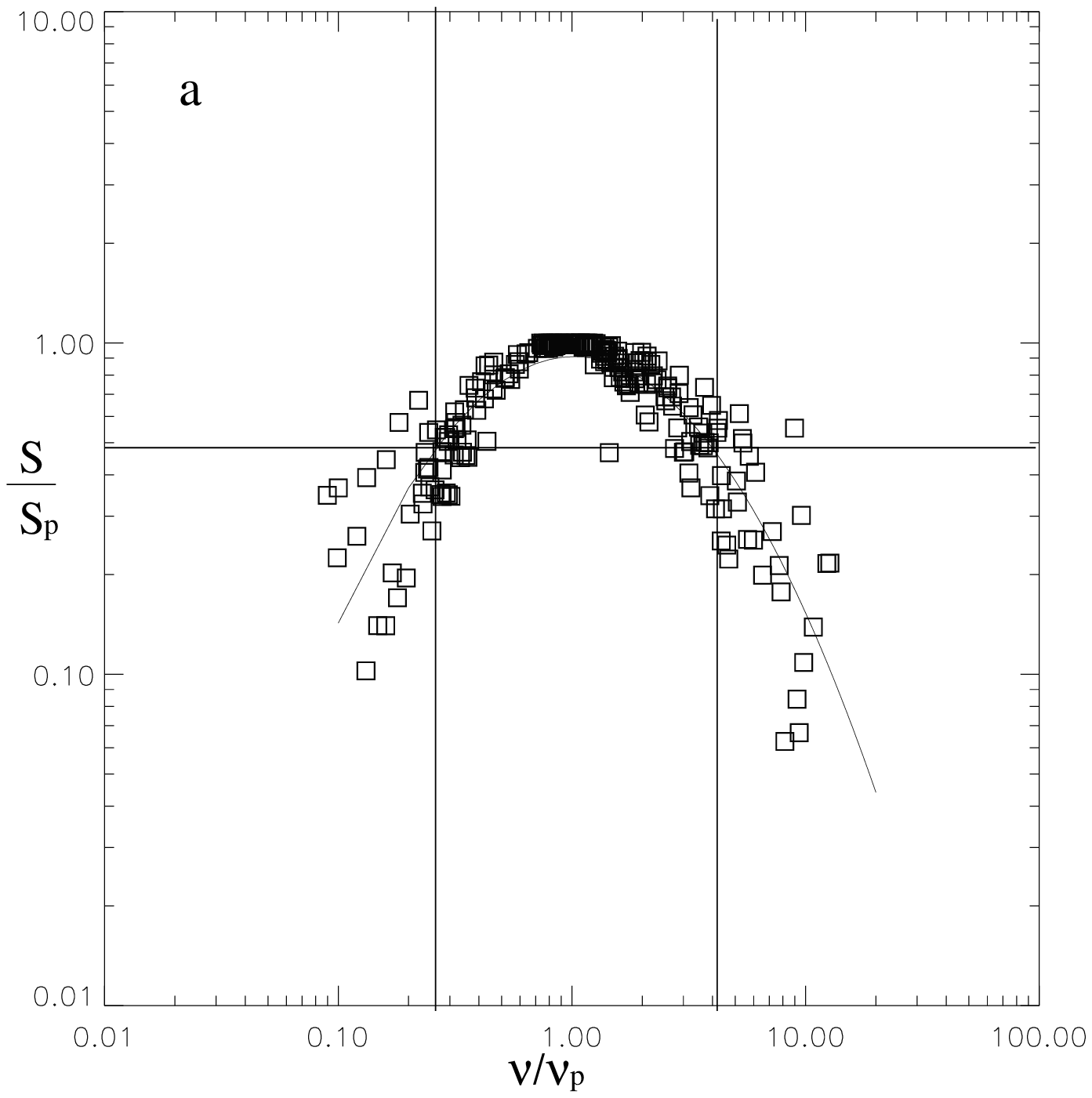}
\includegraphics{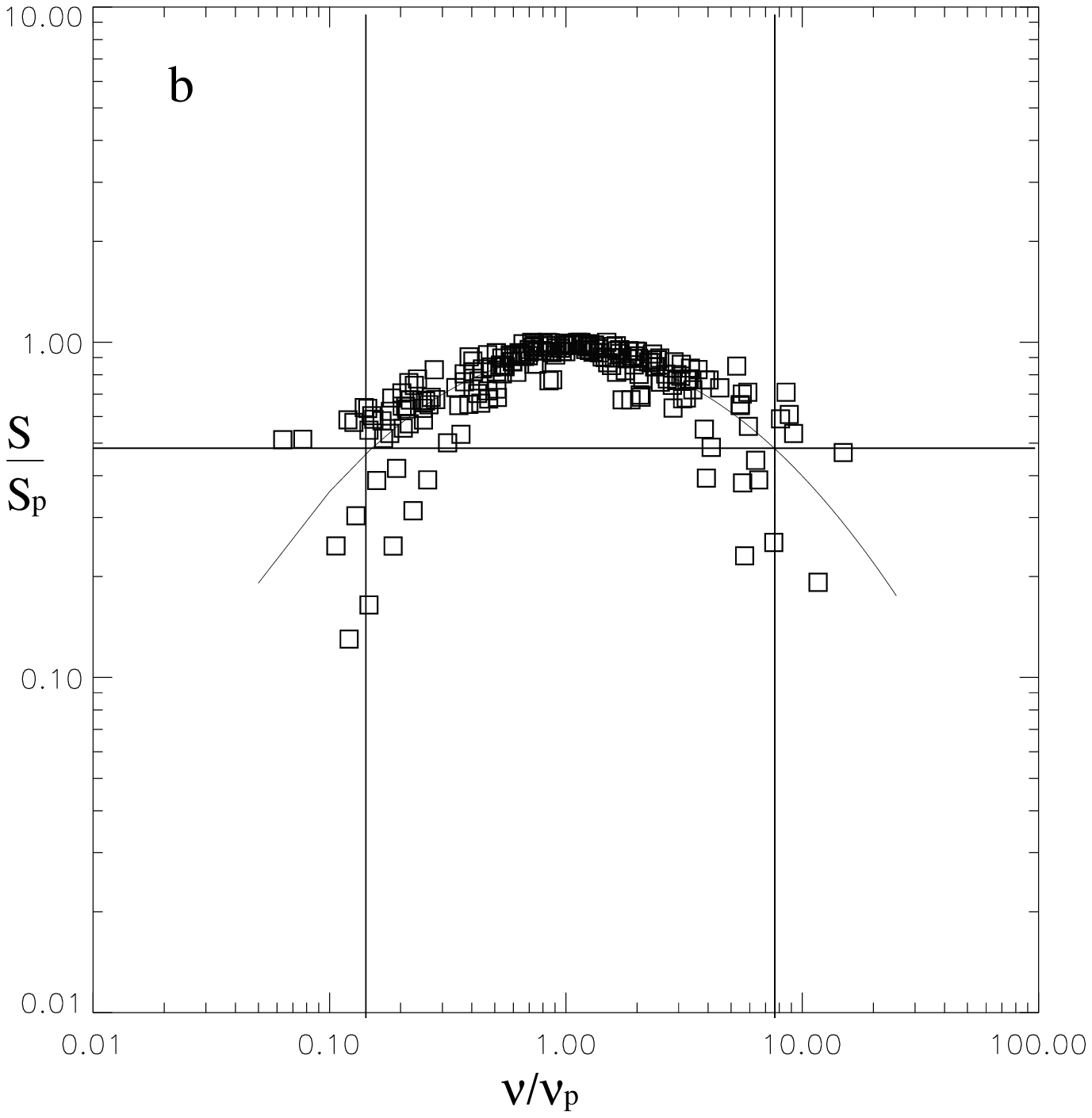}
\vspace{12.0cm}
\caption{Canonical radio spectra computed for radio
  sources with $V$$<$3 ({\it top panel}); radio sources with $V$$>$3
  ({\it bottom panel}). The
  continuous line indicates the least-square fit to the data.}
\label{synt}
\end{center}
\end{figure}

On the other hand, among the 14 sources with $V$$>$3 (Section 3.2) we
find 3 objects in which the
turnover frequency shows a remarkable change, although the overall
radio spectrum maintains a convex shape. 
Two of these sources (the quasars
J1645+6330 and J2024+1718)
show a large drop in the peak frequency.
We estimate the size of the emitting region by two independent methods.
In one case, we use the
relationship from O'Dea (\cite{odea98}):\\

${\rm log} \nu_{\rm p} = -0.21 -0.65 {\rm log} LS$\\

\noindent which relates the intrinsic spectral peak and the linear size. 
In the other one, we assume synchrotron self-absorption theory
(Kellerman \& Pauliny-Toth \cite{kpt81}),
in which:\\

\begin{equation}
\theta = B^{\frac{1}{4}}S_{p}^{\frac{1}{2}} \nu_{p}^{- \frac{5}{4}}(1+z)^{\frac{1}{4}} 
\label{self}
\end{equation}

\noindent
where $B$ is the magnetic field, $S_{p}$ the peak
flux density, $\nu_{p}$ the turnover frequency and $z$ the redshift.
For the magnetic field we consider the values reported by Orienti
et al. (\cite{mo06}), obtained assuming equipartition
conditions. \\
As will be discussed in more detail in the following section, if
we assume that the magnetic field is frozen within an
adiabatically-expanding homogeneous region, its value can be
considered constant during the 5 years that elapsed between the most
distantly separate
observing runs.\\
The increment in size 
obtained in both ways are in good agreement,
and corresponds to an
expansion velocity $\geq$ $c$, which is clearly unrealistic for
unboosted young objects. We conclude that such sources are
beamed objects and we reject them as genuine candidate
HFPs.\\
The other object, the BL Lac J1751+0939, shows an increment of the peak
frequency (from 8.5 to 29 GHz), which can be interpreted in terms of
different knots in the jet base dominating the radio emission at
the two epochs. This source has a well known history of flux density
and spectral variability, and therefore there is no doubt about its
blazar nature.\\ 
In general, the observed peak frequency of galaxies and quasars are
similar as a consequence of the selection criteria. However, if we
consider the intrinsic turnover frequency, the KS test finds a clear
difference ($>$99\%) concerning:\\
a) galaxies and quasars;\\
b) the parsec-scale structure depending on its having the CSO-like
morphology typical of young radio sources (Orienti et
al. \cite{mo06}).\\
%between galaxies and quasars and between sources with or
%without CSO-like morphology (Orienti et al. \cite{mo06}). 
While in the former case 
the different distribution is easily explained in terms of redshift
(quasars are found at higher redshifts than galaxies), in the latter, 
such a segregation is likely to be indicating two different source
populations.\\
In order to analyze the spectral shape of our sample sources, we
construct a canonical radio spectrum for sources with either $V$$<$3
(Fig. \ref{synt}a)
or $V$$>$3 (Fig. \ref{synt}b).\\    
Following the work of de Vries et al. (\cite{dv97}), 
the canonical radio spectra have been obtained by normalizing the
observed frequencies and fluxes by the source peak frequency and the
peak flux density, computed averaging all the epochs:\\

\begin{equation} 
\nu = \frac{1}{m} \sum_{i=1}^{m} \frac{\nu_{i}}{\nu_ {p\,i}}
\label{nu_synt}
\end{equation}

\begin{equation} 
S = \frac{1}{m} \sum_{i=1}^{m} \frac{S_{i}}{S_ {p\,i}}
\label{s_synt}
\end{equation}

where $\nu$ and $S$ are the average normalized frequency and flux density
respectively. $\nu_{i}$ and $\nu_{p\, i}$ are the observed frequency
and the observed turnover frequency at the {\it i-}th epoch, while 
$S_{i}$ and $S_{p\, i}$ are the observed flux density and the observed
peak flux density at the {\it i-}th epoch, and $m$ the epochs available.\\ 
Sources with $V$$<$3 or $V$$>$3 display different canonical spectral
shape. The former have a quite convex spectrum, with a narrow width
(FWHM $\sim$1.7). The spectral indices are: $\alpha_{\rm thick}$
$\sim$-0.9$\pm$0.1 and $\alpha_{\rm thin}$ $\sim$ 0.7$\pm$0.1.\\
The latter have a flatter spectral shape, with a FWHM $\sim$3.6 and
spectral indices: $\alpha_{\rm thick}$
$\sim$-0.4$\pm$0.1 and $\alpha_{\rm thin}$ $\sim$ 0.4$\pm$0.1.\\

\begin{table*}
\begin{center}
\begin{tabular}{ccccccccccc}
\hline
Source&Morph.&$\nu_{\rm ep1}$&$\nu_{\rm ep2}$&$\nu_{\rm ep3}$&$\nu_{\rm ep4}$&V$_{\rm
  ep1}$&V$_{\rm ep2}$&Var.\\  
(1)&(2)&(3)&(4)&(5)&(6)&(7)&(8)&(9)\\
\hline
&&&&&&&&&&\\
{\bf J0003+2129}&CSO&  5.7$\pm$0.1&  5.4$\pm$0.1 &  5.2$\pm$0.9&5.29$\pm$0.02&   2.67d&  3.32e&   H\\
{\bf J0005+0524}&CSO&  4.13$\pm$0.09&  3.40$\pm$0.09&  3.6$\pm$0.7&  3.5$\pm$0.7&  1.65c&  2.1e6&  H\\
{\bf J0037+1109}&CSO&  5.9$\pm$0.1&  6.2$\pm$0.2&  6.2$\pm$0.7&  6.0$\pm$0.7&  1.03c&  1.47e&  H\\
{\bf J0111+3906}&CSO&  4.76$\pm$0.06&  4.68$\pm$0.07&  4.66$\pm$0.19&     & 4.63c&   &      H\\
{\bf J0116+2422}&Un&  5.1$\pm$0.1&  6.3$\pm$0.3&  6.0$\pm$0.8&  6.0$\pm$1.1&  4.93c& 3.73e &  V\\
J0217+0144&Un& 18$\pm$1& flat& flat& flat&94.72c&94.10e&  F\\
J0329+3510&CJ&  6.7$\pm$0.3& flat&  flat& flat&
4.31d& 24.43g&  F \\
J0357+2319&Un& 12$\pm$1& flat& flat& flat&  12.67d&
42.48g&      F \\
{\bf J0428+3259}&CSO&  7.3$\pm$0.2&  6.8$\pm$0.2&6.9$\pm$0.5& 6.55$\pm$0.38& 1.12d&2.40g& H\\
J0519+0848&Un& $>$22 &7.4$\pm$0.6&7.26$\pm$0.01&flat&13.28c&19.84g&  F\\
J0625+4440&Un&13$\pm$2&7.4$\pm$1.0&flat &flat &18.18c&62.38g&    F\\
{\bf J0638+5933}&CSO&12$\pm$2&9.2$\pm$0.7&11.35$\pm$0.30&10.22$\pm$0.20&1.79c&3.74g&    H\\
{\bf J0642+6758}&Un&4.5$\pm$0.1&4.08$\pm$0.08&4.41$\pm$0.46&4.41$\pm$0.54&0.71c&9.18g&   V\\
{\bf J0646+4451}&MR&15$\pm$2&11.2$\pm$0.6&10.36$\pm$0.09&10.65$\pm$0.05&19.19c&32.44g&   V\\
{\bf J0650+6001}&CSO&7.6$\pm$0.3&5.2$\pm$0.2&5.23$\pm$0.18&5.45$\pm$0.15&2.89c&3.12g&   V\\
J0655+4100&Un&7.8& & flat & flat &0.46c&3.73g&  F\\
{\bf J0722+3722}&MR&4.3& &4.0$\pm$0.7& 3.4$\pm$0.6& 2.67d &4.76g & H\\
J0927+3902&CJ&6.9& &8.3$\pm$0.1& &15.45d & &V \\
J1016+0513&Un&7.1& &flat& &72.80d & &F \\
{\bf J1045+0624}&Un&3.7& &4.7$\pm$0.5& &5.75d & & H\\
{\bf J1148+5254}&CSO&8.7& &7.9$\pm$0.7& &9.74a& & V\\
{\bf J1335+4542} &CSO&5.1$\pm$0.1&4.9$\pm$0.1&5.1$\pm$0.4& &4.64a& & H\\
{\bf J1335+5844} &CSO&6.0$\pm$0.2&5.5$\pm$0.1&6.50$\pm$0.34& &2.84a& & H\\
{\bf J1407+2827} &CSO&5.34$\pm$0.05&5.01$\pm$0.08&4.95$\pm$0.01& &2.02b& & H\\
{\bf J1412+1334} &Un&4.7$\pm$0.1&4.18$\pm$0.09&4.3$\pm$0.5& &0.97b& & H\\
{\bf J1424+2256} &Un&4.13$\pm$0.07&3.94$\pm$0.06&3.7$\pm$0.3& &8.80b& & V\\
{\bf J1430+1043} &MR&6.5$\pm$0.2&5.7$\pm$0.1&6.2$\pm$0.1& &1.37b& & H\\
J1505+0326 &Un&7.1$\pm$0.4&6.8$\pm$0.4& flat & &30.13b& & F\\
{\bf J1511+0518} &CSO&11.1$\pm$0.4&10.8$\pm$0.4&10.10$\pm$0.04& &10.24b& & V\\
{\bf J1526+6650}&MR&5.7$\pm$0.1&5.5$\pm$0.1&5.5$\pm$0.9& &2.04d & & H \\
{\bf J1623+6624} &Un&6.0$\pm$0.2&6.0$\pm$0.2&5.1$\pm$0.8&4.5$\pm$0.5&
0.87d&11.16i& V \\
J1645+6330 &Un&14$\pm$2&10.1$\pm$0.7&6.0$\pm$0.1&6.8$\pm$0.3& 12.38d&21.61i& V\\
J1717+1917&Un&11.5& &flat& &81.23h& & F\\
{\bf J1735+5049}&CSO&6.4$\pm$0.2&6.3$\pm$0.3&5.6$\pm$0.2&5.6$\pm$0.2&0.27h&2.07i&H\\
J1751+0939&CJ&8.5& &29.02$\pm$0.04& &8.75h& & V\\
{\bf J1800+3848} &Un&17$\pm$3&13$\pm$1&13.1$\pm$0.1& &1.92f& & H\\
J1840+3900 &Un&5.7$\pm$0.5&5.2$\pm$0.4&flat& &6.87f& & F\\
{\bf J1850+2825} &MR&9.1$\pm$0.3&9.5$\pm$0.3&9.9$\pm$0.4& &2.15f& & V\\
{\bf J1855+3742} &CSO&4.00$\pm$0.07&3.81$\pm$0.06&4.0$\pm$0.5& &1.32f& & H\\
{\bf J2021+0515} &CJ&3.75$\pm$0.08&4.5$\pm$0.1&3.7$\pm$0.3& &9.94f& & V\\
J2024+1718 &Un&14$\pm$2&8.6$\pm$0.4&6.6$\pm$0.4& &14.51f& & V\\
J2101+0341&Un&17$\pm$2&3.7$\pm$0.2&flat&flat&28.29c&17.00f& F\\
J2114+2832&CJ&9.8& &flat& &34.76d & & V \\
J2123+0535 &CJ&18$\pm$4&flat&flat& &112.74f& & F\\
&&&&&&&&&&\\
\hline
\end{tabular} 
\vspace{0.5cm}
\end{center}   
\caption{Peak frequencies and flux density variability between 
the different epochs of the 51 sources observed with the VLA
  during the observing runs presented in this paper. The variability
indices have been computed by comparing the flux densities 
measured at each epoch with the mean value obtained by averaging all the
available observing epochs, as described in Section 3.2 (see Col. 7 and 8). 
Col. 1: source name (J2000); 
Col. 2: Morphological classification from pc-scale information
(Orienti et al. \cite{mo06})
Col. 3 and 4: peak
frequency of the first two epochs (Dallacasa et al. \cite{dd00}; Tinti et al. \cite{st05}); Col. 5
and 6: peak frequency from these new VLA data. 
Col. 7 and 8: the variability index computed by comparing the flux
  densities measured in
  2003/2004 and presented in this paper with those
  averaged on all the available epochs, as described in Section 3.2. 
  Since the data flux densities used in Column 7 and 8 were
  collected during different observing runs, near the value of the
  variability we report also the observation code from Table
  $\ref{obslog}$, which indicates for each source the date
  of the observations.
Col. 9: The classification
of the source spectra:
H= sources without significant variability, see Section 3.2; 
V= variable (i.e. with a variability index V$>$3, as
  defined in Section 3.2 at least during one epoch); F= flat spectrum
(i.e. sources with a flat spectrum at least during one epoch; see Section
  3.1). Sources marked in boldface are those still considered candidate HFPs.}
\label{variability}
\end{table*}  

\addtocounter{table}{-1}
\begin{table*}
\begin{center}
\begin{tabular}{ccccccccccc}
\hline
Source&Morph.&$\nu_{\rm ep1}$&$\nu_{\rm ep2}$&$\nu_{\rm ep3}$&$\nu_{\rm ep4}$&V$_{\rm
  ep1}$&V$_{\rm ep2}$&Var.\\  
(1)&(2)&(3)&(4)&(5)&(6)&(7)&(8)&(9)\\
\hline
&&&&&&&&&&\\
J2136+0041&CJ&5.0& &5.52$\pm$0.01&5.73$\pm$0.02&2.88c&0.37f &H\\
{\bf J2203+1007} &CSO&4.86$\pm$0.07&5.0$\pm$0.1&4.6$\pm$0.7& &1.94c& & H\\
J2207+1652 &Un&7.4$\pm$0.3&3.5$\pm$0.3&flat&flat& 36.01d&35.45e& F\\
J2212+2355 &Un&13$\pm$2&9$\pm$1&flat&flat& 43.10d&40.29e& F\\
J2257+0243 &Un& $>$22&$>$22& flat &13.6$\pm$0.3&3.27c&1.92e& F\\
J2320+0513 &Un&5.4$\pm$0.2&flat&9.2$\pm$0.2&flat&23.84c&5.75e& F\\
J2330+3348&MR&5.6$\pm$0.3&flat&flat& &18.84d & & F\\
&&&&&&&&&&\\
\hline
\end{tabular} 
\vspace{0.5cm}
\end{center}   
\caption{Continued.}
\label{variability}
\end{table*} 

\section{Discussion}

A multi-epoch monitoring of the spectral characteristics provides
information on the evolution of the radio spectra.
From the analysis of the spectral shape we
  find 12 sources (see Section 3.1) showing a flat spectrum during
  at least one of the observing epochs presented in this paper, 
  in addition to the 7 flat-spectrum objects
  already rejected by Tinti et al. (\cite{st05}). By comparing the
  frequency break obtained at different epochs we also find 
  that in 3 sources  
  (see Section 3.3) the break has noticeably shifted. Such a change can be
  explained only assuming a beamed nature for these sources, and
  therefore we reject them from the sample of genuine young HFPs.
  If we also consider the 7 sources (5 of them even with a flat
  spectrum) found with a Core-Jet morphology
  by Orienti et al. (\cite{mo06}) and then rejected,
  we obtain that only 31 of the 55 sources of the sample 
  can still be
  considered HFP candidates, and they are marked in boldface in
  Table $\ref{variability}$.

Although the identification of contaminant objects is the main goal,
a multi-epoch monitoring program can provide important information on
the spectral evolution in young radio sources.\\
We assume that young radio sources are described by a continuous
injection model, where the radio emission is
continuously replenished by a
constant flow of fresh relativistic particles with a power-law
energy distribution. However, the overall
spectra of such sources, characterized by a convex shape, 
show significant departure from the expected classical
power-law (S $\propto$ $\nu^{- \alpha}$). 
The deviations are well explained by two different
phenomena: 
synchrotron self-absorption causes the rising spectrum (S
$\propto$ $\nu^{5/2}$) at
frequencies below the peak, while synchrotron losses 
steepen the spectrum (S $\propto$ $\nu^{-(\alpha +0.5)}$) at high frequencies
(Pacholczyk \cite{pacho70}). In very young radio sources such as
the HFPs, such a steepening of the spectrum occurs at very high
frequency implying that synchrotron losses cannot be investigated with our
VLA observations.\\ 
In the following discussion we investigate the nature of the radio
spectrum and its evolution.\\

\subsection{Adiabatic expansion}

At frequencies below the turnover, 
the adiabatic expansion is the main mechanism that influences the
spectral evolution.\\
In this regime we have:\\

\begin{equation}
S( \nu) \propto B^{- \frac{1}{2}} \nu^{ \frac{5}{2}} \theta^{2}
\label{flux}
\end{equation}

where $B$ is the magnetic field, $\nu$ the frequency and $\theta$ the
angular size of the emitting region.\\
As the radio source adiabatically expands, the opacity decreases, the
turnover moves to lower frequency and the flux density increases.\\
Is it possible to detect such an increment in our multi-epoch spectra?\\
First, we assume that the radio emission is due to a
homogeneous component that is adiabatically expanding at a constant
rate:\\ 
\begin{equation}
\theta = \theta_{0} \left( \frac{t_{0}+ \Delta t}{t_{0}} \right)
\label{expansion}
\end{equation}

where $\theta_0$ is the angular size at the time $t_{0}$, 
and $\theta$ at the time $t_{0}$+$\Delta t$,
and the magnetic field is frozen in the
plasma:\\

\begin{equation}
B = B_{0} \left( \frac{t_{0}}{t_{0}+ \Delta t} \right)^{2}
\label{magnetic}
\end{equation}

where $B_{0}$ is the magnetic field at the epoch $t_{0}$, 
while $B$ is at the time $t_{0}+ \Delta t$.\\
From Equation \ref{flux}, the flux density at the frequency $\nu$ at
a given time $t$ is:\\

\begin{equation}
S \propto B_{0}^{- \frac{1}{2}} \nu^{ \frac{5}{2}} \theta_{0}^{2}
\left( \frac{t_{0}+ \Delta t}{t_{0}} \right)^{3}.
\end{equation}

The turnover frequency is:\\

\begin{equation}
\nu_{\rm t,0} \propto B_{0} E_{0}^{2}
\label{synchro}
\end{equation}

\noindent where $E_{0}$ is the energy of the relativistic particles.  
During the expansion of the homogeneous synchrotron-emitting region
$E \propto t^{-1}$ (Pacholczyk \cite{pacho70}), which implies that\\

\begin{equation}
\nu_{\rm t,1} = \nu_{\rm t,0} \left( \frac{t_{0}}{t_{0} + \Delta t}
\right)^{4}.
\label{shift_peak}
\end{equation}

If $t-t_{0}$=$\Delta t$ $\ll$ $t_{0}$, the flux density
and the turnover frequency can be considered approximately constant.
We consider $\Delta t$ $\sim$ 5 years (the time elapsed between the
first and last observing run), 
and $t_{0}$ $>$ 15 years, since
all the sources are part of the 87GB sample. If $t_{0}$ = 20
years, we 
expect a flux density increment of a factor of $\sim$ 2 in the optically
thick part of the spectra. Such an increment becomes undetectable for
sources with ages of $\sim$ 100 years. 
On the other hand, the $\nu_{t}$ should have decreased by about 70\%
of its previous value.
If $t_{0}$ $\sim$ 100 years, the decrement should be less than
20\% of its previous value, 
i.e. a turnover frequency of 22 GHz moves to 18 GHz, making the
detection rather difficult due to the poor frequency coverage available. 
This result is based on the strong assumptions that the magnetic field is
frozen in a homogeneous component which is adiabatically expanding.
However, it is possible that either the time-variable magnetic field is not
frozen in the plasma, or the radio emitting region is not
homogeneous. In both cases the flux density increment and the
frequency decrement are expected to
be even less significant.\\
Therefore, we conclude that adiabatic expansion is not able to
produce any significant changes in the optically thick
part of the spectrum detectable in our multi-epoch observations,
perhaps with the exception of a few sources, such as J0650+6001
(Section 4.3) where a tentative flux density increment at a frequency
below the spectral peak has been found, or the ``faint'' HFP J1459+3337
(Orienti \& Dallacasa, submitted). \\

\subsection{The nature of the overall spectra}    

\begin{figure}
\begin{center}
\includegraphics{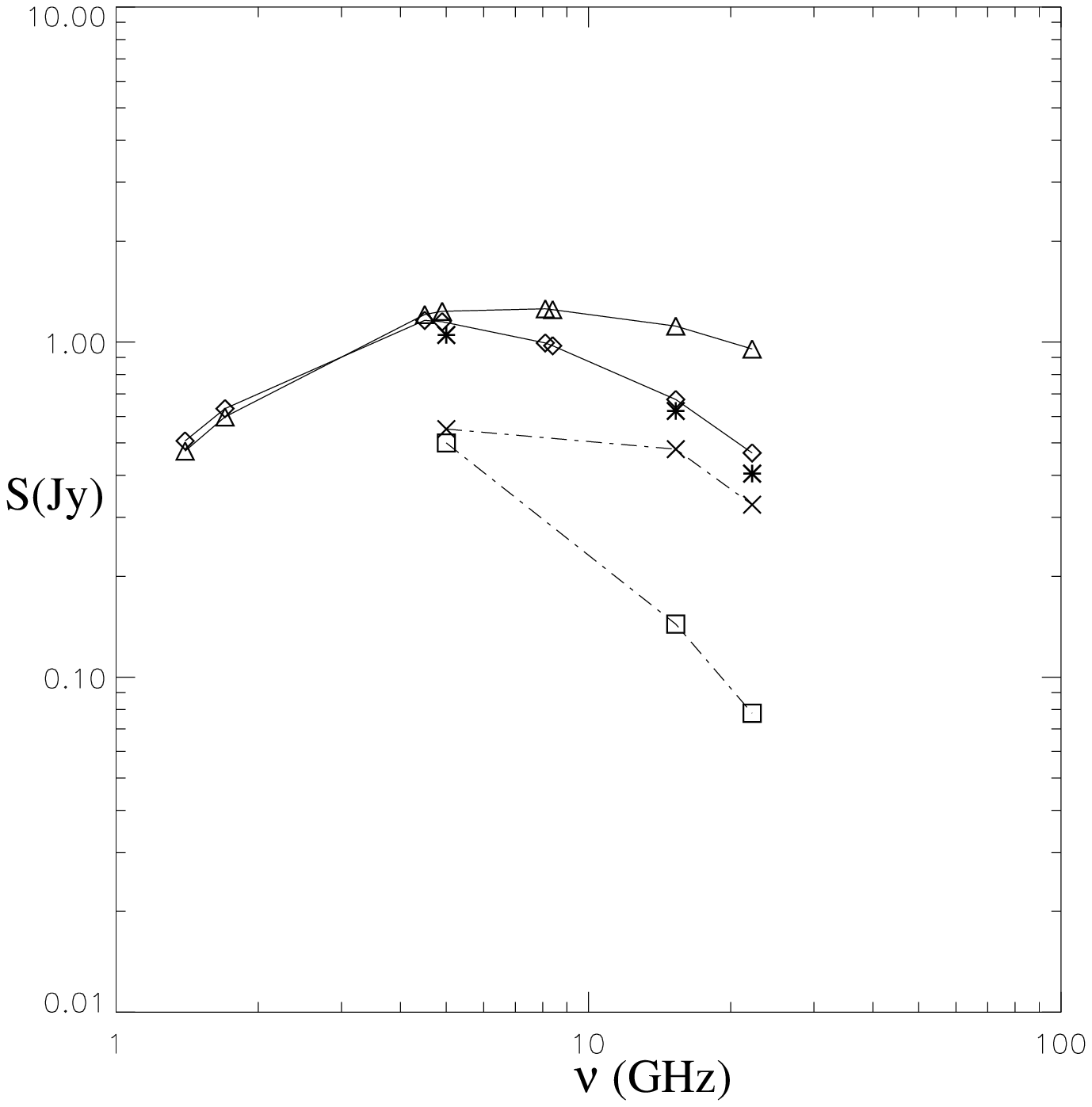}
\includegraphics{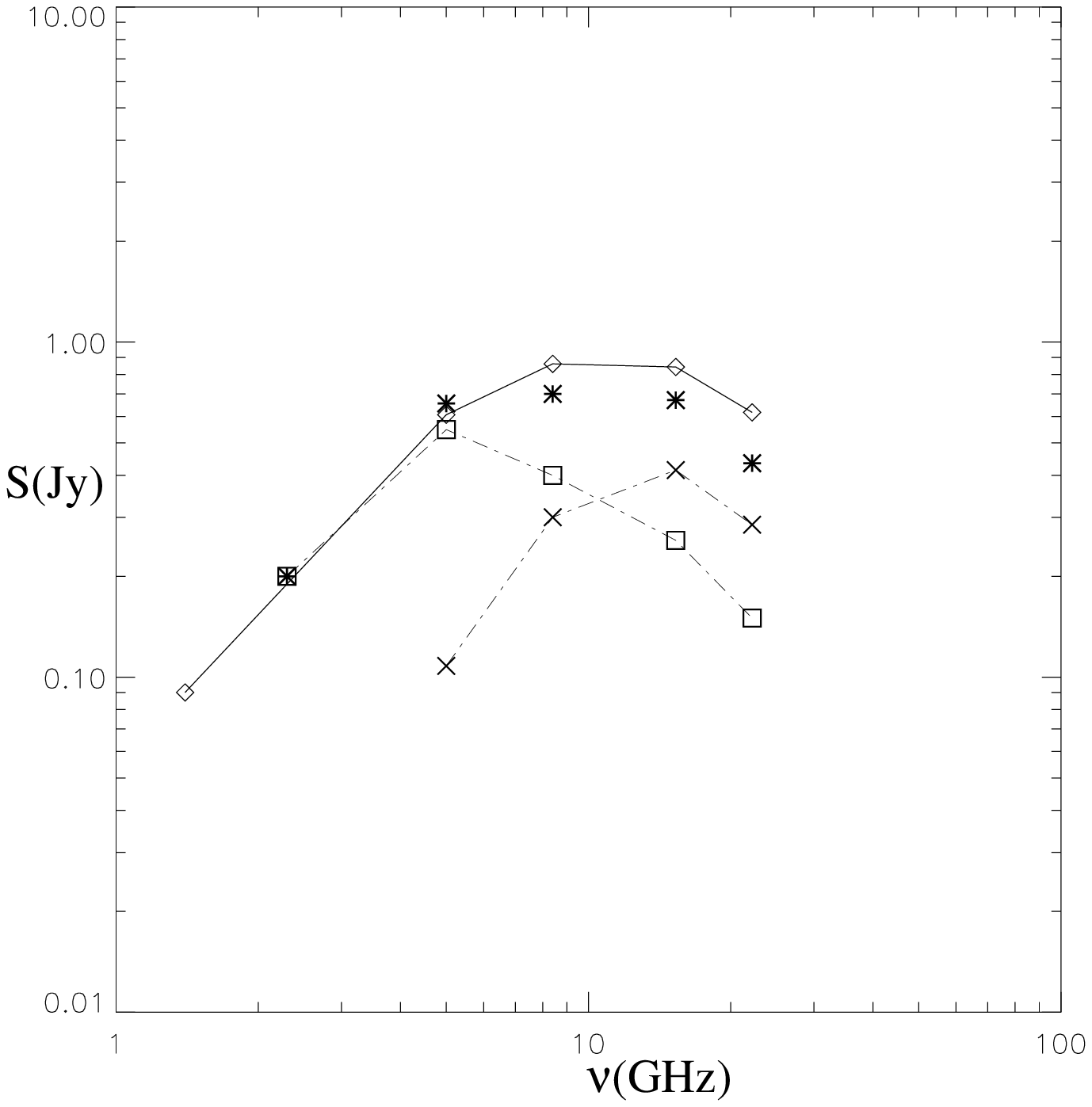}
\vspace{13.0cm}
\caption{The spectra of the sources J0650+6001 ({\it top}) and
  J1511+0518 ({\it bottom}): diamonds represent the overall spectrum
  obtained with these new VLA observations,
  while crosses and squares represent the spectra of the two source
  components (Xiang et al. \cite{xiang02}; Xu et
  al. \cite{xu95}; Orienti et
  al. \cite{mo06}). In the case of J0650+6001, triangles show the overall
  spectrum from Dallacasa et al. (\cite{dd00}). The asterisks represent
  the sum of the spectra of the single components. In the case of
  J1511+0518 not all the flux density could be recovered by VLBI
  images (Orienti et al. \cite{mo06}).}
\label{0650}
\end{center}
\end{figure}

In the previous Sections we have assumed that the radio emission is due to
a partially opaque electron synchrotron radiation which originates
within a homogeneous region. Variations in the opacity throughout the
emitting region lead to an overall spectrum 
which is the superposition of the spectra of many single regions. \\ 
Figure \ref{0650} shows the overall spectrum (diamonds) of
J0650+6001 and J1511+0518, 
together with the spectra of each single
component (squares and crosses), as revealed by VLBI observations
(Orienti et al. \cite{mo06}). In both sources, pc-scale resolution
images show that the radio emission originates within two main
components, which can be interpreted as the lobes of a mini radio source.\\
The overall
spectra are easily explained as the result of the superposition of 
the spectra of individual components. In both
sources, the two radio emitting regions have different turnover
frequencies, and, thus,
different parts of the overall spectra are influenced more by one
component, rather than the other, as is easily seen in the case of
J1511+0518. As a consequence, changes in the total spectra
may strongly depend on the evolution of the emission of each single
component.\\
The variation of the total spectrum that occurred over 5 years makes
J0650+6001 an interesting case study.\\ 
If we compare the overall spectra measured at different epochs, we
find that at 1.4 GHz, well below the peak frequency, the flux density
appears to be increasing, although within the errors, as expected by
adiabatic expansion.\\
On the other hand, at frequencies higher than the peak the flux
density decreases.
During the first epoch of observation, the optically thin spectrum 
was described
by a power-law  with $\alpha$ $\sim$
0.3, while during the subsequent observing runs it displayed $\alpha$
$\sim$ 0.7. Such a steepening  
cannot be directly related to a shift of
the break frequency of the overall spectrum, but is rather due to
the evolution of the spectra of the individual components.\\

\section{Conclusions}
We have presented the results of new epochs of high-sensitivity
simultaneous multi-frequency VLA observations of a sample of young 
HFP candidates.
By considering the spectral shape and the change of the frequency break
we find 15 sources which turned out to be blazar
objects, in addition to the 7 already found with a flat
spectrum by Tinti et al. (\cite{st05}). 
If we consider also the 7 sources rejected by
Orienti et al. (\cite{mo06}) on the basis of their Core-Jet
morphology, and 5 of them also with a flat spectrum,
we find that 24 of the 55 sources from the HFP sample are contaminant
blazar objects, and thus only 31 objects ($\sim$56\%)
can still be considered HFP candidates.\\
The comparison of the variability properties with the optical
identification have shown that quasars and galaxies display different
characteristics. If in the case of the turnover frequency distribution 
such a segregation is due to redshift ranges (quasars
are found at systematically higher redshifts than galaxies), the
difference in the variability index likely reflects two kinds of 
radio source populations. An even stronger segregation in the
flux-density and spectral-shape variability is found between objects
with or without a CSO-like morphology.\\  
Since polarization properties are another effective tool to discriminate
young radio sources from contaminant blazars, polarimetric data
concerning the sources of the ``bright'' HFP sample have
been studied, and they will be presented in a companion paper
(Orienti \& Dallacasa 2007, Paper II).\\  
Making use of the several epochs available, we have tried to
investigate the spectral evolution of the radio emission of
genuine young radio sources.\\ 
At frequencies below the peak the spectral evolution is dominated by
adiabatic expansion, and we would expect a flux density
increment, while at frequencies above the spectral peak, 
in addition to the decrease of the flux density due to adiabatic
expansion, synchrotron losses also may play a role.
Considering the time elapsed between the two farthest spaced observing runs
(5 years), we conclude that a variation in the optically-thick part of
the overall spectrum
would have been detected only
if the radio emission would have started very recently,
i.e. $\sim$ 20 years, which is statistically unlikely for the sources
in the sample.\\
On the other hand, as seen in the analysis of the spectra of individual
objects, there are a few
cases, such as the source J0650+6001, with spectral variations. 
However, if we consider that the radio emission comes from a
single homogeneous region, 
we are not able to explain such changes. Combining
the low-resolution, flux density information provided by VLA
observations with the pc-scale resolution from VLBA data, we can 
describe the total radio spectrum  
as the result of the superposition of the spectra of different 
emitting regions.\\ 
In general, to determine the evolution of the radio emission, we must
resolve each single source component. \\
For this, new simultaneous multi-frequency VLBA
observations in both the optically thick and thin parts of the spectrum
have been carried out for 5 genuine HFPs, with a CSO-like
morphology. The analysis of the radio spectrum of each single
component will enable us to set strong constraints on the fate of the
radio emission.\\

\begin{acknowledgements}
We thank the referee Merja Tornikoski for carefully reading the
manuscript and valuable suggestions.
The VLA is operated by the U.S. National Radio Astronomy Observatory
which is a facility of the National Science Foundation operated under
a cooperative agreement by Associated Universities, Inc.
This work has made use of the
NASA/IPAC Extragalactic Database (NED), which is operated by the Jet
Propulsion Laboratory, California Institute of Technology, under
contract with the National Aeronautics and Space Administration.\\
\end{acknowledgements}

\end{document}